\documentclass{aa}  
\usepackage{graphicx}
\usepackage{txfonts}
\usepackage{float}
\usepackage{subfig}
\defcitealias{Quirola2022}{Paper~I}
\begin{document} 

   \titlerunning{Extragalactic FXT Candidates Discovered by \emph{Chandra} (2014--2022)}
   \authorrunning{Quirola-V\'asquez et al.}

   \title{Extragalactic Fast X-ray Transient Candidates Discovered by \emph{Chandra} (2014--2022)}
   
   \author{J. Quirola-V\'asquez
          \inst{1,2,3,4},
          F. E. Bauer\inst{1,2,5},
          P.~G. Jonker\inst{3,6},
          W. N. Brandt\inst{7,8,9},
          G. Yang\inst{10,11},
          A. J. Levan\inst{3,12},
          Y. Q. Xue\inst{13,14},
          D. Eappachen\inst{6,3},
          E. Camacho\inst{2,1},
          M. E. Ravasio\inst{3,15},
          X. C. Zheng\inst{16},
         \and
          B. Luo\inst{17,18}
          }
   \institute{Instituto de Astrof\'isica, Pontificia Universidad Cat\'olica de Chile, Casilla 306, Santiago 22, Chile\\
              \email{jaquirola@uc.cl}
         \and
             Millennium Institute of Astrophysics (MAS), Nuncio Monse$\tilde{\rm n}$or S\'otero Sanz 100, Providencia, Santiago, Chile
        \and
             Department of Astrophysics/IMAPP, Radboud University, P.O. Box 9010, 6500 GL, Nijmegen, The Netherlands
        \and
             Observatorio Astron\'omico de Quito, Escuela Polit\'ecnica Nacional, 170136, Quito, Ecuador
        \and
            Space Science Institute, 4750 Walnut Street, Suite 205, Boulder, Colorado 80301, USA
        \and
            SRON Netherlands Institute for Space Research, Niels Bohrweg 4, 2333 CA Leiden, The Netherlands
        \and
            Department of Astronomy \& Astrophysics, 525 Davey Laboratory, The Pennsylvania State University, University Park, PA 16802, USA
        \and
            Institute for Gravitation and the Cosmos, The Pennsylvania State University, University Park, PA 16802, USA
        \and
            Department of Physics, 104 Davey Laboratory, The Pennsylvania State University, University Park, PA 16802, USA
        \and
            Kapteyn Astronomical Institute, University of Groningen, P.O. Box 800, 9700 AV Groningen, The Netherlands
        \and
            SRON Netherlands Institute for Space Research, Postbus 800, 9700 AV Groningen, The Netherlands
        \and
            Department of Physics, University of Warwick, Coventry, CV4 7AL, UK
        \and
            CAS Key Laboratory for Research in Galaxies and Cosmology, Department of Astronomy, University of Science and Technology of China, Hefei 230026, China
        \and
            School of Astronomy and Space Science, University of Science and Technology of China, Hefei 230026, China
        \and
            INAF – Brera Astronomical Observatory, via Bianchi 46, I–23807 Merate (LC), Italy
        \and
            Leiden Observatory, Leiden University, PO Box 9513, NL-2300 RA, Leiden, the Netherlands
        \and
            School of Astronomy and Space Science, Nanjing University
        \and
            Key Laboratory of Modern Astronomy and Astrophysics (Nanjing University), Ministry of Education, Nanjing 210093, China
             }

   \date{Received January 13, 2023; accepted XX XX, 2023}

  \abstract
    {Extragalactic fast X-ray transients (FXTs) are short flashes of X-ray photons of unknown origin that last a few minutes to hours.}
    { We extend the search for extragalactic FXTs from Quirola et al. 2022 (Paper~I; based on sources in the \emph{Chandra} Source Catalog 2.0, CSC2, using data taken between 2000--2014) to further \emph{Chandra} archival data between 2014--2022.}
   { We extract X-ray data using a method similar to that employed by CSC2 and apply identical search criteria as in Paper~I. }
    {We report the detection of eight FXT candidates, with peak 0.3--10 keV fluxes between 1$\times$10$^{-13}$ to 1$\times$10$^{-11}$~erg~cm$^{-2}$~s$^{-1}$ and $T_{90}$ values from 0.3 to 12.1~ks.
    This sample of FXTs has likely redshifts between 0.7 to 1.8. Three FXT candidates exhibit light curves with a plateau (${\approx}$1--3~ks duration) followed by a power-law decay and X-ray spectral softening, similar to what was observed for a few previously reported FXTs in Paper~I. In light of the new, expanded source lists (eight FXTs with known redshifts from Paper~I and this work), we update the event sky rates derived in Paper~I, finding 36.9$_{-8.3}^{+9.7}$~deg$^{-2}$~yr$^{-1}$ for the extragalactic samples for a limiting flux of ${\gtrsim}$1${\times}$10$^{-13}$~erg~cm$^{-2}$~s$^{-1}$, calculate the first FXT X-ray luminosity function, and compare the volumetric density rate between FXTs and other transient classes.}
   {Our latest \emph{Chandra}-detected extragalactic FXT candidates boost the total \emph{Chandra} sample by $\sim$50\%, and appear to have a similar diversity of possible progenitors.}

    
   \maketitle
%

\section{Introduction}\label{sec:intro}

The last decades have seen remarkable progress in understanding the time-resolved sky. Wide-field optical and near-infrared (NIR) surveys identified thousands of supernovae (SNe) and related sources. In the gamma-ray regime, the progenitors of both long- and short-duration gamma-ray bursts (LGRBs and SGRBs, respectively) have been identified, while in the radio bands decisive inroads have been made into the nature of the fast radio bursts (FRBs). Perhaps surprisingly, our understanding of sources with similar behavior observed in soft X-rays with \emph{Chandra}, \emph{XMM-Newton} and \emph{Swift}-XRT remains relatively poor. 
Phenomenologically, we define extra-galactic fast X-ray transients (FXTs) as non-Galactic sources that manifest as non-repeating flashes of X-ray photons in the soft X-ray regime $\sim$0.3--10~keV, with durations from minutes to hours \citep[e.g.,][]{Alp2020,Quirola2022}. Unfortunately, they still lack a concise or singular physical explanation \citep[e.g.,][]{Soderberg2008,Jonker2013,Glennie2015,Irwin2016,Bauer2017,Lin2018,Lin2019,Xue2019,Yang2019,Alp2020,Novara2020,Lin2020,Ide2020,Pastor2020,Lin2021,Lin2022,Quirola2022}.

Critically, while of order 30 FXTs have been identified to date, both serendipitously and through careful searches, only in one case, XRT~080109/SN~2008D \citep{Soderberg2008,Mazzali2008,Modjaz2009}, has there been a detection of a multi-wavelength counterpart after the outburst. This is because, in the vast majority of cases, the transients themselves have only been identified long after the outburst via archival data mining \citep[e.g.,][]{Alp2020,DeLuca2021,Quirola2022}, so that timely follow-up observations were not possible. Notably, the most stringent limits come from deep optical VLT imaging serendipitously acquired 80~minutes after the onset of XRT~141001 \citep[$m_R{>}$25.7~AB~mag;][]{Bauer2017}. Moreover, only a handful of FXTs have had clear host-galaxy associations and even fewer have firm distance constraints \citep[e.g.,][]{Soderberg2008,Irwin2016,Bauer2017,Xue2019,Novara2020,Lin2022,Eappachen2022,Quirola2022,Eappachen2023a}. Hence, it is not trivial to discern their energetics and distance scale and, by extension, their physical origin.

A variety of different physical mechanisms have been proposed for the origin of FXTs, such as: 
$i)$ stochastic outbursts associated with X-ray binaries in nearby galaxies [XRBs; including subclasses such as Ultra-luminous X-ray sources (ULXs), soft gamma repeaters (SGRs), and anomalous X-ray pulsars (AXPs)] providing possible explanations of FXTs with $L_{\rm X,peak}{\lesssim}$10$^{42}$~erg~s$^{-1}$ (see \citealp[][]{Colbert1999,Kaaret2006,Woods2006,Miniutti2019}; and references therein);
$ii)$ X-ray emission generated from the shock breakout (SBO; $L_{\rm X,peak}{\sim}$10$^{42}$--10$^{45}$~erg~s$^{-1}$) of a core-collapse supernova (CC-SN) once it crosses the surface of the exploding star \citep[e.g.,][]{Soderberg2008,Nakar2010,Waxman2017,Novara2020, Alp2020};
$iii)$ off-axis GRBs could explain FXTs ($L_{\rm X,peak}{\lesssim}$10$^{45}$~erg~s$^{-1}$) where the X-ray emission is produced by a wider, mildly relativistic cocoon jet \citep[Lorentz factor of ${\lesssim}$100;][]{Zhang2004b}, once it breaks through the surface of a massive progenitor star \citep{Ramirez2002,Zhang2004b,Nakar2015,Zhang_book_2018,Delia2018};
$iv)$ Tidal disruption events (TDEs; $L_{\rm X,peak}^{\rm no-Jet}{\lesssim}$10$^{43}$ and $L_{\rm X,peak}^{\rm Jet}{\sim}$10$^{43}$--10$^{50}$~erg~s$^{-1}$ considering non- and jetted emission, respectively) involving a white dwarf (WD) and an intermediate-mass black hole (IMBH), whereby \hbox{X-rays} are produced by the tidal disruption and subsequent accretion of part of the WD in the gravitational field of the IMBH \citep[e.g.,][]{Jonker2013,Glennie2015}; and
$v)$ Mergers of binary neutron stars, \citep[BNS; $L_{\rm X,peak}{\sim}$10$^{44}$--10$^{51}$~erg~s$^{-1}$ considering jetted and line-of-sight obscured emission; e.g.,][]{Dai2018,Jonker2013,Fong2015,Sun2017,Bauer2017,Xue2019}, whereby the X-rays are created by the accretion of fallback material onto the remnant black hole (BH), a wider and mildly relativistic cocoon, or the spin-down magnetar emission \citep{Metzger2014,Sun2017,Sun2019,Metzger2018b}.

In previous work, \citet[][hereafter Paper~I]{Quirola2022} conducted a systematic search for FXTs in the \emph{Chandra} Source Catalog \citep[Data Release 2.0; 169.6~Ms over 592.4~deg$^{2}$ using only observations with $|b|{>}10^{\circ}$ and until 2014;][]{Evans2010,Evans2019,Evans2020}, using an X-ray flare search algorithm and incorporating various multi-wavelength constraints to rule out Galactic contamination. Paper~I reported the detection of 14 FXT candidates \citep[recovering five sources previously identified and classified as FXTs by][]{Jonker2013,Glennie2015,Bauer2017,Lin2019} with peak fluxes ($F_{\rm peak}$) from 1$\times$10$^{-13}$ to 2$\times$10$^{-10}$~erg~cm$^{-2}$~s$^{-1}$ (at energies of 0.5--7 keV) and $T_{90}$ (measured as the time over which the source emits the central 90\%, i.e. from 5\% to 95\% of its total measured counts) values from 4 to 48~ks. Intriguingly, the sample was sub-classified into two groups: six \emph{"nearby"} FXTs that occurred within $d{\lesssim}$100~Mpc and eight \emph{"distant"} FXTs with likely redshifts $\gtrsim$0.1. Moreover, after applying completeness corrections, the event rates for the \emph{nearby} and \emph{distant} samples are 53.7$_{-15.1}^{+22.6}$ and 28.2$_{-6.9}^{+9.8}$~deg$^{-2}$~yr$^{-1}$, respectively. However,
Paper~I does not analyze \emph{Chandra} observations beyond 2014, implying that several intriguing FXTs likely remain undiscovered.

In this paper, we extend the selection of Paper~I to public \emph{Chandra} observations between 2014--2022 using a nearly identical methodology. As in Paper~I, this work focuses only on the non-repeating FXTs, to help reduce sample contamination. We further caution that the sparse nature of repeat X-ray observations means that we cannot rule out that some current FXTs could be repeating FXTs. The study of repeating FXTs is beyond the scope of this manuscript.

The manuscript is organized as follows. We explain the methodology and selection criteria in Sect.~\ref{sec:methodology}. We present the results of a search and cross-match with other catalogs in Sect.~\ref{sec:results}, a spectral and timing analysis of our final candidates in Sect.~\ref{sec:time_spectra_prop}, and the properties of the identified potential host galaxies in Sect.~\ref{sec:counterpart_SED}. In Sect.~\ref{sec:flux}, we discuss possible interpretations of some FXTs. We derive local and volumetric rates for the FXTs in Sect.~\ref{sec:rates}, and compare them to those of other X-ray transients. Finally, we present comments and conclusions in Sect.~\ref{sec:conclusion}. Throughout the paper, a concordance cosmology with parameters $H_0{=}$70~km~s$^{-1}$~Mpc$^{-1}$, $\Omega_M{=}$0.30, and $\Omega_\Lambda{=}$0.70 is adopted. Magnitudes are quoted in the AB system. Unless otherwise stated, all errors are at 1$\sigma$ confidence level.

\section{Methodology and Sample Selection}\label{sec:methodology}

\subsection{Identification of X-ray sources} \label{sec:id}

Paper~I used as an input catalog of the X-ray sources detected by the CSC2. This is not available for \emph{Chandra} observations beyond the end of 2014, so a crucial first step is to generate a comparable source detection catalog for the \emph{Chandra} observations used in this work (see Sect.~\ref{sec:data}), upon which we will apply our FXT  candidate selection algorithm (Sect.~\ref{sec:algorithm}). 

To generate robust X-ray source catalogs, we use the {\sc ciao}
source detection tool \texttt{wavdetect} \citep{Freeman2002}. It detects possible source pixels using a series of ``Mexican Hat'' wavelet functions with different pixel bin sizes to account for the varying PSF size across the detector. The \texttt{wavdetect} tool identifies all point sources above a threshold significance of $10^{-6}$ (which corresponds to one spurious source in a 1000$\times$1000 pixel map) and a list of radii in image pixels from 1 to 32 (to avoid missing detections at large off-axis angles). To avoid erroneous detections, we create exposure and PSF maps, which enable refinement of the source properties. The exposure maps are created by running the \texttt{fluximage} script with the 0.5--7~keV band \citep{Fruscione2006}, while the PSF map, which provides information on the size of the PSF at each pixel in the image, is made using the \texttt{mkpsfmap} task; the PSF size corresponds to the 1$\sigma$ integrated volume of a 2D Gaussian \citep{Fruscione2006}. The output of the {\sc ciao} tool \texttt{wavdetect} is a catalog with essential information about the X-ray sources such as the positions (RA and DEC), positional uncertainty, and significance.

\subsection{Transient-Candidate Selection Algorithm} \label{sec:algorithm}

We adopt the same algorithm as presented in Paper~I, which augments somewhat the initial version presented in \citet[][see their sect.~2.1 for more details]{Yang2019}. This method depends on the total ($N_{\rm tot}$) and background ($N_{\rm bkg}$) counts of the source, working on an unbinned \emph{Chandra} light curve, which is advantageous because it does not depend on the light curve shapes. The algorithm splits the light curves into different segments in two passes: $i)$ in two halves and $ii)$ in three regions, covering the entire \emph{Chandra} observation. FXT candidates are selected when: $i)$ $N_{\rm tot}{>}$5-$\sigma$ Poisson upper limit of $N_{\rm bkg}$ (to exclude low signal-to-noise ratio, S/N, sources); $ii)$ the counts in the different segments ($N_i$) are statistically different at a ${>}$~4$\sigma$ significance level (to select robust detections of short-duration variable sources); and $iii)$ $N_i{>}5\times N_j$ or $N_j{>}5\times N_i$ (to select large-amplitude number of counts-variations).

Finally, to mitigate the effect of background (especially for sources with long exposure times and large instrumental off-axis angles), we additionally chop each light curve into 20~ks segments (or time windows $T_{\rm window}{=}$~20~ks), and re-apply the conditions explained above. This reduces the integrated number of background counts per PSF element and thus enables identification of fainter sources at larger instrumental off-axis angles. To maintain an efficient selection of transients across the gaps between these arbitrary windows, we sequence through the entire light curve in three iterations: a forward division in 20~ks intervals, a backward division in 20~ks intervals, and finally, a forward division with a 10~ks shift in 20~ks intervals to cover gaps.

Based on simulations of the CDF-S XT1 and XT2 fiducial light curves \citep{Bauer2017,Xue2019}, Paper~I derived an efficiency of the method of ${\gtrsim}$~90\% for sources with $\log(F_{\rm peak})~{>}-12.6$ located at off-axis angles ${<}$~11\farcm0, with a relatively sharp decline in efficiency for FXTs with lower fluxes; e.g., ${\approx}$50\% and ${\approx}$5\% efficiencies for $\log(F_{\rm peak})~{=}-12.8$ and $\log(F_{\rm peak})~{=}-13.0$, respectively, at ${\approx}$~11\farcm0. This instrumental off-axis angle limit is enforced because \emph{Chandra}'s detection sensitivity (as measured by, e.g., effective area and PSF size) drops significantly beyond this limit \citep{Vito2016,Yang2016}. Importantly, this algorithm successfully recovered all previously reported sources \citep[XRT~000519, XRT~030511, XRT~110103, XRT~110919, and XRT~141001;][]{Jonker2013,Glennie2015,Bauer2017,Lin2019,Quirola2022}, and thus is flexible enough to recognize FXTs with different light-curve shapes. We stress that this is a key advantage compared to matched filter techniques that assume an underlying light curve model profile.

\begin{figure}
    \centering
    \includegraphics[scale=0.7]{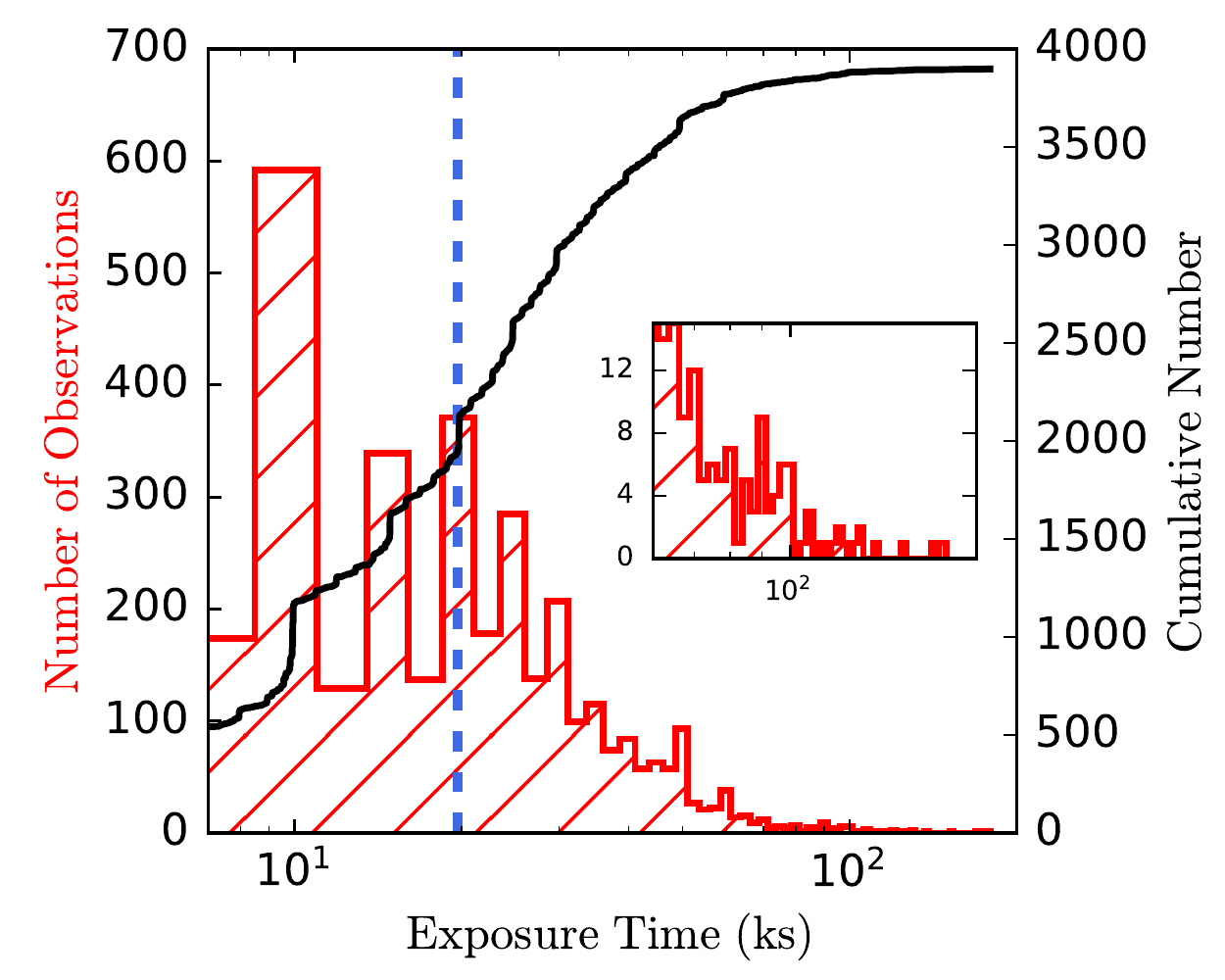}
    \vspace{-0.3cm}
    \caption{Histogram (\emph{red}; \emph{left} Y-axis) and cumulative (\emph{black}; \emph{right} Y-axis) distributions of the exposure time of the 3899 \emph{Chandra} observations used in this work. The inset provides a zoom-in to show the high-exposure time tail of the distribution. The \emph{dashed vertical blue line} indicates the median exposure time (${=}~19.7$~ks) of the total sample. 
    }
    \label{fig:OBS_features}
\end{figure}

\subsection{Data selection}\label{sec:data}

To extend the previous search for extragalactic FXTs in Paper~I beyond the \emph{Chandra Source Catalog 2.0} (CSC2) limit of 2014, we conducted a search through all \emph{Chandra} ACIS imaging observations (science and calibration observations) made publicly available between 2015-01-01 and 2022-04-01. This includes 3899 individual \emph{Chandra}-ACIS observations, outside the Galactic plane at $|b|{>}$10~deg, or ${\approx}$88.8~Ms, 264.4~deg$^2$ conforming to the following criteria. For uniformity, we consider only ACIS observations in the energy range 0.5--7.0~keV, noting that HRC-I observations comprise only a few percent of the overall observations and have poorer/softer response and limited energy resolution compared with the ACIS detector. The \emph{Chandra} observations target a wide variety of astronomical objects, from galaxy clusters to stellar objects. Based on the nature of the extragalactic FXTs identified systematically in Paper~I and the potential sources of contamination, we limit our initial light-curve search to sources with Galactic latitudes $|b|{>}~10$~deg to reduce the expectedly high contamination rate from flaring stars. An additional benefit of considering objects outside the Galactic plane is that it helps to minimize the effects of Galactic extinction in characterizing the spectral properties of our candidates.

To facilitate our search, we use the full-field per-observation \texttt{event files} available from the \emph{Chandra} Data Archive products.\footnote{https://cda.harvard.edu/chaser/} Figure~\ref{fig:OBS_features} shows the cumulative and histogram distributions of exposure time of the \emph{Chandra} observations used in this work.

\subsection{Generation of light curves} \label{sec:LC}

The \texttt{event file} contains the relevant stored photon event data, such as photon arrival time, energy, position on the detector, sky coordinates, observing conditions, and the good time interval (GTI) tables listing the start and stop times. 

To generate light curves, we take X-ray photons in the \hbox{0.5--7.0}~keV range from each \texttt{event file} using an aperture of 1.5$\times R_{90}$, where $R_{90}$ is the radius encircling 90\% of the \hbox{X-ray} counts. Based on simulations developed by \citet{Yang2019}, the aperture of 1.5$\times R_{90}$ encircles ${\gtrsim}$98\% of X-ray counts and depends on the instrumental off-axis angle \citep[and depends on the photon energy; for more details, see][]{Vito2016,Hickox2006}. We compute $N_{\rm bkg}$ taking into account an annulus with inner and outer aperture radius of 1.5${\times}R_{90}$ and 1.5${\times}R_{90}$+20 pixels, respectively. In the particular case where the background region overlaps with a nearby X-ray source, we mask the nearby source (using a radius of 1.5${\times}R_{90}$), and do not include the masked area to estimate $N_{\rm bkg}$. Also, we weigh $N_{\rm bkg}$ by the source-to-background area ratio to correct the light curve of the sources. 

\subsection{Astrometry of X-ray sources} \label{sec:astrometry}

To improve upon the nominal absolute astrometric accuracy of {\it Chandra} [0\farcs8 (1\farcs4) at 90\% (99\%) uncertainty]\footnote{https://cxc.harvard.edu/cal/ASPECT/celmon/}, we cross-match the detected X-ray sources to optical sources from either the \emph{Gaia} Early Data Release 3 \citep[\emph{Gaia}-EDR3;][]{Gaia_DR3_2020} 
or Sloan Digital Sky Survey Data Release 16 \citep[SDSS--DR16;][]{Ahumada2019}
catalogues, using the \texttt{wcs\_match} script in {\sc ciao}. \texttt{wcs\_match} compares two sets of source lists from the same sky region and provides translation, rotation and plate-scale corrections to improve the X-ray astrometric reference frame. We adopt a 2\farcs0 matching radius (i.e, $\leq$8 image pixels), eliminating any source pairs beyond this limit. We typically achieve an accuracy of 0\farcs08--1\farcs64 (90\% quantile range). This improves our ability to discard contaminants (stellar flares, essentially) and eventually measure projected offsets between X-ray sources and host galaxies (in the case of the final sample of FXT candidates). We combine in quadrature all astrometric errors into the X-ray source positional uncertainty.

\begin{figure*}
    \centering
    \includegraphics[scale=0.65]{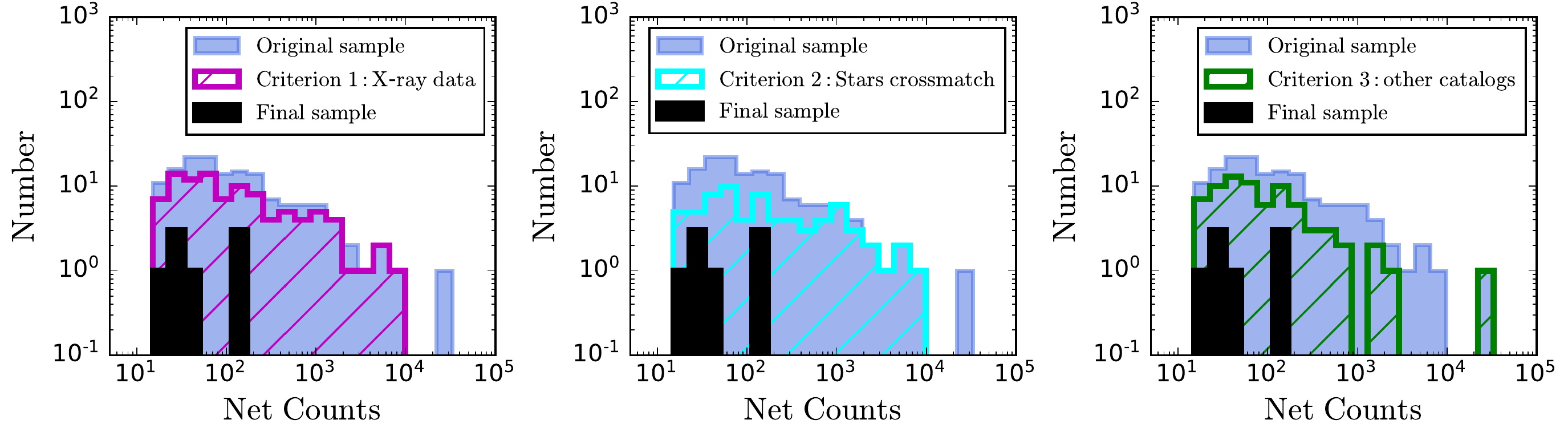}
    \caption{Comparison of 0.5--7.0~keV net-count distributions for the initial (\emph{filled blue histograms}) and final (\emph{filled black histograms}) FXT samples, as well as subsets covered by various purity criteria (\emph{colour non-filled histograms}) for the sample. Net counts are provided by the same regions defined in Sect.~\ref{sec:LC}.}
    \label{fig:histo}
\end{figure*}

\begin{table*}
    \centering
    \begin{tabular}{lllll}
    \hline\hline
        Criterion & \multicolumn{4}{c}{Candidates} \\
                  & \# constrained & \# total removed & \# uniquely removed & \# remaining \\ \hline
         $(a)$ & $(b)$ & $(c)$ & $(d)$ & $(e)$ \\ \hline
        1) Archival \hbox{X-ray} data & 134$^*$ & 53 & 20 & 98 \\ 
        2) Cross-match with stars/\emph{Gaia} & 151 & 83 & 42 & 42 \\
        3) NED + SIMBAD + VizieR & 151 & 75 & 33 & 9 \\
        \hline
        4) Archival images$^{\dagger}$ & -- & -- & -- & 9 \\
        5) Instrumental and variability effects$^{\dagger}$ & -- & 1 & 1 & 8 \\
    \hline
    \end{tabular}
    \caption{Breakdown of FXT candidates as a function of the selection criteria proposed in Sect. \ref{sec:purity_criteria}. 
    \emph{Column $(a)$:} Criterion. 
    \emph{Column $(b)$:} Number of candidates constrained by this criterion.
    \emph{Column $(c)$:} Number of candidates removed that would be cut at this stage if we disregard all previous stages.
    \emph{Column $(d)$:} Number of candidates that are solely removed by this criterion, and not any other.
    \emph{Column $(e)$:} Running total number of candidates that remain after applying this criterion. \\
    $^*$Candidates with additional \emph{Chandra}-ACIS, \emph{XMM-Newton}, or \emph{Swift}-XRT observations. \\
    $\dagger$Note that criteria 4 and 5 are only applied to the sources that remain after the first three criteria are applied.}
    \label{tab:criteria}
\end{table*}

\subsection{Initial candidate results} \label{sec:initial_results}

As a summary, we apply the FXT detection algorithm to the 0.5--7.0 keV light curves of X-ray sources outside of the Galactic plane ($|b|>$10~deg), resulting in 151 FXT candidates. This parent sample has total net counts and instrumental off-axis angles spanning ${\approx}$15--33,000 (mean value of 590) and ${\approx}$0.12--14.0 (mean value of 5.2)~arcmin, respectively. As expected, our selection method identifies FXTs with a wide range of light curve shapes.

\subsection{Initial Purity Criteria}\label{sec:purity_criteria}

As highlighted in both \citet{Yang2019} and Paper~I, our search method does not ensure the unique identification of real extragalactic FXTs. Therefore, it is mandatory to adopt additional criteria considering archival X-ray data and multiwavelength counterparts to differentiate real extragalactic FXTs from Galactic variables and transients among the sample of 151 FXT candidates. We describe and report these additional criteria in Sect. \ref{sec:previous_criteria}--\ref{sec:inst_effects} and summarize the number and percentage, relative to the total, of sources that pass criteria (\emph{column 5}), as well as ignoring all previous steps (\emph{column 4}) in Table~\ref{tab:criteria}. Finally, we discuss the completeness of our search and selection criteria in Sect. \ref{sec:completness}.

\subsubsection{Criterion 1: Archival X-ray data} \label{sec:previous_criteria}

To confirm the transient nature of the FXT candidates, a non-detection in prior and/or subsequent X-ray observations is important. In this way, we consider different observations from \emph{Chandra}, based on other observations in the CSC2 and individual observations \citep{Evans2010}; \emph{XMM-Newton}, based on individual observations of sources in the Serendipitous Source \citep[4XMM-DR11;][]{Webb2020} and Slew Survey Source Catalogues \citep[XMMSL2;][]{Saxton2008}; and the Living \emph{Swift} XRT Point Source Catalogue (LSXPS) based on observations between 2005-01-01 and 2023-02-12 \citep{Evans2023}. We impose that the FXT candidate remain undetected (i.e., consistent with zero net counts) at 3$\sigma$ confidence in all X-ray observations, aside from the \emph{Chandra} observation in which the FXT candidate is detected. This requirement is useful especially to exclude a large number of Galactic stellar flares, but it also may discard FXTs associated with hosts with AGNs, as well as long-lived or recurring X-ray transients (e.g., from SNe in strongly star-forming galaxies). The success of this criterion is related to the number of times a particular field is visited by X-ray facilities. 

To discard candidates with prior and subsequent X-ray observations with \emph{Chandra}, we used the CSC2 or in the cases of candidates with more recent archival observations we downloaded and extracted photometry for these sources, adopting consistent source and background regions and aperture corrections compared to those used in Sect.~\ref{sec:LC}. In total, 127 FXT candidates were observed in multiple \emph{Chandra} observation IDs, while 24 candidates have no additional \emph{Chandra} observations.

To identify additional \emph{XMM-Newton} and \emph{Swift}-XRT detections, we adopt a search cone radius equivalent to the 3$\sigma$ combined positional errors of the \emph{Chandra} detection and tentative \emph{XMM-Newton} or \emph{Swift}-XRT matches from the 4XMM-DR11, XMMSL2 and LSXPS catalogs, respectively. We additionally search the X-ray upper limit servers:
Flux Limits from Images from \emph{XMM-Newton} using DR7 data (FLIX),\footnote{https://www.ledas.ac.uk/flix/flix.html} LSXPS,\footnote{https://www.swift.ac.uk/LSXPS/}
and the HIgh-energy LIght curve GeneraTor (HILIGT) upper limit servers.\footnote{http://xmmuls.esac.esa.int/upperlimitserver/} 
It is important to mention that HILIGT provides upper limits for several X-ray observatory archives (including \emph{XMM-Newton} pointed observations and slew surveys; R\"ontgen Satellite (\emph{ROSAT}) pointed observations and all-sky survey; \emph{Einstein} pointed observations), while LSXPS generates \emph{Swift}-XRT upper limits.\footnote{We used the 0.2--12 keV energy band, which we then converted to 0.5--7.0~keV assuming the default spectral parameters $\Gamma{=}2$ and $N_H{=}3{\times}10^{20}$~cm$^{-2}$.}

We found that the reported detections are not always reliable (e.g., inconsistencies between catalogs using the same observations or failure to confirm upon visual inspection), and hence we require detections to be ${\geq}$5$\sigma$.
We found that: 
72 candidates are observed in \emph{XMM-Newton} 4XMM-DR11, with 12 candidates detected;
%
%
65 candidates are observed in \emph{Swift}-XRT LSXPS, with 4 candidates detected;
1 candidate is observed in \emph{ROSAT} pointed observations, with a clear detection;
%
%
finally, all candidates are observed in the \emph{ROSAT} All-Sky Survey, with 5 candidates detected. Also, zero candidates are observed in \emph{XMM-Newton} XMMSL2 and the \emph{Einstein} pointed observations. 
The upper limits from \emph{Chandra} and \emph{XMM-Newton} pointed observations are similar to or lower than our FXT candidate peak fluxes. So, we can conclude that similar transient episodes would have been detectable in such observations if present. 

In total, 134 candidates have multiple hard X-ray observations by \emph{Chandra}, \emph{XMM-Newton}, and/or \emph{Swift}-XRT, of which 127 candidates have been visited more than once by \emph{Chandra}. This implies re-observed fractions of at least ${\approx}$84\% among the candidate sample (a large fraction of these 84\% of sources lie in fields intentionally observed multiple times; for instance, in the vicinity of the Orion Nebula or M101). The high X-ray re-detection fraction indicates that this is a very effective criterion if additional \emph{Chandra}, \emph{XMM-Newton} or \emph{Swift} observations are available.

In summary, 98 candidates pass this criterion (see Table~\ref{tab:criteria}), albeit largely because they lack multiple sensitive X-ray observations. We note that 20 candidates are discarded by this criterion but not by the others (see Table~\ref{tab:criteria}). The \emph{left panel} of Fig.~\ref{fig:histo} shows the net-count distribution for all the sources that pass this criterion. To conclude, this criterion appears to be an extremely effective means to identify persistent or repeat transients, when data are available.

\subsubsection{Criterion 2: Optical detections in \emph{Gaia}} \label{sec:gaia}

In previous works \citep[e.g., Paper~I and][]{Yang2019}, an important fraction of FXT candidates had a Galactic origin, especially related to relatively bright stars. To identify these, we cross-match with the \emph{Gaia} Early Data Release 3 \citep[\emph{Gaia} EDR3; employing the \texttt{VizieR} package;][]{Gaia_DR3_2020} catalog, which contains photometric and astrometric constraints for sources in the magnitude range $G{=}3$--21~mag including accurate positions, parallaxes, and proper motions throughout the Local Group \citep{Lindegren2018,Gaia2018}. 
We adopt the 3$\sigma$ positional uncertainty (obtained by the {\sc ciao} \texttt{wavdetect} task) associated with each candidate as our cone search radius. In general, this radius is sufficiently small to find a unique counterpart given the high spatial resolution and astrometric precision of \emph{Chandra} \citep[][]{Rots2010}; 9 candidates show multiple \emph{Gaia} sources in their cone search area, for which we adopt the nearest \emph{Gaia} source. 

From our initial sample of 151 FXT candidates, 107 sources have cross-matches in \emph{Gaia EDR3}. Nevertheless, we only discard FXT candidates matched to ``stellar'' \emph{Gaia} EDR3 optical detections, where stellar is taken to mean those with nonzero proper motion and/or parallax detected at $>$3$\sigma$ significance; this amounts to 83 candidates from the initial sample. These likely stellar sources cover a wide range in magnitude $G{=}$9.2--20.1~mag ($\overline{G}{\approx}$16.4~mag) and proper motion $\mu{=}$0.7--154.5~mas~yr$^{-1}$ ($\overline{\mu}{\approx}$22.1~mas~yr$^{-1}$).

The \emph{middle panel} of Fig.~\ref{fig:histo} shows the net-count distribution of the 68 sources that pass this criterion. Among the total sample, ${\approx}$55\% are associated with stellar flares of bright stars. Moreover, this criterion discards 42 FXT candidates that the additional criteria do not (see Table~\ref{tab:criteria}). However, because of the magnitude limit and optical window of \emph{Gaia}, this criterion may not identify all persistent or recurring transient Galactic objects, which we return to in the next subsection. As a running total, only 42 candidates successfully pass both this and the previous criterion (see Table~\ref{tab:criteria}).

\subsubsection{Criterion 3: NED, SIMBAD, and VizieR Search} \label{sec:catal_cross}

To identify known Galactic and Local Group contaminating objects not detected by \emph{Gaia}, we search for counterparts (or host galaxies) in large databases using the \texttt{astroquery} package: the NASA/IPAC Extragalactic Database \citep[NED;][]{Helou1991}, the Set of Identifications, Measurements, and Bibliography for Astronomical Data \citep[SIMBAD;][]{Wenger2000}, and VizieR \citep[which provides the most complete library of published astronomical catalogs;][]{Ochsenbein2000}.

We perform a cone search per FXT candidate, using a circular region with a radius of 3$\sigma$ based on the X-ray positional uncertainty from the {\sc ciao} \texttt{wavdetect} task to find associated sources. These three databases contain many catalogs across the electromagnetic (EM) spectrum, which permit us to rule out candidates in our sample associated with previously classified stars, young stellar objects (YSOs) embedded inside nebulae (where the absorption and obscuration do not permit \emph{Gaia} detections), globular clusters, or high-mass X-ray binaries (HMXBs) in either our Galaxy or the Local Group. 
This criterion is important in our analysis because ${\approx}$80\% (i.e., 121 FXT candidates) of the initial sample show associated sources with the SIMBAD and NED databases. We uniquely identify 33 objects, either as YSOs embedded in nebulae or stars identified by other catalogs, for instance, the VISTA Hemisphere Survey (VHS), the United Kingdom InfraRed Telescope (UKIRT) Infrared Deep Sky Survey, the Sloan Digital Sky Survey (SDSS), or the all-sky Wide-field Infrared Survey Explorer (WISE) CatWISE source catalog at 3.4 and 4.6 $\mu$m \citep{McMahon2013,Dye2018,Marocco2021}. It is important to mention that 33 FXT candidates are discarded solely by this criterion (see Table~\ref{tab:criteria}).

The \emph{right panel} of Fig.~\ref{fig:histo} shows the net-count distribution for the 76 FXT candidates that pass this criterion. Applying all criteria thus far, the sample is reduced to nine candidates.

\subsubsection{Archival Image Search} \label{sec:archive_images}

To rule out still fainter stellar counterparts, we carried out a search of ultraviolet (UV), optical, NIR, and mid-infrared (MIR) image archives. We perform a cone search within a radius equivalent to the 3$\sigma$ \emph{Chandra} positional uncertainty of the respective FXTs for the following archives: 
the Hubble Legacy Archive;\footnote{https://hla.stsci.edu/hlaview.html}
the Pan-STARRS archive \citep{Flewelling2016};\footnote{http://ps1images.stsci.edu/cgi-bin/ps1cutouts}
the National Science Foundation's National Optical-Infrared Astronomy Research (NOIR) Astro Data Lab archive,\footnote{https://datalab.noao.edu/sia.php} which includes 
images from the Dark Energy Survey \citep[DES;][]{Abbott2016} and
the Legacy Survey (DR8);
the Gemini Observatory Archive;\footnote{https://archive.gemini.edu/searchform}
the National Optical Astronomy Observatory (NOAO) science archive;\footnote{http://archive1.dm.noao.edu/search/query/}
the ESO archive science portal;\footnote{http://archive.eso.org/scienceportal}
the VISTA Science Archive;\footnote{http://horus.roe.ac.uk/vsa/}
the Spitzer Enhanced Imaging Products archive \citep{Teplitz2010};\footnote{https://irsa.ipac.caltech.edu/data/SPITZER/Enhanced/SEIP/}
the UKIRT/Wide Field Camera (WFCAM) Science Archive;\footnote{http://wsa.roe.ac.uk/}
and the WISE archive \citep[][]{Wright2010}.

For images obtained under good seeing (${<}$~1$"$) and weather conditions, we inspect visually for counterparts or host galaxies in the 3$\sigma$ uncertainty X-ray location of the FXT. We only apply this additional criteria for the FXT candidates that remain after the previous three criteria (see Sect. \ref{sec:previous_criteria}--\ref{sec:catal_cross}). If a source is found, we identify it as a star if it is consistent with the spatial resolution of the imaging, we quantify its significance and assess its extent and radial profile visually. We confirm that none of the nine candidates is associated with stellar sources, leaving the number of candidates unchanged.

\begin{table}
    \centering
    \advance\leftskip-0.5cm
    \scalebox{1.0}{
    \begin{tabular}{lccc}
    \hline\hline
     FXT & Odds ratio & Prob. &  Var. Index \\ \hline
     (1) & (2) & (3) & (4) \\ \hline
    15 &  10.18 & 0.99 & 9 \\
    16 &  93.73 & 1.0 & 10 \\
    17 &  9.09 & 0.99 & 8 \\
    18 &  8.67 & 0.99 & 8 \\
    19 &  167.0 & 1.0 & 10 \\
    20 &  34.47 & 1.0 & 10 \\
    21 &  6.19 & 0.99 & 8 \\ 
    22 &  29.56 & 1.0 & 9 \\ \hline
    \end{tabular}
    }
    \caption{Variability properties of the extragalactic FXT candidates detected and/or discussed in this work obtaining by the G-L method, ordered by subsample and date.
    \emph{Column 1:} shorthand identifier (FXT~\#) used throughout this work. 
    \emph{Column 2:} logarithmic odds ratio (ratio of obtaining the observed distribution versus obtaining a flat distribution) for variability signal. 
    \emph{Column 3:} variable signal probability (the probability that the flux calculated from the source region is not constant throughout the observation).
    \emph{Column 4:} variability index (ratio of obtaining the observed distribution versus obtaining a flat distribution).}
    \label{tab:variability}
\end{table}

\subsubsection{Instrumental and variability effects} \label{sec:inst_effects}

Finally, we visually check the X-ray data to rule out false-positive candidates that may arise from background flares, bad pixels or columns, or cosmic-ray afterglows rather than intrinsic variability. Again, we only undertake this last criteria for the remaining nine candidates after Sect.\ref{sec:archive_images}.

First, we use the \texttt{glvary} tool to confirm variability using the Gregory-Loredo (G-L) algorithm. The Gregory-Loredo variability algorithm is a commonly used test to detect time variability in sources \citep{Gregory1992}.\footnote{The G-L method splits the sources into multiple time bins and looks for significant deviations between them. The tool assigns a variability index based on the total odds ratio and the corresponding probability of a variable signal.} This adds a second criterion for variability, increasing the probability that the light curves of our candidate FXTs show strong variability during the observation. Applying the G-L task to our sample of nine FXT candidates, we found that one of them (identified in the \emph{Chandra} ObsId 16302 at $\alpha$=13$^{\text{h}}$56$^{\text{m}}$01.\!\!$^{\text{s}}$10, $\delta$=--32$^\circ$35$^\prime$15.95$^{\prime\prime}$) has a low probability to be a variable source (${\approx}$0.1) with a variability index of 2.\footnote{Although our algorithm is designed to select only sources with large amplitude variations in the number of counts (see Sect.~\ref{sec:algorithm} or sect.~2.1 in Paper~I), this source does not vary. The peculiar light curve of this source erroneously allows it pass our initial method. The light curve is split into different time windows, then erroneously our method selects this source since in one window the light curve contains one of the two peaks and a quiescent phase (mimicing the light curve of a transient source). Thus, the G-L test is necessary to rule out any such source.} These results guarantee that this source is inconsistent with flux variability throughout the observation. The remaining eight sources show a clear variability throughout the \emph{Chandra} observation according to their variable probability (${\gtrsim}$0.99) and variability index (${\gtrsim}$8) (see Table~\ref{tab:variability} for more details).

Finally, to reject possible strong background flaring episodes in the 0.5--7~keV band, we employ the \texttt{dmextract} and \texttt{deflare} tools to examine the evolution of the background count rate during the observations. None of the FXT candidates is affected by background flares.  Furthermore, we confirm visually that the counts from all sources are detected in dozens to hundreds of individual pixels (discarding association with bad columns or hot pixels) tracing out portions of \emph{Chandra}'s Lissajous dither pattern (appearing as a sinusoidal-like evolution of $x$ and $y$ detector coordinates as a function of time; see Appendix Fig.~\ref{fig:lissajaus}) over their duration, reinforcing that they are astrophysical sources. Therefore, we have a final sample of eight FXTs.

\subsubsection{Completeness}\label{sec:completness}

It is important to keep in mind that real FXTs may have been ruled out erroneously by the criteria above. To roughly estimate this, we compute the probability that a FXT candidate overlaps with another X-ray source and/or star by chance. Assuming Poisson statistics (i.e., $P(k,\lambda)$), the probability of one source ($k{=}1$) being found by chance inside the 3$\sigma$ localization uncertainty region of another is given by
\begin{equation}
    P(k{=}1,\lambda){=}\frac{e^{-\lambda}\lambda^k}{k!},
    \label{eq:001}
\end{equation}
where $\lambda$ is the source density of X-ray sources and/or stars on the sky multiplied by the 3$\sigma$ \emph{Chandra} localization uncertainty area. As a reference, the mean density of X-ray sources detected by \emph{Chandra}, \emph{XMM-Newton} and \emph{Swift}-XRT is 0.36, 1.68, and 0.07~arcmin$^{-2}$, respectively, while the mean density of optical sources detected by \emph{Gaia} is 2.0~arcmin$^{-2}$. We use the X-ray detections from the CSC2, 4XMM-DR11 and 2SXPS catalogs \citep{Evans2010,Webb2020,Evans2020b}, and the \emph{Gaia EDR3} catalog for stars \citep{Gaia_DR3_2020} to determine the X-ray or optical source densities, respectively. The probability is 0.0024 and 0.0029 for X-ray and optical sources, respectively. Taking into account just the X-ray sources discarded solely by {Criteria 1 or 2}, 20 and 42 X-ray sources (see Table~\ref{tab:criteria}), respectively, we expect ${\ll}$1 of these to be ruled out wrongly. 
If we consider only the 109 X-ray sources which were discarded by both {Criteria 1 and 2}, the combined probability is 1${\times}$10$^{-5}$, and thus the expected number of erroneously dismissed sources is also ${\ll}$1. 

Considering the densities of X-ray sources in individual \emph{Chandra} fields, they span a minimum-maximum density range between 0.0042--2.302~arcmin$^{-2}$, yielding a probability range of $1.4{\times}10^{-6}$ to 0.0505. Thus, under these extreme density conditions, the number of X-ray sources discarded wrongly by {Criteria 1} is ${\approx}0.1$. This value is relatively low, and thus reinforces the idea that an erroneous rejection is unlikely even in extreme conditions.
As an extreme example, we can consider the X-ray positions of CDF-S XT2 and source XID$_\text{4Ms}$256 \citep[${\approx}$30 photons detected during the 4~Ms exposure, classified as a normal galaxy;][]{Xue2011,Luo2017}, which differ by only ${\approx}2\farcs0$. Upon further investigation of the flux and position of the X-ray variability, it was realized that XID$_\text{4Ms}$256 and CDF-S XT2 are distinct sources \citep{Xue2019}. The X-ray source density (at ${\sim}$2\farcm0 off-axis angle) in the \emph{Chandra Deep Field South} at 7~Ms is ${\approx}$5.6~arcmin$^{-2}$ \citep{Luo2017}, leading to a chance alignment probability (using Eq.~\ref{eq:001}) of 0.019 between CDF-S XT2 and ID$_\text{4Ms}$256. Although this value is low, it is non-zero, and thus care should be given to the spatial/temporal alignment of X-ray sources, so as to not discard candidates erroneously.

It is not easy to assess the contribution by {Criterion 3} to the completeness given the highly disjoint nature of the databases. Similar to Paper~I, we assume that {Criterion 3} does not disproportionately discard real FXTs. In aggregate, we conclude that our rejection criteria do not apparently impact on the completeness of our FXT candidate sample.

\begin{table*}
    \centering
    \advance\leftskip-0.3cm
    \scalebox{0.85}{
    \begin{tabular}{llrclrrrrccr}
    \hline\hline
     FXT & Id & ObId & Exp. (ks) & Date & $T_{90}$ (ks) & RA (deg.) & DEC (deg.) & Off. Ang. & Pos. Unc. & HR & S/N\\ \hline
     (1) & (2) & (3) & (4) & (5) & (6) & (7) & (8) & (9) & (10) & (11) & (12) \\ \hline
     15 & XRT~140507 & 16093 & 68.8 & 2014-05-07 & 4.8$_{-5.5}^{+3.6}$ & 233.73496 &  23.46849 & 2\farcm2 & 0\farcs8 & $-1.5\pm0.21$ & 6.9 \\ 
     16 & XRT~150322/ & 16453 & 74.7 & 2015-03-21/22 & 10.3$_{-6.3}^{+9.0}$ &  53.07672 & -27.87345 & 4\farcm3 & 0\farcs32 & $-0.33\pm0.08$ & 35.2 \\ 
      & CDF-S~XT2$\dagger$ &  &  &  &  &  &  &  &  &  &  \\
     17 & XRT~151121 & 18715 & 24.5 & 2015-11-20/21 & 5.5$_{-4.2}^{+1.2}$ & 40.82972 & 32.32390 & 8\farcm6 & 2\farcs0 & $-0.17\pm0.20$ & 6.4 \\ 
     18 & XRT~161125 & 19310 & 6.1 & 2016-11-25 & 2.9$_{-2.4}^{+0.8}$ &  36.71489 & -1.08317 & 12\farcm1 & 5\farcs0 & $-0.64\pm0.18$ & 4.3 \\ 
     19 & XRT~170901$\dagger$ & 20635 & 77.0 & 2017-08-31/09-01 & 3.9$_{-0.7}^{+1.2}$ & 356.26437 & -42.64494 & 3\farcm1 & 0\farcs14 & $+0.00\pm0.08$ & 43.7 \\
     20 & XRT~191127 & 21831 & 22.5 & 2019-11-26/27 &  0.3$\pm$0.2 & 207.34711 & 26.58421 & 5\farcm9 & 1\farcs2 & $-0.34\pm0.17$ & 8.6 \\
     21 & XRT~191223 & 23103 & 18.1 & 2019-12-23 & 3.7$_{-0.5}^{+2.4}$ &  50.47516 &  41.24704 & 3\farcm5 & 0\farcs44 & $-0.16\pm0.20$ & 9.7 \\ 
     22 & XRT~210423$\dagger$ & 24604 & 26.4 & 2021-04-23/24 &  12.1$_{-4.1}^{+4.0}$ & 207.23523 & 26.66230 & 7\farcm5 & 0\farcs6 & $-0.18\pm0.10$ & 12.4  \\ \hline
    \end{tabular}
    }
    \caption{Properties of the extragalactic FXT candidates detected and/or discussed in this work, ordered by date.
    \emph{Column 1:} shorthand identifier (FXT~\#) used throughout this work. 
    \emph{Column 2:} X-ray transient identifier (XRT~date), plus previous name when available. 
    \emph{Columns 3, 4, and 5:} \emph{Chandra} observation ID, exposure time in units of ks, and date. 
    \emph{Column 6:} $T_{90}$ duration, which measures the time over which the source emits the central 90\% (i.e., from 5\% to 95\%) of its total measured counts, in units of ks.
    \emph{Columns 7 and 8:} Right Ascension and Declination in J2000 equatorial coordinates. 
    \emph{Column 9:} Off-axis angle of the FRXT candidates, with respect to the \emph{Chandra} aimpoint, in units of arcminutes. 
    \emph{Column 10:} Estimated 2$\sigma$ X-ray positional uncertainty, in units of arcseconds (see Sect.~\ref{sec:astrometry}).
    \emph{Column 11:} Hardness ratio (HR) and 1$\sigma$ uncertainty, defined as HR=(H$-$S)/(H$+$S) where H=2--7~keV and S=0.5--2~keV energy bands, using the Bayesian estimation of \citet{Park2006}. 
    \emph{Column 12:} Approximate signal-to-noise (S/N).\\
    $\dagger$ Previously reported as FXTs by \citet{Xue2019} for FXT~16, \citet{Lin2019} for FXT~19, and \citet{Lin2021} in the case of FXT~22.}
    \label{tab:my_detections}
\end{table*}

\begin{figure*}[h!]
    \centering
    \includegraphics[scale=0.65]{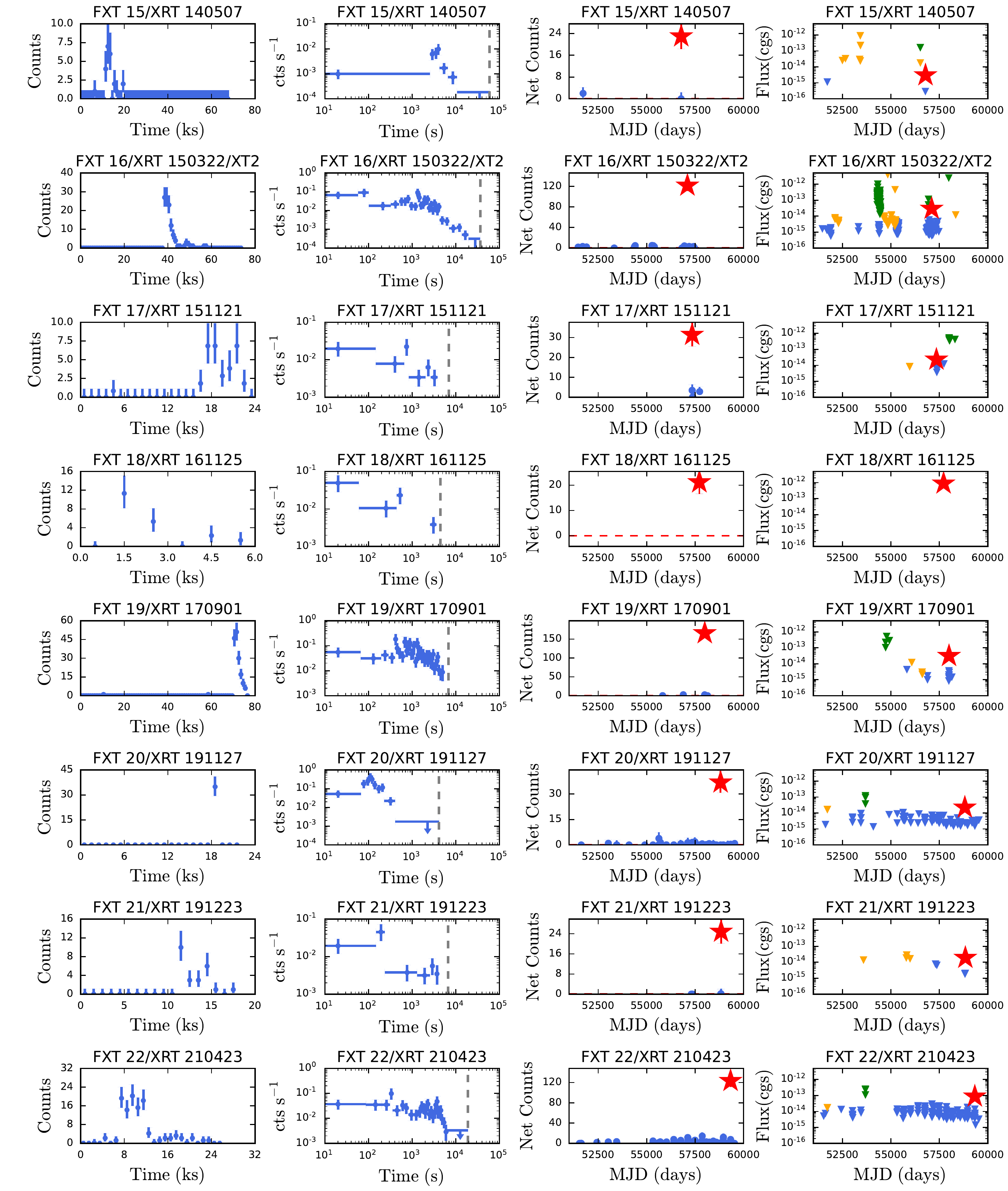}
    \vspace{-0.3cm}
    \caption{0.5--7~keV light curves for each FXT candidate: (\emph{1st column}) full exposure, in units of counts; (\emph{2nd column}) zoom in, from the detection of first photon to the end of the exposure, in units of count rate (cts~s$^{-1}$), with log-log scaling and 5 counts per bin. The \emph{gray dashed lines} show the stop-time per observation regarding the beginning of the transient; (\emph{3rd column}) long-term light curve, with each point representing individual \emph{Chandra} exposures (cyan circles with 1-$\sigma$ error bars) to highlight the significance of detections and non-detections, in units of counts; (\emph{4th column}) long-term light curve, with each point representing individual \emph{Chandra} (\emph{cyan}), {\it XMM-Newton} (\emph{orange}) and \emph{Swift-XRT} (\emph{green}) exposures in units of flux (erg~s$^{-1}$~cm$^{-2}$). For the long-term light curves, the observation including the transient is denoted by a \emph{large red star} (1-$\sigma$ error bars), while triangles denote observations with (3-$\sigma$) upper limits. All fluxes are reported in the 0.5--7~keV band in the observer's frame.}
    \label{fig:light_curves_1}
\end{figure*}

\subsubsection{Summary}\label{sec:filter_summary}

We identify eight FXT candidates, three of them have been previously discovered and classified as FXTs by \citet{Xue2019}, \citet{Lin2021} and \citet{Lin2022}; see Sect.~\ref{sec:results} for more details. Table~\ref{tab:my_detections} shows important information of the final sample: the coordinates, duration ($T_{90}$), instrumental off-axis angle, positional uncertainty, hardness ratio \citep[HR; computed following][]{Park2006}, and S/N ratio (computed using the \texttt{wavdetect} tool). Figure~\ref{fig:light_curves_1} shows the background-subtracted 0.5--7.0~keV light curves of our final sample of FXT candidates: short-term, in units of counts (\emph{first column}) and logarithmic count rates (\emph{second column}); long-term in units of net-counts for \emph{Chandra} only (\emph{third column}) and flux to compare uniformly \emph{Chandra}, \emph{XMM-Newton} and \emph{Swift}-XRT data (\emph{fourth column}).
It is important to mention that the first three criteria considered (X-ray archival data, \emph{Gaia} detection cross-match, and NED/SIMBAD/VizieR catalogs) contribute significantly and in complementary ways to clean the sample (especially for discarding stellar contamination). 

Finally, we label each FXT candidate by ``XRT'' followed by the date, where the first two numbers correspond to the year, the second two numbers to the month, and the last two numbers to the day (see Table~\ref{tab:my_detections}, \emph{second column}). Nevertheless, similar to Paper~I, to identify each source quickly throughout this manuscript we also denominate them by ``FXT''+\# (ordered by date; see Table~\ref{tab:my_detections}, \emph{first column}) from 15 to 22, because this work is a sequel paper to Paper~I where FXTs were labeled until FXT~14. Furthermore, FXT~18 does not have additional \emph{Chandra}, \emph{XMM-Newton} or \emph{Swift}-XRT observations to ensure its transient nature, however, we keep it to be consistent with the selection criteria of this work.

\begin{figure*}
    \centering
    \includegraphics[scale=0.7]{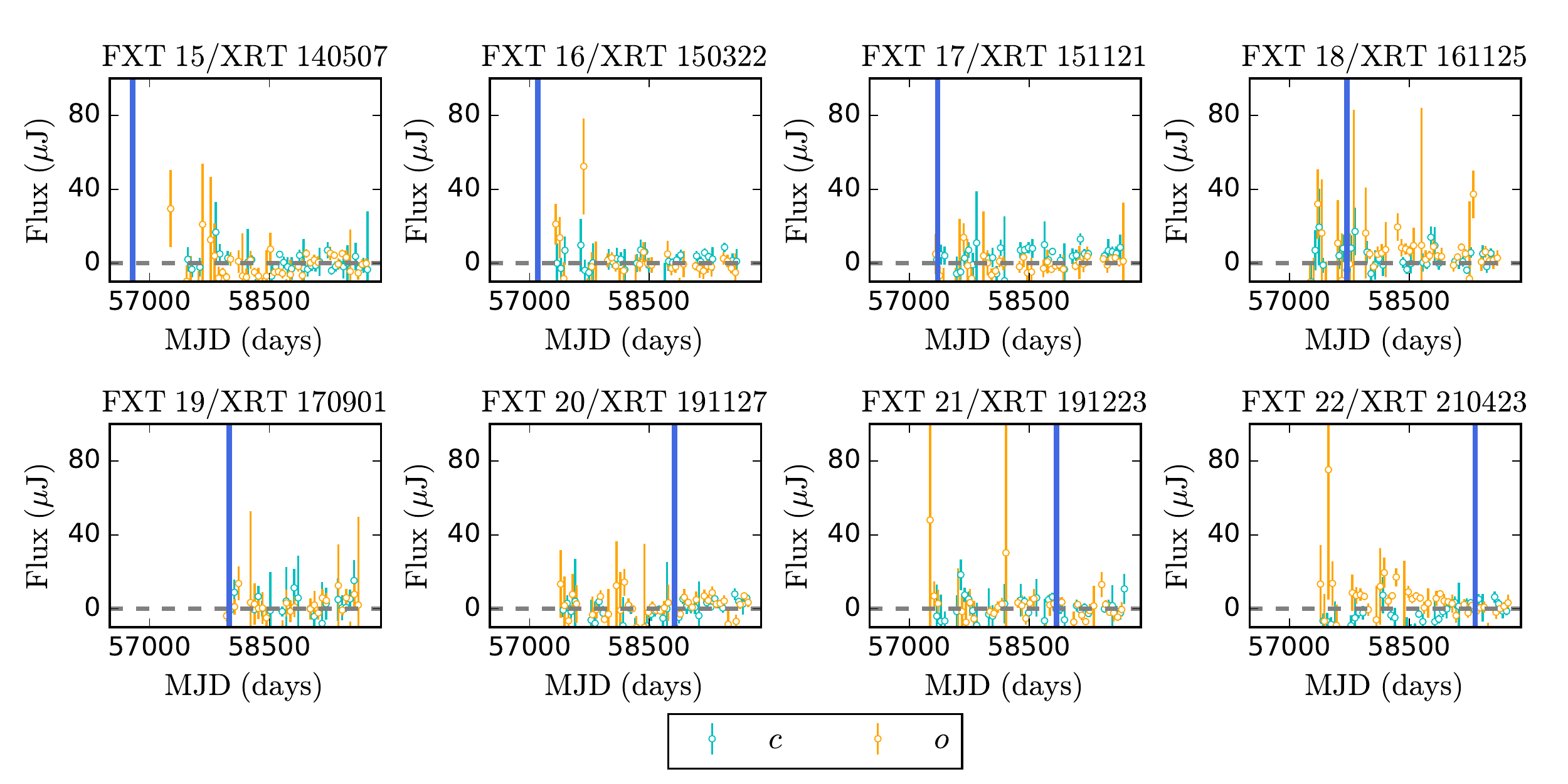}
    \vspace{-0.2cm} \caption{ATLAS differential photometry of the \emph{cyan} ($c$) and \emph{orange} ($o$)-bands light curves performed at the position of the FXT candidates. $<5\sigma$ data points are shown in hollow circles, while $>5\sigma$ data points are shown in solid circles. The \emph{blue vertical lines} show the epochs when the FXT candidates were detected by \emph{Chandra}, while the \emph{dashed gray lines} represent the zero flux.}
    \label{fig:forzed_photometry_AT}
\end{figure*}

\begin{figure*}
    \centering
    \includegraphics[scale=0.7]{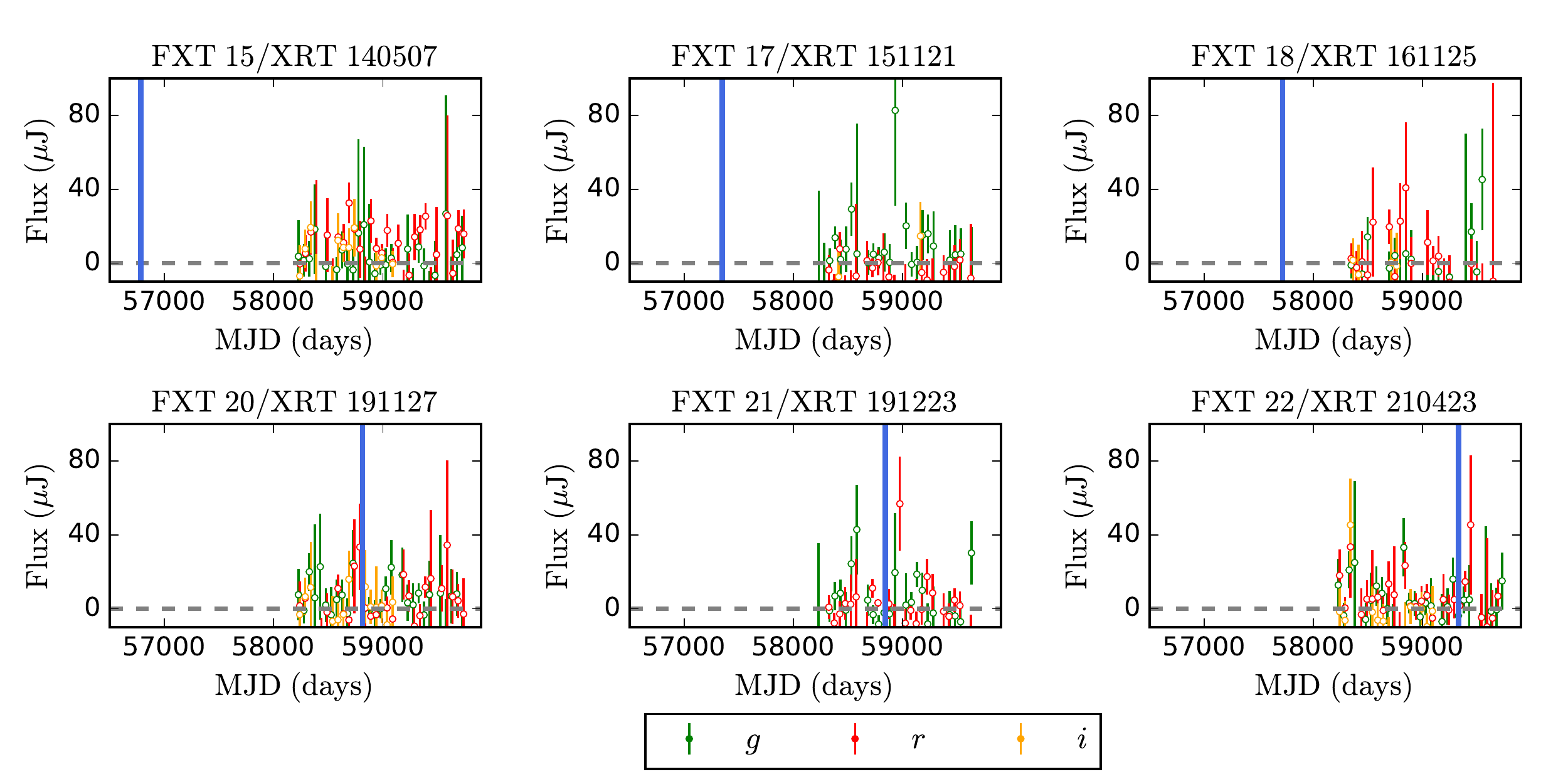}
    \vspace{-0.2cm} \caption{ZTF differential photometry of the $g$ (\emph{green points}), $r$ (\emph{red points}) and $i$-bands (\emph{orange points}) light curves performed at the position of the FXT candidates. $<5\sigma$ data points are shown in hollow circles, while $>5\sigma$ data points are shown in solid circles. The \emph{blue vertical lines} shows the epochs when the FXT candidates were detected by \emph{Chandra}, while the \emph{dashed gray lines} represents the zero flux.}
    \label{fig:forzed_photometry_ZTF}
\end{figure*}

\begin{figure*}
    \centering
    \includegraphics[width=18cm,height=2.3cm]{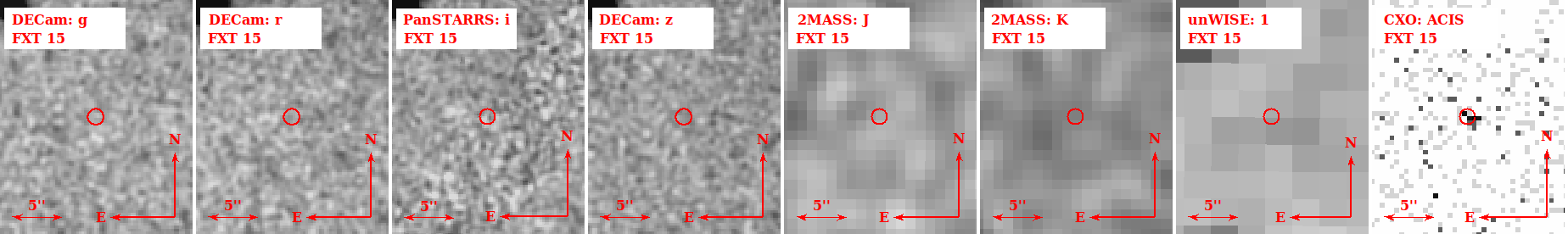}
    \includegraphics[width=18cm,height=2.3cm]{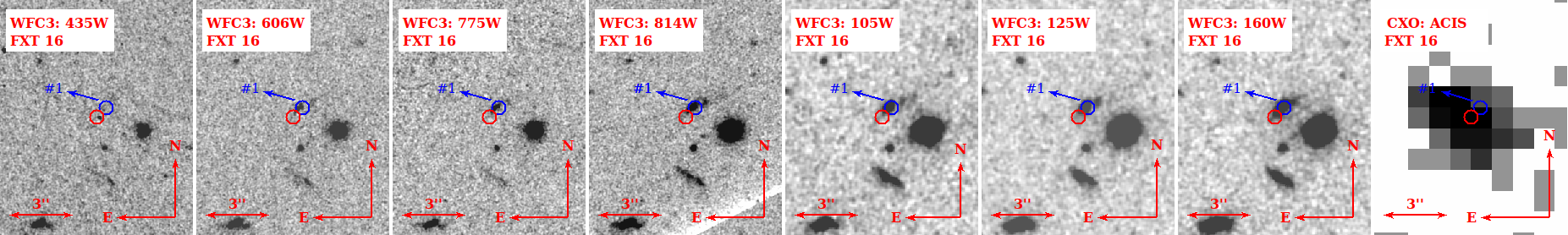}
    \includegraphics[width=18cm,height=2.3cm]{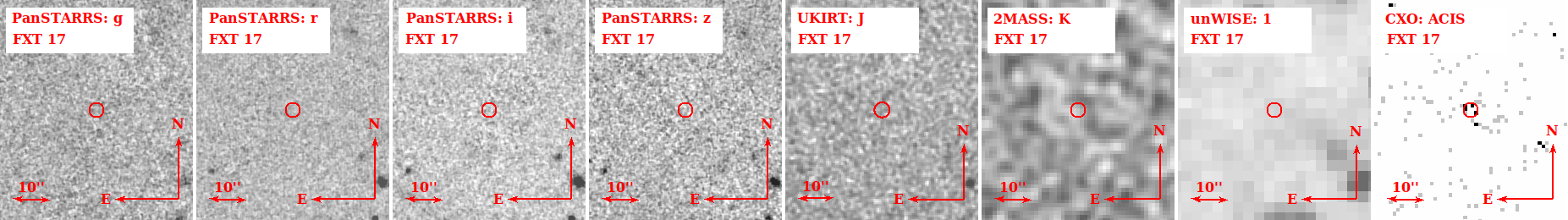}
    \includegraphics[width=18cm,height=2.3cm]{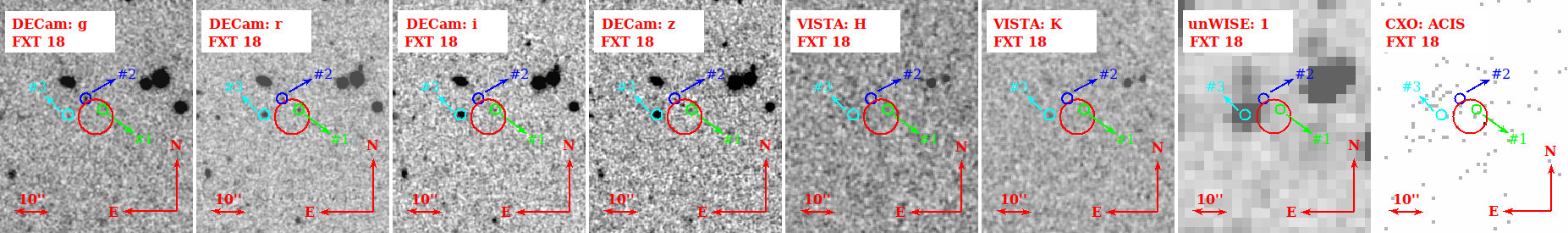}
    \includegraphics[width=18cm,height=2.3cm]{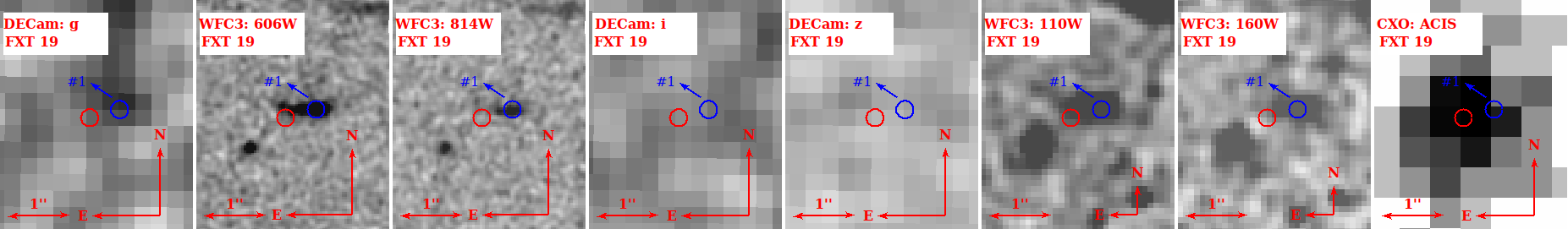}
    \includegraphics[width=18cm,height=2.3cm]{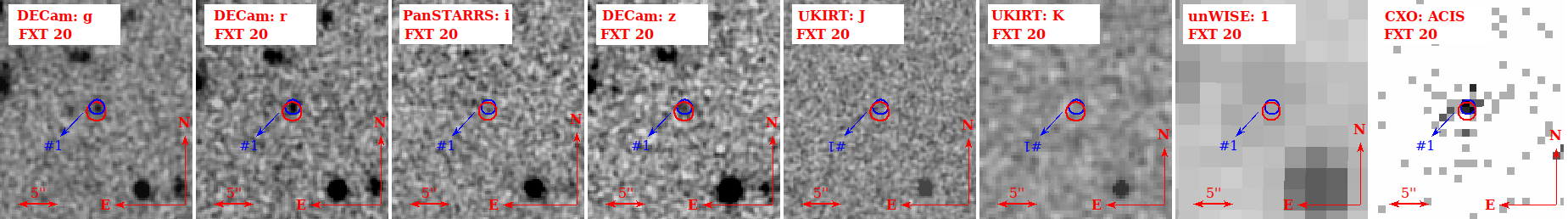}
    \includegraphics[width=18cm,height=2.3cm]{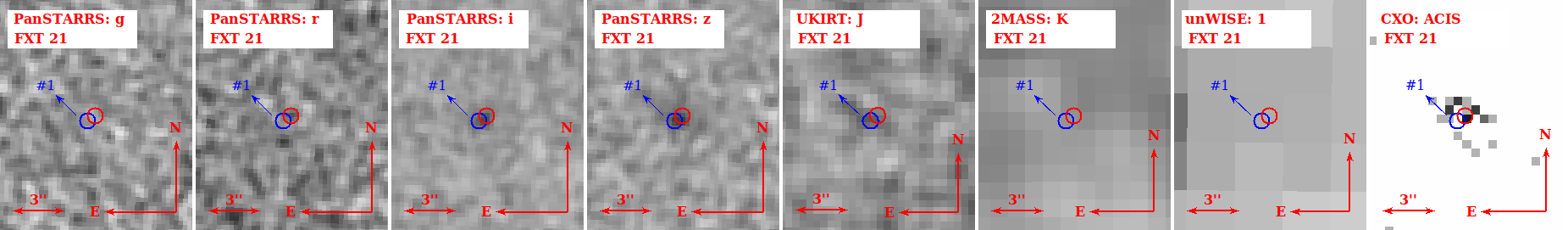}
    \includegraphics[width=18cm,height=2.3cm]{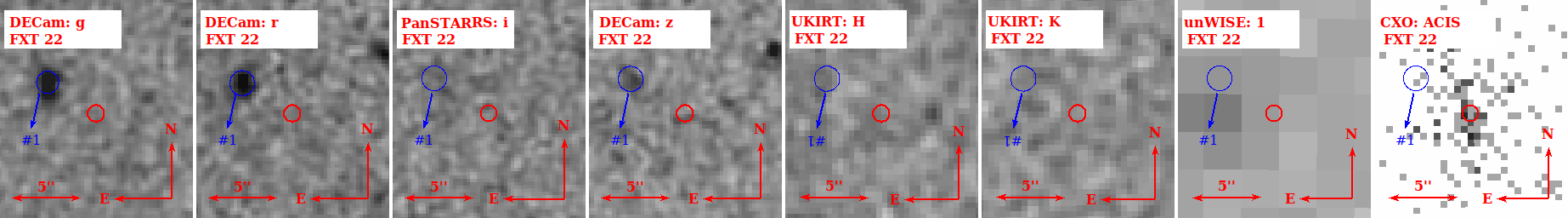}
    \caption{Archival optical, near-infrared, mid-infrared and \hbox{X-ray} images of extragalactic FXT candidates; the telescope/instrument +filter and FXT ID name are shown in the upper-left corner. Each cutout is centered on the \hbox{X-ray} position, while red circles denote 2$\sigma$ \emph{Chandra} errors in the source localisation. \emph{(1st, 2nd, 3rd and 4th columns)} optical band (DECam, Pan-STARRS and HST) images. \emph{(5th and 6th columns)} near-infrared $J$ or $H$ and $K$ (2MASS, UKIRT or VISTA) images; \emph{(7th column)} 3.4$\mu$m (unWISE) images; and \emph{(8th column)} \hbox{X-ray} \emph{Chandra} (ACIS) 0.5--7 keV images. The \emph{coloured arrows and circles} show the localization of the possible host/counterparts of the FXT candidates. HST images were aligned using the astrometry of Gaia.}
    \label{fig:image_cutouts}
\end{figure*}

\subsection{Fainter Electromagnetic detections}\label{sec:results}

We now focus on a detailed multiwavelength search (in Sects. \ref{sec:radio} to \ref{sec:gamma} from radio to gamma rays) of each candidate for a contemporaneous counterpart\footnote{Hereafter, "counterpart" refers to the emission relating to the FXT candidate, not its host galaxy, during epochs close to the X-ray trigger.} and host galaxy using several archival datasets to understand their origin.

\subsubsection{Radio Emission} \label{sec:radio}

We search for any possible radio emission associated to our FXT candidates using the \emph{RADIO--Master Radio Catalog}, which is a revised master catalog with select parameters from a number of the \texttt{HEASARC} database tables. It holds information on radio sources across a wide range of telescopes/surveys [e.g, the Very Long Baseline Array, the Very Large Array (VLA), and the Australia Telescope Compact Array] and frequencies (from 34~MHz to 857~GHz). Because of the poor angular resolution of some associated radio catalogs, we perform an initial cone search for radio sources with a radius of 60 arcseconds. Following this initial 60 arcseconds cut, we repeat a search using limiting radii consistent with the combined radio + X-ray 3$\sigma$ positional errors. The current version of the master catalog does not yet contain the recent VLA Sky Survey (VLASS)\footnote{https://science.nrao.edu/vlass/} and Rapid ASKAP Continuum Survey \citep[RACS;][]{Hale2021} catalogs, so we additionally query these using resources from the Canadian Astronomy Data Centre\footnote{https://www.cadc-ccda.hia-iha.nrc-cnrc.gc.ca/en/} interface. Unfortunately, our search returns no matches indicating that none of the final sample of FXT candidates or host sites is unambiguously detected at radio wavelengths.

\subsubsection{Optical and Mid-Infrared Counterpart Emission} \label{sec:OIR_forzed}

In order to explore possible optical and MIR contemporaneous counterparts of our final sample, we examine forced differential photometry taken from the Zwicky Transient Facility \citep[ZTF;][]{Bellm2019,Graham2019,Masci2019}, the Asteroid Terrestrial-impact Last Alert System \citep[ATLAS;][]{Tonry2018,Smith2020}, and a visual inspection of images obtained during epochs around the X-ray trigger obtained from the unWISE time-domain \citep{Meisner2022} and the Legacy Surveys DR10  catalogs \citep{Dey2019}.

ZTF is a wide-field (field of view of 55.0~deg$^2$) time-domain survey, mapping the sky every few nights with custom $g$, $r$, and $i$ filters to 5-$\sigma$ limiting magnitudes of $\approx$20.8, 20.6, 19.9~AB mag, respectively \citep{Bellm2019}. ATLAS is a four-telescope asteroid impact early warning system, which observes the sky several times every night in custom \emph{cyan} ($c$; 420--650~nm) and \emph{orange} ($o$; 560--820~nm) filters to 5-$\sigma$ limiting magnitudes of $\approx$19.7~AB~mag  \citep{Tonry2018,Smith2020}.

Figures~\ref{fig:forzed_photometry_AT} and \ref{fig:forzed_photometry_ZTF} show the differential photometry light curves taken from ZTF ($gri$) and ATLAS ($co$), respectively, for the FXT candidates.  For visual clarity, the ZTF and ATLAS photometry are binned by 50 days, with the errors added in quadrature. The locations of all eight FXT candidates have been observed by ATLAS (see Fig.~\ref{fig:forzed_photometry_AT}), although FXT~15 and FXT~16 were not visited by ATLAS around the time of the X-ray detection (highlighted by the vertical blue lines). ZTF, on the other hand, has only observed six FXT candidate locations (see Fig.~\ref{fig:forzed_photometry_ZTF}), of which FXT~15, FXT~17 and FXT~18 fields not being observed around the time of the \emph{Chandra} detection. Notably, the most recent FXTs (FXTs~20, 21 and 22) have forced differential photometry light curves from ZTF and ATLAS, covering a wide epoch both before and after the \emph{Chandra} X-ray detections.
Overall, none of the FXT candidates exhibits significant (${>}5{\sigma}$) detections of optical variability or flares by ZTF and ATLAS around the time of the FXT candidate X-ray trigger, nor are there any robust detections in any previous or subsequent epochs.
We derive 3$\sigma$ upper limits from the closest observation taken by ZTF and ATLAS (for the available filters and FXTs), as listed in Table~\ref{tab:upper_limits}.

To check if the forced photometry is consistent with zero flux (around the X-ray trigger), we use the statistical test {\sc ConTEST} \citep{Stoppa2023}, developed explicitly to compare the consistency between the observations and a constant zero flux model. We adopt a methodology identical to that discussed in Eappachen et al. (2023, submitted). We applied this test considering two-time windows, [-10;20] and [-10;100] days, with respect to the X-ray trigger, because possible optical counterparts have timescales from days (e.g., the afterglow of GRBs) to weeks/months (e.g., CC-SNe emission). We concluded that for all the sources, the model of zero flux density detected by both periods is consistent with the observations.

Furthermore, the DESI Legacy Imaging Surveys (DR10) combine three major public projects plus additional archival data to provide imaging over a large portion of the extragalactic sky visible from the Northern and Southern Hemispheres in at least three optical bands ($g$, $r$, and $z$). The sky coverage (${\sim}$30,000~deg$^2$) is approximately bound by $-90^\circ{<}\delta{<}+84^\circ$ in celestial coordinates and $|b|{>}15^\circ$ in Galactic coordinates \citep{Dey2019}. Thus, the Legacy Imaging survey observes most FXT locations (except for FXTs~17 and 21). We explore visually each individual imaging epoch provided by the Legacy survey in $g$-, $r-$, $i-$, and $z$-bands to identify potential optical contemporaneous counterparts of the FXTs. However, no contemporaneous optical counterparts are identified around the X-ray trigger time after a visual inspection.

The Wide-field Infrared Survey Explorer \citep[WISE;][]{Wright2010} provides an unprecedented time-domain view of the MIR sky at W1${=}$3.4~$\mu m$ and W2${=}$4.6~$\mu m$ due to the NEOWISE mission extension \citep{Mainzer2011,Mainzer2014}. WISE has completed more than 19 full-sky epochs over a ${>}$12.5~year baseline, with each location having been observed for ${\gtrsim}$12 single exposures \citep{Meisner2022}. In order to search for a potential counterpart inside the WISE and NEOWISE images, we use the time-domain API tools provided by the {\it unTimely Catalog}, which considers data from 2010 to 2020 \citep{Meisner2022}. We 
visually inspect each single-epoch image of each FXT field (for FXT~22, only up to ${\sim}$1.5~years before the X-ray trigger). Unfortunately, none of the FXT candidates shows significant detections of variability or flares around the time of the X-ray trigger.

\subsubsection{Ultraviolet, Optical, and Infrared Host Galaxy Identification} \label{sec:UOIR_host}

To search for UV, optical, NIR and MIR emission associated with any possible host galaxy in the vicinity of each FXT candidate, we perform a cone search within a radius equivalent to the 3$\sigma$ \emph{Chandra} error position (see Table~\ref{tab:my_detections}) in the following catalogs: 
GALEX Data Release 5 \citep[GR5;][]{Bianchi2011}, 
Pan-STARRS Data Release 2 \citep[Pan-STARRS--DR2;][]{Flewelling2018}, 
the DES Data Release 2 \citep[DES--DR2;][]{Abbott2021}, 
the SDSS Data Release 16 \citep[SDSS--DR16;][]{Ahumada2019},
the NOAO Source Catalog Data Release 2 \citep[NSC--DR2;][]{Nidever2020},
the \emph{Hubble} Source Catalog version 3 \citep[HSCv3;][]{Whitmore2016},
the UKIRT InfraRed Deep Sky Survey Data Release 11+\citep[UKIDSS--DR11+;][]{Warren2007},
the UKIRT Hemisphere Survey Data Release 1 \citep[UHS--DR1;][]{Dye2018},
the Two Micron All Sky Survey \citep[2MASS;][]{Skrutskie2006}, 
the VHS band-merged multi-waveband catalogs Data Release 5 \citep[DR5;][]{McMahon2013}, 
the Spitzer Enhanced Imaging Products Source List \citep{Teplitz2010}, 
and the unWISE \citep[unWISE;][]{Schlafly2019} and CatWISE \citep{McMahon2013,Dye2018,Marocco2021} catalogs, 
as well as the 
ESO Catalogue Facility and 
the NED \citep{Helou1991}, 
SIMBAD \citep{Wenger2000}, and
VizieR \citep{Ochsenbein2000} databases. We supplement this by including any extended sources found during our archival image analysis in Sect. \ref{sec:archive_images}. We assume that uncertainties in the UV through MIR centroid positions contribute negligibly to the overall error budget.
Figure~\ref{fig:image_cutouts} shows image cutouts of the localization region of the FXTs (one per row), typically from Pan-STARRS, DECam, or {\it HST} in the optical (\emph{1st--4th columns}, using $g$, $r$, $i$ and $z$ or the corresponding {\it HST} filters), VISTA, UKIRT, 2MASS or HST in the NIR (\emph{5th and 6th columns}, using $J$, $H$, $K$ or the corresponding {\it HST} filters), unWISE in the MIR (\emph{7th column}, in the 3.6$\mu$m filter) band, and the \emph{Chandra}-ACIS image (\emph{8th column}, in the 0.5--7.0~keV band).

FXT~15 has no optical and NIR sources detected within the 3$\sigma$ X-ray positional uncertainty of this source in the {\it HST}, DECam, 2MASS, or unWISE images (see Fig.~\ref{fig:image_cutouts}). Upper limits are given in Table~\ref{tab:photometric_data}.
    
FXT~16/CDF-S~XT2 (identified previously by \citealt{Zheng2017} and analyzed in detail by \citealt{Xue2019}) was detected by \emph{Chandra} with a 2$\sigma$ positional uncertainty of 0\farcs32 (see Table~\ref{tab:my_detections}). This accurate \emph{Chandra} X-ray position allows us to identify the host galaxy, which lies at an offset of $\approx0$\farcs44 (i.e., a projected distance of ${\approx}$3.3~kpc) using \emph{HST} images (see Fig.~\ref{fig:image_cutouts}). The galaxy has a spectroscopic redshift of $z_{\rm spec}{=}$0.738. The probability of a random match between FXT~16 and a galaxy as bright as or brighter than $m_{\rm F160W}{\approx}$24~AB~mag within 0\farcs44 is $\approx$0.01 \citep{Xue2019}.

FXT~17 does not have optical and NIR sources detected within the 3$\sigma$ X-ray error region of this source in the Pan-STARRS, 2MASS, or unWISE images (see Fig.~\ref{fig:image_cutouts}). Upper limits are given in Table~\ref{tab:photometric_data}.

For FXT~18, one faint source with $m_{r}{\approx}24.2$~AB~mag (see Fig.~\ref{fig:image_cutouts}, \emph{source \#1}) appears inside the large localization region ($r{\approx}$7\farcs5 at 3$\sigma$) in the DECam $g$, $r$, $i$ and $z$-band images with an off-set angle of ${\approx}$2\farcs6 from the X-ray center position; it has a chance association probability of ${<}$0.095. Two other sources lie slightly outside the X-ray uncertainty region, \emph{sources \#2} and \emph{\#3}, with chance probabilities of 0.582 and 0.363, respectively (see Fig.~\ref{fig:image_cutouts}); such high probabilities suggest an association with either one of them is unlikely.

FXT~19 (reported previously by \citealt{Lin2019} and analyzed in detail by \citealt{Lin2022}) lies close to a faint ($m_{\rm F606W}{\approx}$24.8, $m_{\rm F814W}{\approx}$24.9, $m_{\rm F110W}{\approx}$24.7 and $m_{\rm F160W}{\approx}$24.3~AB~mag, using aperture photometry) and extended optical and NIR source in \emph{HST} imaging (see Fig.~\ref{fig:image_cutouts}, \emph{source \#1}) with an angular offset ${\approx}$0\farcs45. The chance probability for FXT~19 and \emph{source \#1} to be randomly aligned in F160W is very low, only 0.005 \citep{Lin2022}.

FXT~20 was detected 6\farcm812 (or ${\approx}$500~kpc in projection) from the center of the galaxy cluster Abell 1795 (located at ${\approx}$285.7~Mpc) during a \emph{Chandra} calibration observation (ObsId 21831). FXT~20 lies close to a faint source $m_{r}{\approx}23.5$~AB~mag (see Fig.~\ref{fig:image_cutouts}, \emph{source \#1}) identified in DECam $g$, $r$, and $z$-bands at an offset angle of ${\approx}$0\farcs6. The probability of a false match is $P{<}0.005$ \citep[adopting the formalism developed by][]{Bloom2002} for such offsets from similar or brighter objects.

FXT~21 has a faint optical source ($m_r{\approx}$25.1~AB~mag) inside the 3$\sigma$ X-ray error position in Pan-STARRS images (see Fig.~\ref{fig:image_cutouts}, \emph{source \#1}), but no source is detected in 2MASS NIR or unWISE MIR images. The offset between the FXT and the optical source position is ${\approx}$0\farcs5, with a false match probability of $P{<}0.0085$ \citep[adopting the formalism developed by][]{Bloom2002} for such offsets from similar or brighter objects.

Finally, FXT~22 \citep[identified previously by][]{Lin2021} was detected 4\farcm079 (or ${\approx}$300~kpc in projection) from the center of the galaxy cluster Abell 1795 (located at ${\approx}$285.7~Mpc) during a \emph{Chandra} calibration observation (ObsId 24604). No sources are detected within the 3$\sigma$ X-ray error region of this source in the DECam optical, VISTA NIR, or unWISE MIR images (see Fig.~\ref{fig:image_cutouts}). However, this source falls close to an extended object, SDSS~J134856.75+263946.7, with $m_r{\approx}$21.4~AB~mag that lies at a distance of ${\approx}$4\farcs5 from the position of the FXT (${\approx}$40~kpc in projection) with a spectroscopic redshift of $z_{\rm spec}{=}$1.5105 \citep{Andreoni2021,Jonker2021,Eappachen2023a}. The probability of a false match is $P{<}0.041$ \citep[adopting the formalism developed by][]{Bloom2002} for such offsets from similar or brighter objects.

To summarize, we conclude that four (FXTs~16, 19, 20 and 21) of the eight FXT candidates have high probabilities of being associated with faint (FXT~20) or moderately bright (FXTs~16, 19, and 21) extended sources within the 3$\sigma$ positional error circle. In the case of FXT~22, it may be associated with the extended source SDSS J134856.75+263946.7 ($z_{\rm spec}{=}$1.5105); nevertheless, a relation with a faint background source cannot be excluded \citep[a faint extended source is in the X-ray uncertainty region;][]{Eappachen2023a}. In the case of FXT~18, its large positional uncertainty does not allow us to determine robustly the counterpart optical or NIR source. 
Finally, two FXT candidates (FXTs~15 and 17) have no associated optical or NIR sources in the available moderate-depth archival imaging, and remain likely extragalactic FXTs. None of the FXT candidates analyzed in this work appear to be associated with a nearby galaxy (${\lesssim}$100~Mpc). In Sect. \ref{sec:galactic}, we explore a scenario where these sources are related to Galactic stellar flares from faint stars.

\begin{table*}
    \centering
    \advance\leftskip-0.3cm
    \scalebox{0.75}{
    \begin{tabular}{llllllllllll}
    \hline\hline
    FXT  & $m_g$ & $m_r$ & $m_i$ & $m_z$ & $m_y$ & $m_Y$ & $m_J$ & $m_H$ & $m_K$ & W1 & W2 \\ \hline
    (1) & (2) & (3) & (4) & (5) & (6) & (7) & (8) & (9) & (10) & (11) & (12)  \\ \hline
    15 &  ${>}$24.0$^{h}$ & ${>}$23.0$^{h}$ & ${>}$22.3$^{h}$ & ${>}$22.2$^{h}$ & ${>}$20.0$^{a}$ & -- & ${>}$16.9$^{f}$ & ${>}$16.6$^{f}$ & ${>}$17.2$^{f}$ & ${>}$20.3$^{c}$ & ${>}$20.6$^{c}$  \\  	
    16(S1) &  26.509$\pm$0.1$^{j}$ & 25.307$\pm$0.1$^{j}$ & 24.598$\pm$0.1$^{j}$ & 24.506$\pm$0.1$^{j}$ & -- & 24.450$\pm$0.1$^{j}$ & 24.300$\pm$0.1$^{j}$ & 23.852$\pm$0.1$^{j}$ & ${>}$20.1$^{g}$ & 22.407$\pm$0.1$^{j}$ & 21.928$\pm$0.1$^{j}$ \\
    17 & ${>}$24.1$^{h}$ & ${>}$22.4$^{h}$ & ${>}$21.8$^{h}$ & ${>}$21.5$^{h}$ & ${>}$20.0$^{a}$ & -- & ${>}$17.1$^{f}$ & ${>}$17.1$^{f}$ & ${>}$17.2$^{f}$ & ${>}$19.8$^{c}$ & ${>}$20.1$^{c}$ \\
    18(S1) & 24.6${\pm}$0.2$^{l}$ & 24.2$\pm$0.1$^{l}$ & 24.0$\pm$0.1$^{l}$ & 23.6$\pm$0.2$^{l}$ & $>$20.1$^{a}$ & -- & ${>}$19.8$^{g}$ & ${>}$19.2$^{g}$ & ${>}$19.2$^{g}$ & ${>}$19.3$^{c}$ & ${>}$20.2$^{c}$ \\
    18(S2) & 24.9$\pm$0.2$^{d}$ & 24.6$\pm$0.1$^{d}$ & 24.2$\pm$0.2$^{d}$ & 25.2$\pm$0.9$^{d}$ & 24.3$\pm$1.1$^{d}$ & -- & ${>}$19.8$^{g}$ & ${>}$19.2$^{g}$ & ${>}$19.2$^{g}$ & ${>}$19.3$^{c}$ & ${>}$20.2$^{c}$ \\
    18(S3) & 25.6${\pm}$0.7$^{h}$ & 23.0$\pm$0.04$^{d}$ & 21.8$\pm$0.2$^{a}$ & 21.2$\pm$0.1$^{a}$ & 21.1$\pm$0.1$^{d}$ & 20.2$\pm$0.2$^{d}$ & 20.1$\pm$0.1$^{g}$ & ${>}$19.8$^{b}$ & 19.7$\pm$0.2$^{b}$ & 19.3$\pm$0.1$^{c}$ & 20.2$\pm$0.3$^{c}$ \\
    19(S1) & ${>}$25.6$^{d}$ & 24.79$\pm$0.04$^{k\ddag}$ & 24.93$\pm$0.05$^{k\ddag}$ & ${>}$24.32$^{d}$ & ${>}$24.04$^{a}$ & -- & 24.67$\pm$0.04$^{k\ddag}$ & 24.33$\pm$0.04$^{k\ddag}$ & ${>}$20.3$^{g}$ & ${>}$20.1$^{c}$ & ${>}$20.3$^{c}$  \\
    20(S1) & 24.3${\pm}$0.2$^{l}$ & 23.5${\pm}$0.2$^{l}$ & ${>}$21.6$^{a}$ & 23.2${\pm}$0.2$^{l}$ & ${>}$19.8$^{a}$ & -- & ${>}$17.9$^{f}$ & ${>}$17.7$^{f}$ & ${>}$17.7$^{f}$ & ${>}$20.2$^{c}$ & ${>}$20.5$^{c}$ \\
    21(S1) & ${>}$23.6$^{h}$ & 25.09$\pm$1.49$^{a}$ & 22.68$\pm$0.13$^{a}$ & 21.73$\pm$0.13$^{a}$ & 21.55$\pm$0.28$^{a}$ & -- & ${>}$17.5$^{f}$ & ${>}$17.4$^{f}$ & ${>}$17.6$^{f}$ & ${>}$19.8$^{c}$ & ${>}$20.3$^{c}$ \\
    22(S1) & 20.9$\pm$0.2$^{h}$ & 21.4$\pm$0.4$^{h}$ & 20.4$\pm$0.3$^{h}$ & 20.5$\pm$0.9$^{h}$ & ${>}$19.7$^{a}$ & -- & ${>}$17.8$^{f}$ & ${>}$17.5$^{a}$ & ${>}$17.7$^{a}$ & ${>}$20.1$^{c}$ & ${>}$20.4$^{c}$ \\\hline
    \end{tabular}
    }
    \caption{Host and/or counterpart's photometric data or upper limits of FXT candidates. All magnitudes are converted to the AB magnitude system using \citet{Gonzalez2018} for VHS and 2MASS data, \citet{Hewett2006} for UKIDSS data, and \citet{Wright2010} for unWISE data. If an optical/NIR counterpart candidate is detected, we list its magnitude and 1-$\sigma$ error, otherwise we provide 3$\sigma$ limits from several catalogs: $^{a}$Pan-STARRS-DR2 \citep{Flewelling2018}, $^{b}$UKIDSS-DR11+ \citep{Warren2007}, $^{c}$unWISE \citep{Schlafly2019}, $^{d}$DES-DR2 \citep{Abbott2021}, $^{e}$NSC-DR2p \citep{Nidever2020}, $^{f}$2MASS \citep{Skrutskie2006}, $^{g}$VHS-DR5 \citep{McMahon2013}, $^{h}$SDSS-DR16 \citep{Ahumada2019}, $^{i}$ INT/CFHT \citep{Jonker2013}, $^{j}$ CANDELS \citep[nearest HST/Spitzer bands substituted: $g{=}F435W$, $r{=}F606W$, $i{=}F814W$, $z{=}F850LP$, $Y{=}F105W$, $J{=}F125W$, $H{=}F160W$, $W1{=}ch1$, $W2{=}ch2$;][]{Guo2013}, $^{k}$ (nearest HST bands substituted: $r{=}F606W$, $i{=}F814W$, $J{=}F110W$, $H{=}F160W$), $^l$ Legacy Surveys DR9 \citep{Dey2019}. \\
    $\dagger$ Photometric data of FXTs with counterpart(s) (S+\# means the \emph{source number}).\\
    $\ddag$ Obtained using a photometric aperture of 0\farcs6.\\
    }
    \label{tab:photometric_data}
\end{table*}

\subsubsection{Higher energy counterparts}\label{sec:gamma}

To explore if hard X-ray and $\gamma$-ray observations covered the sky locations of the FXTs, we developed a cone search in the Nuclear Spectroscopic Telescope Array \citep[\emph{NuStar};][]{Harrison2013}, \emph{Swift}-Burst Alert Telescope \citep[\emph{Swift}-BAT;][]{Sakamoto2008}, INTErnational Gamma-Ray Astrophysics Laboratory \citep[\emph{INTEGRAL};][]{Rau2005}, High Energy Transient Explorer 2 \citep[\emph{HETE-2};][]{Hurley2011}, \emph{InterPlanetary Network} \citep{Ajello2019}, and \emph{Fermi} \citep{von_Kienlin2014,Narayana2016} archives. We adopt a 10\farcm0 search radius for the \emph{INTEGRAL}, \emph{Swift}-BAT, \emph{HETE-2} and \emph{Interplanetary Network} Gamma-Ray Bursts catalogs, while for the Gamma-ray Burst Monitor (\emph{GBM}) and the Large Area Telescope (LAT) \emph{Fermi} Burst catalogs we consider a cone search radius of 4~deg \citep[which is roughly the typical positional error at 1$\sigma$ confidence level for those detectors;][]{Connaughton2015}. Additionally, we implement a time constraint criterion of $\pm$15~days in our search between Gamma-ray and FXT triggers.

To further probe whether there may be weak $\gamma$-ray emission below the trigger criteria of \emph{Fermi}-GBM at the location of the FXTs, we investigated the \emph{Fermi}-GBM daily data, the \emph{Fermi} position history files\footnote{https://fermi.gsfc.nasa.gov/ssc/data/access/gbm/}, and the \texttt{GBM Data Tools} \citep{GbmDataTools}\footnote{https://fermi.gsfc.nasa.gov/ssc/data/analysis/gbm/gbm\_data\_tools/gdt-docs/install.html}. We confirmed that FXTs~15, 16, 17, 20, and 21 were in the FoV of \emph{Fermi}-GBM instruments during the X-ray trigger time${\pm}50$~s, while FXTs~18, 19, and 22 were behind the Earth around the X-ray burst trigger time; thus, their fields were not visible. Table~\ref{tab:GBM} summarizes the visibility of the sources and the \emph{Fermi}-GBM instruments covering the fields around the X-ray trigger time (at a distance of ${\lesssim}60$~degrees).
In summary, we find no hard X-ray or $\gamma$-ray counterparts associated with \emph{NuSTAR}, \emph{INTEGRAL}, \emph{Swift}-BAT, \emph{HETE-2}, \emph{Interplanetary Network} and the GBM and LAT \emph{Fermi} Burst catalogs, but cannot rule out weak $\gamma$-ray emission for FXTs~18, 19, and 22.

\begin{figure*}
    \centering
    \includegraphics[scale=0.75]{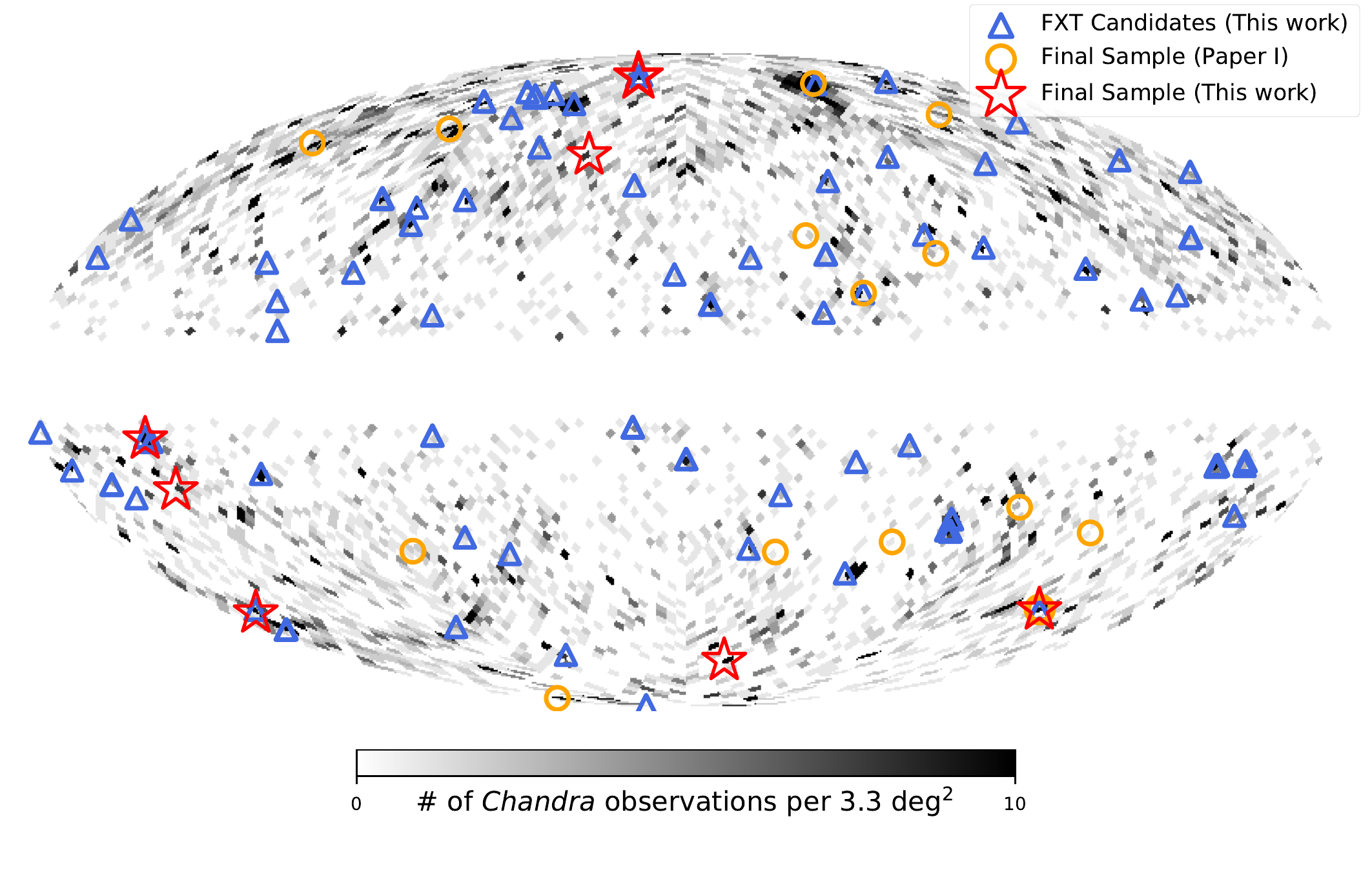}
    \vspace{-0.3cm}
    \caption{Sky positions, in Galactic coordinate projection, of FXT candidates: the initial 151 FXT candidates of this work are represented by \emph{blue triangles} (see Sect.~\ref{sec:initial_results}; some symbols overlap on the sky);  the \emph{final sample} of eight extragalactic FXT candidates from this work are denoted by \emph{large red stars} (FXTs~20 and 22 overlap on this scale); and the final sample of 14 extragalactic FXTs analyzed in Paper~I are shown as \emph{orange circles} (FXTs~14 and 16 overlap on this scale). The background grey scale encodes the location and number of distinct co- or closely-located observations among the combined 3899 and 5303 \emph{Chandra} observations used in this work and Paper I, respectively.}
    \label{fig:positions}
\end{figure*}

\begin{table*}
\centering
\scalebox{0.9}{
    \begin{tabular}{lllllllll}
    \hline\hline
    FXT & $T_0$(UTC) & Model & T$_{\text{break}}$(ks) & $\tau_1$ & $\tau_2$ & $F_0$ (erg~cm$^{-2}$~s$^{-1}$) & $\ln{\mathcal{L}}$(dof) & BIC \\ \hline
    (1) & (2) & (3) & (4) & (5) & (6) & (7) & (8) & (9)  \\ \hline
    15 &  2014-05-07 04:26:03.74 & BPL & 3.7$\pm$0.7 & -0.4$\pm$0.1 & 2.9$\pm$1.2 & (6.8$\pm$1.9)$\times$10$^{-14}$ & 1.3(2) & -2.1 \\
      & & PL & -- & 0.0$\pm$0.02 & -- & (1.0$\pm$1.1)$\times$10$^{-14}$ & 9.1(4) & 6.1 \\ \hline
    16 & 2015-03-22 07:02:29.30 & BPL & 2.2$\pm$0.3 & 0.09$\pm$0.1 & 2.0$\pm$0.3 & (4.1$\pm$0.8)$\times$10$^{-13}$ & 26.5(26) & 9.8 \\
      &  & PL & -- & 0.6$\pm$0.1 & -- & (1.0$\pm$0.1)$\times$10$^{-11}$ & 79.3(28) & 35.9 \\ \hline
    17 & 2015-11-21 03:26:16.08 & BPL & 0.4$\pm$2.5 & 0.4$\pm$0.1 & 0.4$\pm$1.0 & (1.8$\pm$3.9)$\times$10$^{-13}$ & 6.8(4) & 6.9 \\
      & & PL & -- & 0.3$\pm$0.1 & -- & (8.6$\pm$6.7)$\times$10$^{-13}$ & 6.3(6) & 2.2 \\ \hline
    18 & 2016-11-25 10:12:22.22 & BPL & -- & -- & -- & -- & -- & -- \\
      & & PL & -- & 0.5$\pm$0.1 & -- & (4.6$\pm$3.8)$\times$10$^{-11}$ & 1.6(2) & -0.8 \\ \hline
    19 & 2017-09-01 13:27:02.10 & BPL &  2.1$\pm$0.3 &  0.1$\pm$0.1 & 1.9$\pm$0.5 & (5.5$\pm$0.7)$\times$10$^{-13}$ & 36.9(38) & 9.5 \\
      & & PL & -- & 0.3$\pm$0.1 & -- & (3.3$\pm$1.5)$\times$10$^{-12}$ & 1.7(40) & 27.7 \\ \hline
    20 & 2019-11-27 02:12:13.71 & BPL & 0.1$\pm$0.01 & -1.0$\pm$0.2 & 2.5$\pm$0.3 & (4.2$\pm$0.5)$\times$10$^{-12}$ & 1.5(6) & -9.7 \\
      & & PL & -- & 0.5$\pm$0.3 & -- & (3.5$\pm$4.6)$\times$10$^{-12}$ & 21.9(8) & 12.5 \\ \hline
    21 & 2019-12-23 10:40:46.48 & BPL & -- & -- & -- & -- & -- & -- \\
      & & PL & -- & 0.4$\pm$0.1 & -- & (1.1$\pm$0.8)$\times$10$^{-12}$ & 4.6(4) & 1.9 \\ \hline
    22 & 2021-04-23 22:15:36.63 & BPL & 4.4$\pm$0.4 & 0.2$\pm$0.1 & 3.8$\pm$1.2 & (3.2$\pm$0.4)$\times$10$^{-13}$ & 17.9(26) & -1.7 \\
      & & PL & -- & 0.4$\pm$0.1 & -- & (3.5$\pm$1.6)$\times$10$^{-12}$ & 46.1(28) & 19.7 \\ \hline
    \end{tabular}
    }
    \caption{Best-fit parameters obtained using either a broken power-law (BPL) or a power-law (PL) model fit the \hbox{X-ray} light curves. 
    \emph{Column 2:} start time when the count rate is $\gtrsim$3-$\sigma$ higher than the Poisson background level.
    \emph{Column 3:} the model used. 
    \emph{Column 4:} the break time for the  BPL model.
    \emph{Columns 5 and 6:} the slope(s) for the BPL or PL model. 
    \emph{Column 7:} the normalization for the BPL or PL model. 
    \emph{Columns 8 and 9:} log-likelihood ($\ln{\mathcal{L}}$)/degrees-of-freedom (dof) and \emph{Bayesian Information Criterion} (BIC) of the fit, respectively. Errors are quoted at the 1-$\sigma$ confidence level.}
    \label{tab:fitting_para}
\end{table*}

\begin{figure*}
    \centering
    \includegraphics[scale=0.8]{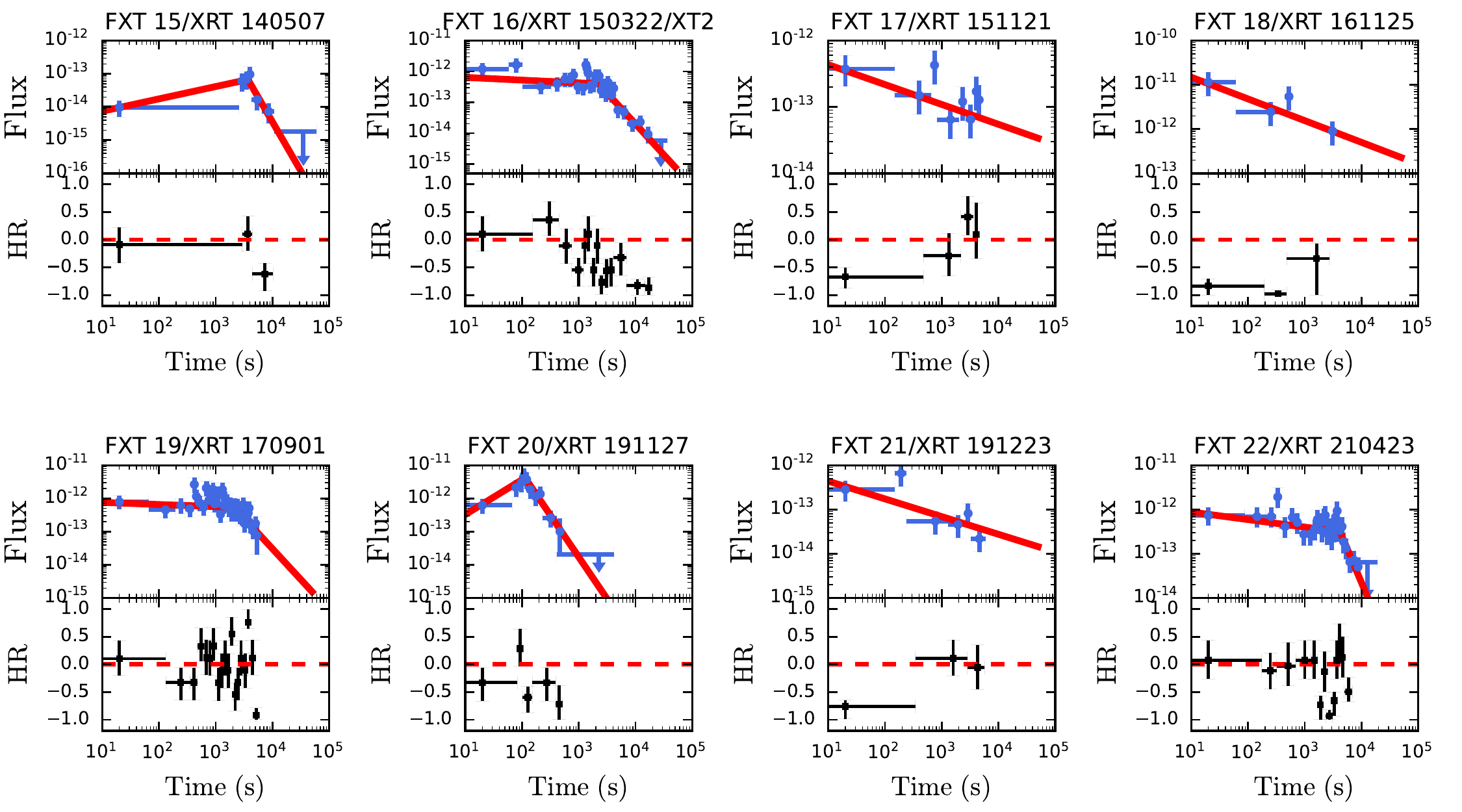}
    \vspace{-0.2cm} \caption{Top panels: the observed 0.5--7.0~keV \hbox{X-ray} light curves in cgs units (\emph{blue points}), starting at $T{=}20$~seconds. We also plot the best-fit broken power-law or simple power-law model (\emph{red solid lines}). The light curves contain 5 counts per bin. \emph{Bottom panels:} the hardness ratio evolution (the soft and hard energy bands are 0.5--2.0 keV and 2.0--7.0 keV, respectively), following the Bayesian method of \citet{Park2006}. The \emph{red dashed line} denotes a hardness ratio equal to zero. $T_0=0$~s is defined here as the time when the count rate is 3$\sigma$ higher than the Poisson background level.}
    \label{fig:models_BPL}
\end{figure*}

\section{Spatial, Temporal and X-ray Spectral properties}\label{sec:time_spectra_prop}

We analyze the spatial distribution of our final sample of FXT candidates in Sect. \ref{sec:spatial}. Furthermore, the time evolution and spectral properties could give important information about the physical processes behind the FXT candidates, and thus we explore and describe these in Sects. \ref{sec:X-ray_LC} and  \ref{sec:X-ray_fitting}, respectively. Finally, we explore a Galactic stellar flare origin of this sample in Sect. \ref{sec:galactic}.

\subsection{Spatial properties}\label{sec:spatial}

If the FXT candidates are extragalactic and arise from large distances, then given the isotropy of the universe on large scales, we expect them to be randomly distributed on the sky. Figure~\ref{fig:positions} shows the locations, in Galactic coordinates, of the final FXT candidates of Paper I and this work, the initial FXT candidates of this work, and the \emph{Chandra} observations analyzed in Paper I and this work. We investigate the randomness of the FXT candidate distribution on the sky compared to all \emph{Chandra} observations considered in this work. For this, we use the non-parametric Kolmogorov--Smirnov (K-S) test \citep{Kolmogorov1933,Massey1951,Ishak2014}. 

We explore the randomness of the spatial distribution of our final sample of eight FXTs. For this, we simulate 10,000 samples of 40,000 random sources distributed over the sky, taking as a prior distribution the \emph{Chandra} sky positions used in this work (which are functions of the pointings and exposures). Out of these fake sources, we randomly select eight sources, which we compare to the spatial distribution of the eight real FXT candidates  using a 2D K-S test (following the methods developed by \citealp{Peacock1983} and \citealp{Fasano1987}). We can reject the $\mathcal{NH}$ that these sources are drawn from the same (random) distribution only in ${\approx}$0.2\% of the draws. Therefore, the positions of the eight FXT candidates are consistent with being randomly distributed over the total \emph{Chandra} observations on the sky.

Intriguingly, FXTs~14 and 16 lie in the same field of view (i.e., in the \emph{Chandra} Deep Field South), as do FXTs~20 and 22 (i.e., in the direction of the galaxy cluster Abell 1795). Thus, we explore the probability that two FXTs occur in the same field, which is given by the Poisson statistic [i.e., $P(k,\alpha)$, using Eq.~\ref{eq:001}], where $k{=}2$ and $\alpha$ is the ratio between the total \emph{Chandra} exposure time in a particular field (for the \emph{Chandra} Deep Field South and the cluster Abell 1795 are 6.8 and 3.1~Ms, respectively\footnote{Both values taken from https://cda.harvard.edu/chaser}) and the total \emph{Chandra} exposure time analyzed in Paper~I and this work (${\approx}$169.6 and 88.8~Ms, respectively) normalized to the total number of FXTs identified (i.e., 22 FXTs). The chance probabilities for FXTs~14, 16 and 20, 22 are 0.115 and 0.029, respectively. We can conclude that the occurrence of two FXTs being found in these particular fields is unusual, but not ruled out at high significance.

\subsection{Temporal properties}\label{sec:X-ray_LC}

To characterize and measure the break times and light-curve slopes in the X-ray light curves of the candidate FXTs, we consider a single power-law (PL) model with index $\tau_1$, or a broken power-law (BPL) model with indices $\tau_1$, $\tau_2$ and break time $T_{\text{break}}$ (for more detail, see Paper~I, sect.~3.2).

Both models describe well the majority of the FXT X-ray light curves in this work. To fit the data, we use the least-square method implemented by the \texttt{lmfit} Python package.\footnote{https://github.com/lmfit/lmfit-py/} The best-fit model parameters and statistics are given in Table~\ref{tab:fitting_para}, while the light curves (in flux units; light curves have five counts per bin) and best-fit models are shown in Fig.~\ref{fig:models_BPL}. We define the light-curve zeropoint ($T_0{=}0$~sec) as the time when the count rate is 3$\sigma$ higher than the Poisson background level (see Table~\ref{tab:fitting_para}). To confirm the zeropoint, we divide the light curves in bins of $\Delta t{=}$100 and 10~seconds, and compute the chance probability that the photons per bin come from the background ($P_{\rm bkg}$)\footnote{We use the Poisson probability mass function, $P_{\rm bkg}{=}\exp{(-\mu)}\frac{\mu^k}{k!}$, where $k$ and $\mu$ are the number of photons per bin and the background rate multiple by the bin time, respectively.}. We found that the bins after $T_0$ have a $P_{\rm bkg}{\lesssim}$0.01, while $P_{\rm bkg}$ immediately before $T_0$ is higher $P_{\rm bkg}{\gtrsim}$0.1--0.2. We use the \emph{Bayesian Information Criterion} (BIC)\footnote{BIC${=}-2\ln{\mathcal{L}}+k\ln{N}$, where $\mathcal{L}$ is the maximum value of the data likelihood, $k$ is the number of model parameters, and $N$ is the number of data points \citep{Ivezic2014}.} to understand which of the two models describes better the data. We consider the threshold criterion of \hbox{$\Delta$BIC${=}$BIC$_{h}$-BIC$_{l}{>}2$} to discriminate when comparing two different models, where BIC$_{\rm h}$ is the higher model BIC, and BIC$_{\rm l}$ is the lower model BIC. The larger $\Delta$BIC, the stronger the evidence against the model with a higher BIC is \citep{Liddle2007}.

The parameters of the best-fitting models of the light curves are listed in Table~\ref{tab:fitting_para}, while Figure~\ref{fig:models_BPL} shows the best-fit broken power-law or simple power-law models. Five sources (FXTs~15, 16, 19, 20, and 22) require a break time (based on the BIC criterion), while three do not (FXTs~17, 18, and 21). In two of the former (FXT~15 and 20), $\tau_1$ is negative, indicating a discernible rise phase; the other three (FXTs~16, 19, and 22) are consistent with an early plateau phase.

\begin{table*}
    \centering
\scalebox{1.0}{
    \begin{tabular}{lllllllll}
    \hline\hline
    FXT & $z$ & $N_{\rm H,Gal}$ & $N_H$& $\Gamma$ & $\log{\rm Norm}$ & Flux & C-stat(dof) & $\ln \mathcal{Z}$ \\ \hline
    (1) & (2) & (3) & (4) & (5) & (6) & (7) & (8) & (9)\\ \hline
    15 & 0.0 & 0.5 & 3.8$_{-3.5}^{+12.5}$ & 2.1$_{-1.2}^{+2.0}$ & $-5.9_{-0.5}^{+0.9}$ & 0.3${\pm}$0.1 & 14.3(20) & $-15.4{\pm}0.01$ \\ 
     & 1.0 & 0.5 & 18.7$_{-18.5}^{+58.9}$ & 2.1$_{-1.2}^{+1.8}$ & $-5.9_{-0.5}^{+0.8}$ & 0.3${\pm}$0.7 & 14.3(20) & $-13.7{\pm}0.01$ \\ 
    16 &  0.0 & 0.2 & 2.4$_{-2.3}^{+7.3}$ & 2.1$_{-0.6}^{+0.7}$ & $-4.9_{-0.2}^{+0.3}$ & 2.9${\pm}$0.3 & 72.8(88) & $-46.8{\pm}0.02$ \\ 
     &  0.738 & 0.2 & 8.0$_{-7.4}^{+13.4}$ & 2.1$_{-0.5}^{+0.7}$ & $-5.0_{-0.2}^{+0.3}$ & 2.9${\pm}$0.3 & 73.1(88) & $-45.8{\pm}0.02$ \\ 
    17 & 0.0 & 0.2 & 18.1$_{-14.0}^{+16.6}$ & 3.4$_{-1.8}^{+1.5}$ & $-4.2{\pm}$0.9 & 2.3$_{-0.3}^{+0.5}$ & 24.9(34) & $-19.4{\pm}0.02$ \\ 
     &  1.0 & 0.2 & 66.8$_{-55.4}^{+31.8}$ & 2.6${\pm}1.4$ & $-4.6_{-0.7}^{+0.6}$ & 2.5$_{-0.3}^{+0.5}$ & 25.0(34) & $-18.4{\pm}0.02$ \\ 
    18 & 0.0 & 0.2 & 1.1$_{-1.0}^{+3.1}$ & ${>}$6.5 & $-4.1_{-0.3}^{+0.5}$ & 88.9$_{-17.7}^{+17.9}$ & 10.9(11) & $-15.4{\pm}0.02$ \\ 
     &  0.35 & 0.2 & 1.3$_{-1.1}^{+5.3}$ & ${>}$6.5 & $-4.1{\pm}$0.3 & 94.9$_{-20.8}^{+23.1}$ & 11.1(11) & $-15.2{\pm}0.02$ \\ 
    19 & 0.0 & 0.5 & 5.3$_{-4.3}^{+6.0}$ & 2.2${\pm}$0.6 & $-4.8_{-0.3}^{+0.4}$ & 3.1${\pm}$0.3 & 99.5(126) & $-59.9{\pm}0.02$ \\ 
     &  1.44 & 0.5 & 47.1$_{-36.2}^{+42.6}$ & 2.2${\pm}$0.6 & $-4.9{\pm}$0.3 & 3.1${\pm}$0.2 & 99.9(126) & $-57.9{\pm}0.02$ \\ 
    20 & 0.0 & 0.3 & 4.7$_{-4.4}^{+11.9}$ & 3.0$_{-1.3}^{+1.8}$ & $-4.7_{-0.5}^{+0.8}$ & 2.2${\pm}$0.5 & 30.6(36) & $-23.4{\pm}0.02$ \\
     &  1.0 & 0.3 & 25.3$_{-23.6}^{+57.9}$ & 3.0$_{-1.3}^{+1.8}$ & $-4.7_{-0.5}^{+0.8}$ & 2.2$_{-0.3}^{+0.5}$ & 30.6(36) & $-21.6{\pm}0.01$ \\
    21 & 0.0 & 0.7 & 2.5$_{-2.3}^{+8.3}$ & 3.1$_{-1.1}^{+1.4}$ & $-4.8_{-0.4}^{+0.6}$ & 1.9${\pm}$0.4 & 16.6(60) & $-17.0{\pm}0.01$ \\
     &  0.85 & 0.7 & 11.6$_{-11.4}^{+40.0}$ & 3.1$_{-1.1}^{+1.5}$ & $-4.8_{-0.4}^{+0.6}$ & 1.9${\pm}$0.5 & 16.8(60) & $-15.6{\pm}0.01$ \\ 
    22 & 0.0 & 0.3 & 1.9$_{-1.8}^{+5.9}$ & 2.3$_{-0.6}^{+0.9}$ & $-4.5_{-0.2}^{+0.4}$ & 8.7$_{-0.7}^{+0.9}$ & 78.1(62) & $-49.4{\pm}0.02$ \\ 
     & 1.5105 & 0.3 & 17.0$_{-15.5}^{+50.1}$ & 2.2$_{-0.5}^{+0.7}$ & $-4.5_{-0.2}^{+0.3}$ & 8.6$_{-1.0}^{+0.9}$ & 78.1(62) & $-47.3{\pm}0.02$ \\ \hline
    \end{tabular}
    }
    \caption{Results of the 0.5--7 keV \hbox{X-ray} spectral fits for the final sample of FXT candidates.
    \emph{Column 2:} Redshift assumed ($z{=}$0, 1, or from Table~\ref{tab:SED_para}). 
    \emph{Columns 3 and 4:} Galactic and intrinsic column density absorption ($\times$10$^{21}$) in units of cm$^{-2}$, respectively. The former is kept fixed during the fit. 
    \emph{Column 5:} Photon index from the power-law model. 
    \emph{Column 6:} Normalization parameter (in units of photons~keV$^{-1}$~cm$^{-2}$~s$^{-1}$). 
    \emph{Column 7:} Absorbed fluxes ($\times$10$^{-14}$) in units of erg~cm$^{-2}$~s$^{-1}$ (0.5--7.0~keV). 
    \emph{Column 8:} C-stat value and the number of degrees of freedom. 
    \emph{Column 9:} the log-evidence ($\ln \mathcal{Z}$) values for each model. The errors are quoted at the 3$\sigma$ confidence level from the posterior distributions obtained by BXA except for the flux (which is quoted at 1$\sigma$).}
    \label{tab:spectral_para}
\end{table*}

\begin{figure*}
    \centering
    \leftskip-0.2cm
    \includegraphics[scale=0.7]{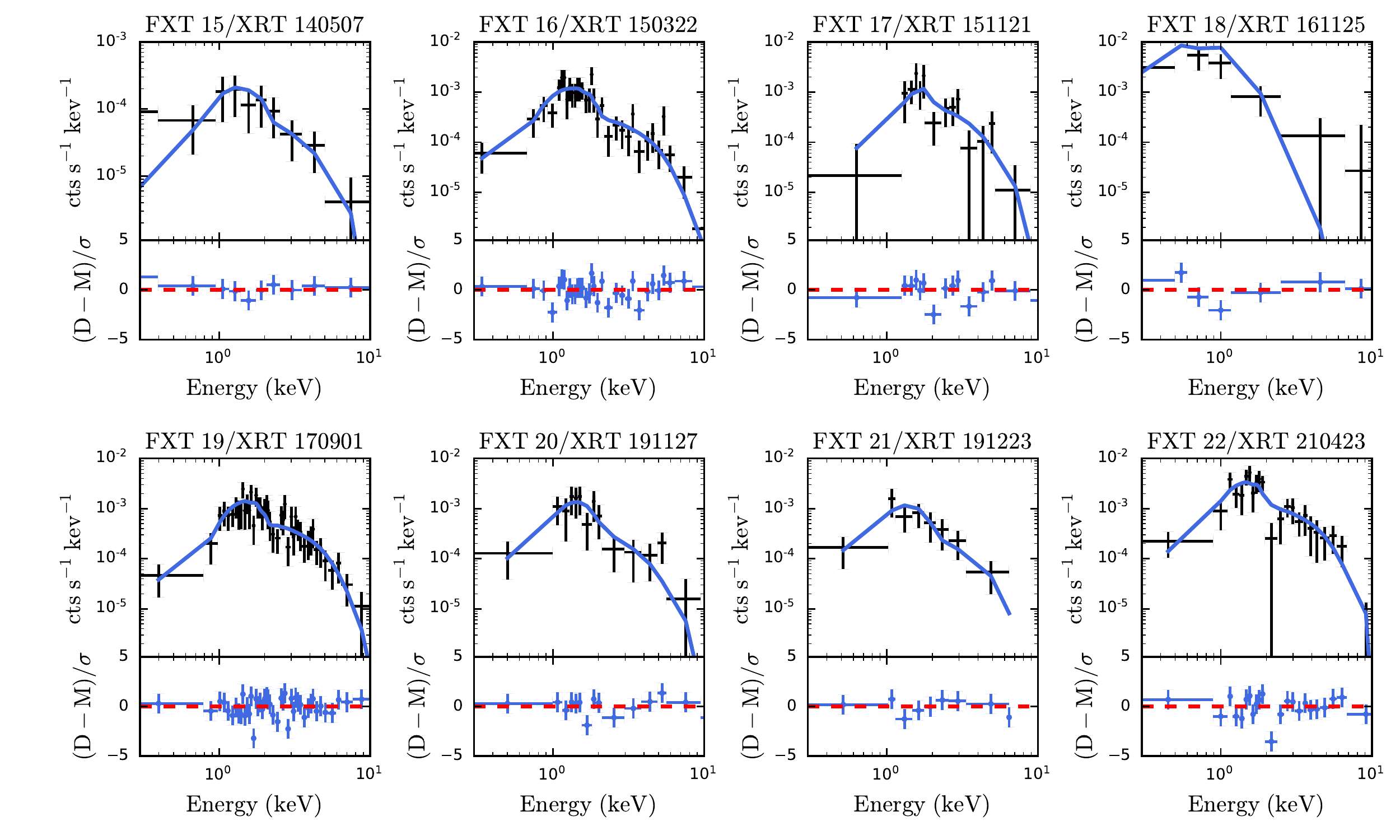}
    \vspace{-0.2cm} \caption{\emph{Top panels:} X-ray spectra (\emph{black crosses}), in units of counts~s$^{-1}$~keV$^{-1}$. We also plot the best-fit absorbed power-law (\emph{blue lines}) spectral model; see Table~\ref{tab:spectral_para} for the corresponding best-fitting parameters. \emph{Bottom panels:} residuals (defined as data-model normalized by the uncertainty) of each spectral model.}
    \label{fig:X_ray_spectra}
\end{figure*}

\subsection{Spectral properties}\label{sec:X-ray_fitting}

Using X-ray spectra and response matrices generated following standard procedures for point-like sources using {\sc ciao} with the \texttt{specextract} script, we analyze the spectral parameters of the FXT candidates. The source and background regions are the same as those previously generated for the light curves (see Sect.~\ref{sec:LC}). To find the best-fit model, because of the low number of counts, we consider the maximum likelihood statistics for a Poisson distribution called Cash-statistics \citep[C-stat;][]{Cash1979}.\footnote{The Cash-statistic, C-stat, is defined as $C{=}-2\ln{L_{\rm Poisson}}{+}{\rm const}$} Because of the Poisson nature of the X-ray spectral data, the C-stat is not distributed like $\chi^2$ and the standard goodness-of-fit is inapplicable \citep{Buchner2014,Kaastra2017a}. Thus, similarly to Paper~I, we use \emph{the Bayesian X-ray Astronomy package} \citep[BXA;][]{Buchner2014}, which joins the Monte Carlo nested sampling algorithm \texttt{MultiNest} \citep{Feroz2009} with the fitting environment of \texttt{XSPEC} \citep{Arnaud1996}. BXA computes the integrals over the parameter space, called the evidence ($\mathcal{Z}$), which is maximized for the best-fit model, and assuming uniform model priors.

We consider one simple continuum model: an absorbed power-law model (\texttt{phabs*zphabs*po}, hereafter the PO model), which is typically thought to be produced by a non-thermal electron distribution. We choose this simple model because we do not know the origin and the processes behind the emission of FXTs. Furthermore, the low number of counts does not warrant more complex models. The spectral absorption components \texttt{phabs} and \texttt{zphabs} represent the Galactic and intrinsic contribution to the total absorption, respectively. During the fitting process, the Galactic absorption ($N_{\rm H,Gal}$) was fixed according to the values of \citet{Kalberla2005} and \citet{Kalberla2015}, while for the intrinsic neutral hydrogen column density ($N_{\rm H}$), we carried out fits for both $z{=}0$ (which provides a lower bound on $N_{\rm H}$ since firm redshifts are generally not known, and is useful for comparison with host-less FXTs) and the redshift values from Table~\ref{tab:SED_para} or fiducial values of $z{=}1$ for host-less sources.

\begin{figure}
    \centering
    \includegraphics[scale=0.8]{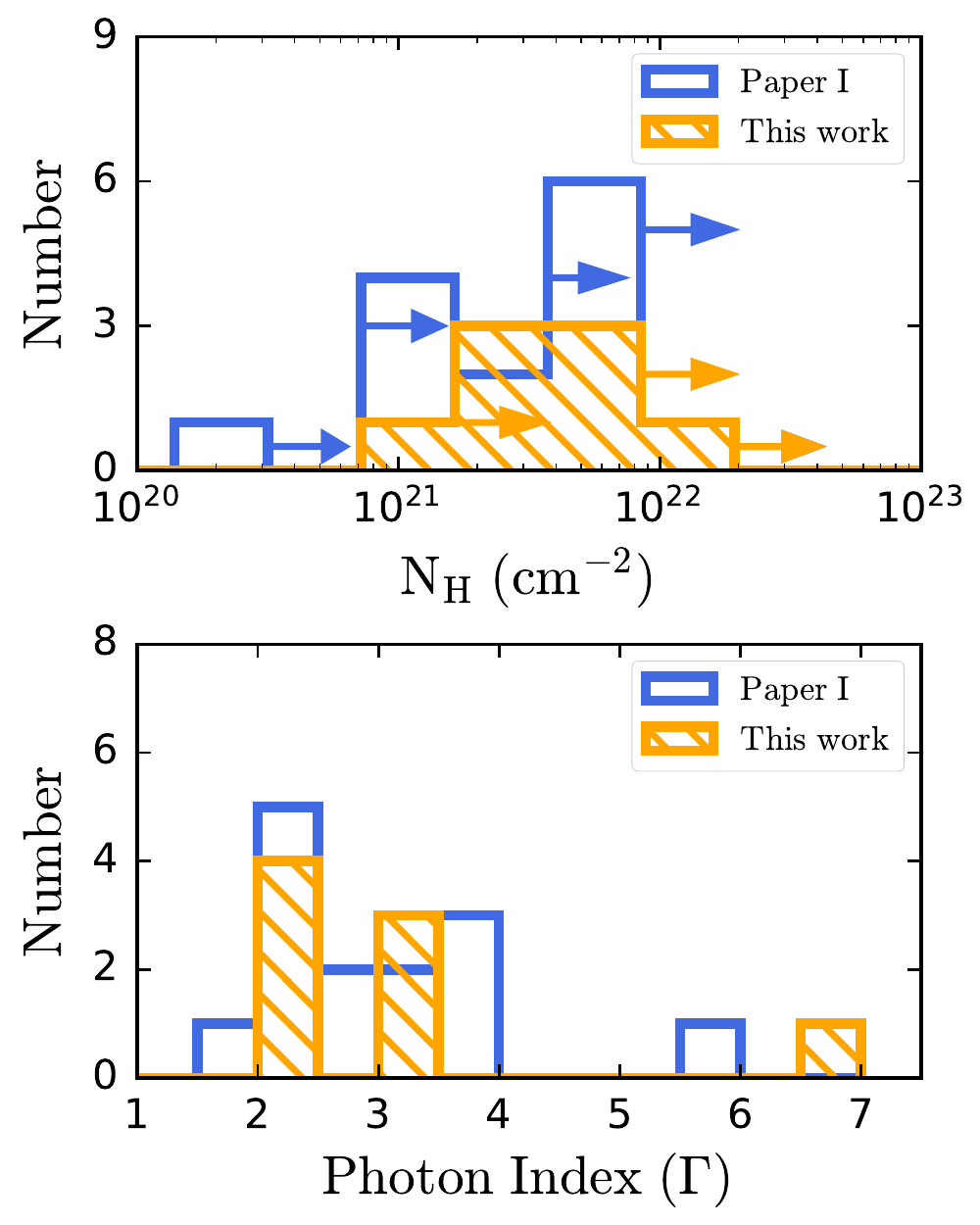}
    \vspace{-0.1cm}
    \caption{\emph{Top panel:} Intrinsic neutral hydrogen column density distribution, evaluated at $z{=}0$ and in units of cm$^{-2}$, obtained using the \emph{power-Law} model for extragalactic FXT candidates from this work (\emph{orange histogram}) and Paper~I (\emph{blue histogram}). The \emph{arrows} indicate that the $z{=}0$ intrinsic hydrogen column densities are lower bounds. \emph{Bottom panel:} Photon index distribution, obtained using a \emph{power-law} model, for FXT candidates from this work (\emph{orange histogram}) and Paper~I (\emph{blue histogram}). Note that the uncertainties on these parameter values for individual sources can be considerable (see Table~\ref{tab:spectral_para}.)}
    \label{fig:X-ray-params}
\end{figure}

The best-fitting absorbed power-law models (and their residuals) and their parameters are provided in Fig.~\ref{fig:X_ray_spectra} and Table~\ref{tab:spectral_para}, respectively; additionally, Fig.~\ref{fig:X-ray-params} shows the histograms of the best-fit intrinsic neutral hydrogen column densities ($N_{\rm H}$; \emph{top panel}) and photon index ($\Gamma$; \emph{bottom panel}) for extragalactic FXTs candidates of this manuscript (\emph{orange histograms}) and from Paper~I (\emph{blue histograms}). The candidates show a range of $N_{\rm H}{\approx}$(1.1--18.1)$\times$10$^{21}$~cm$^{-2}$ (assuming $z{=}0$), and a mean value of $\overline{N}_{\rm H}{\approx}$5.0$\times$10$^{21}$~cm$^{-2}$, consistent with the range for sources reported by Paper~I (see Fig.~\ref{fig:X-ray-params}, \emph{top panel}). We note that in all cases here, the best-fit $N_{\rm H}$ is higher than the $N_{\rm H,Gal}$ estimates from \citet{Kalberla2005} and \citet{Kalberla2015} by a factor of $\approx$4--90. In every case, intrinsic absorption and the Galactic component are needed, with at least ${\approx}95$\% confidence, and in some cases even ${\approx}99$\% confidence level. Therefore, two absorption components are needed in the fitting process in general.

Furthermore, excluding the soft candidate FXT~18, the best-fit power-law photon index ranges between $\Gamma{\approx}$2.1--3.4 for the candidate FXTs, with a mean value of $\overline{\Gamma}{=}$2.6. FXT~18 is an exceptionally soft source ($\Gamma{\gtrsim}$6.5) compared to both this sample and the FXT candidates presented in Paper~I (see Fig.~\ref{fig:X-ray-params}, \emph{bottom panel}). Finally, FXTs~17, 18, and 21, whose light curves are best-fitted by a PL model, have some of the softest photon indices ($\Gamma{\gtrsim}$3).

\begin{figure}
    \centering
    \includegraphics[scale=0.75]{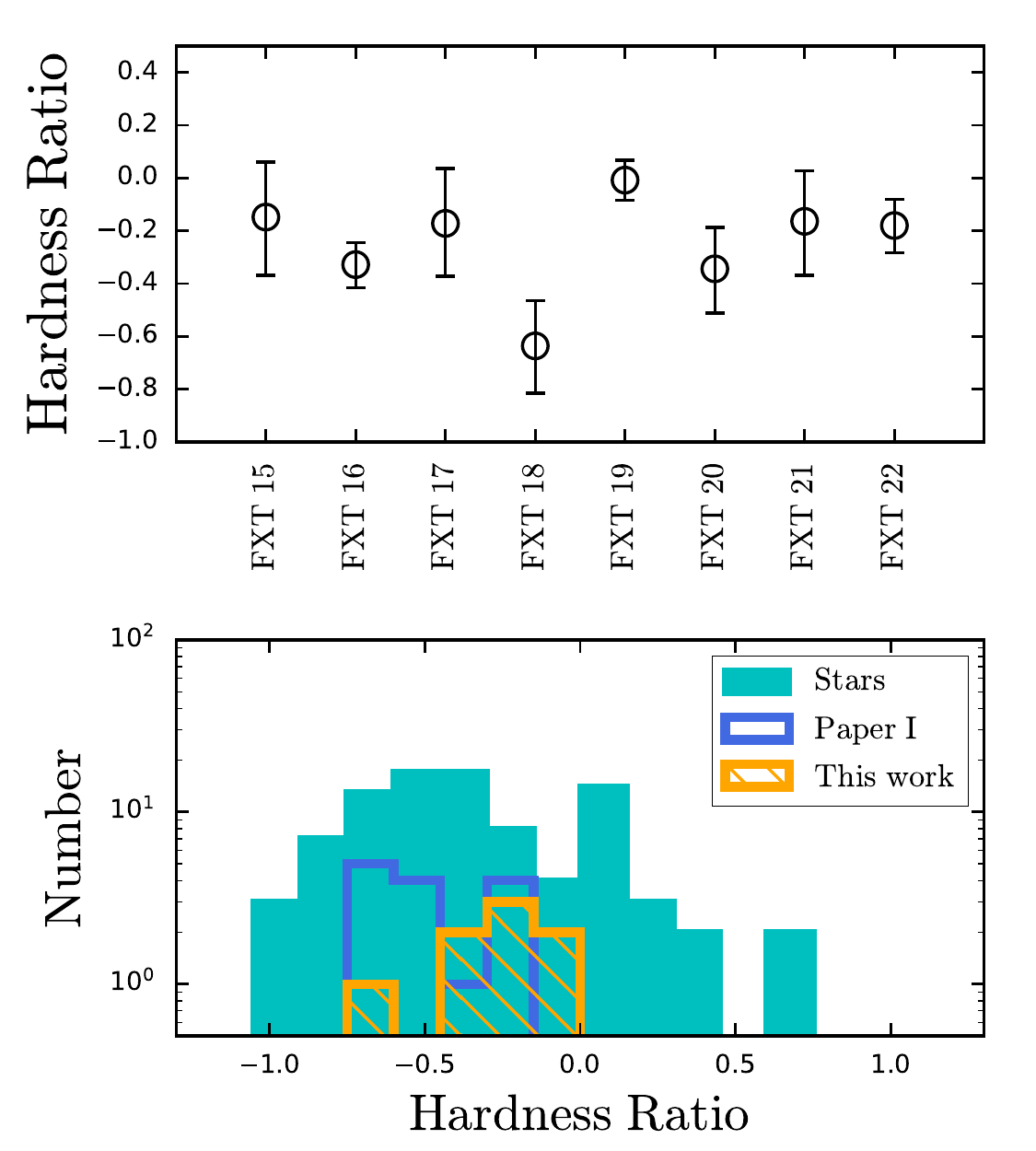}
    \vspace{-0.3cm}
    \caption{\emph{Top panel:} Hardness ratio of each FXT candidate \citep[using the Bayesian \texttt{BEHR} code;][]{Park2006} at 1$\sigma$ confidence level. \emph{Bottom panel:} Hardness-ratio distributions of our final samples of FXTs (\emph{orange} histogram), compared to the X-ray transients classified as ``stars'' by \emph{Criterion 2} using \emph{Gaia} (filled \emph{cyan} histogram) and the sources identified previously as \emph{distant} FXTs (\emph{blue histogram}) by Paper~I. 
    }
    \label{fig:hardness}
\end{figure}

\subsubsection{Hardness ratio and photon index evolution}\label{sec:hardness_ratio}

The hardness ratio (HR) can be used to distinguish between \hbox{X-ray} sources, and permit us to explore their spectral evolution, especially in cases with low-count statistics \citep[e.g.,][]{Lin2012,Peretz2018}. In this work, the HR is defined as:
\begin{equation}
    HR=\frac{H-S}{H+S},
\end{equation}
where $H$ and $S$ are the number of X-ray photons in the \hbox{0.5--2.0}~keV soft and 2.0--7.0~keV hard energy bands. For each source candidate, we calculate the HR using the Bayesian code \texttt{BEHR} \citep[][]{Park2006}, which we list in Table~\ref{tab:my_detections}, \emph{column 11}, and plot in Fig.~\ref{fig:hardness} (\emph{top panel}).

We compare the HR of the 90 objects identified as ``stars'' by \emph{Criterion 2} (see Fig.~\ref{fig:hardness}, \emph{bottom panel, cyan histogram}) in Sect.~\ref{sec:gaia} with the final sample of FXTs in this work (\emph{orange histogram}) and the sample of FXTs reported by Paper~I (\emph{blue histogram}). Stars typically have very soft X-ray spectra \citep{Gudel2009}, confirmed by the fact that ${\approx}$90\% of the star candidates strongly skew toward soft HRs (${\lesssim}$0.0). Clearly, Fig.~\ref{fig:hardness} shows that FXTs do not stand out in the HR plane; thus, HR is not a useful discriminator on its own between stellar contamination and extragalactic FXTs.

We also analyze how the HR and power-law index of the X-ray spectrum change with time. To this end, we compute the time-dependent HR, with the requirement of 10 counts per time bin from the source region (to improve the statistics), which we show in the  \emph{lower panels} of Fig.~\ref{fig:models_BPL}. For sources that are well-fit by a BPL model, we also split the \texttt{event files} at $T_{\rm break}$ and extract and fit the spectra to compute spectral slopes "before" and "after" $T_{\rm break}$ ($\Gamma_{\rm before}$ and $\Gamma_{\rm after}$, respectively; see Table~\ref{tab:fitting_para_Tbreak}) using the value for the absorption derived from the fit to the full spectrum (see Table~\ref{tab:spectral_para}). We fit both spectral intervals together assuming fixed, constant $N_{\rm H,Gal}$ and $N_H$ (taken from Table~\ref{tab:spectral_para}).

The spectra of FXT~16 clearly softens after the plateau phase (Fig.~\ref{fig:models_BPL} and Table~\ref{tab:fitting_para_Tbreak}) at $>$90\% confidence. Similar spectral evolution was also seen from previous FXT candidates XRT~030511 and XRT~110919 (Paper~I). FXTs~15 and 20 exhibit similar spectral softening trends, with $T_{\rm break}$ as a pivot time, although with only marginal significance, while the rest show no obvious evidence of such trends (Fig.~\ref{fig:models_BPL} and Table~\ref{tab:fitting_para_Tbreak}). Finally, it is important to mention that the FXTs whose light curves follow a PL model (FXTs~17, 18, and 21) show hardening trends in their HR evolution (see Fig.~\ref{fig:models_BPL}).

\begin{table}
    \centering
    \begin{tabular}{lll}
    \hline\hline
    FXT & $\Gamma_{\rm before}$($T{<}T_{\rm break}$) & $\Gamma_{\rm after}$($T{\geq}T_{\rm break}$)\\ \hline
    (1) & (2) & (3) \\ \hline
    15 & 1.2$_{-1.0}^{+1.0}$ & 2.8$_{-1.1}^{+1.2}$ \\
    16 & 1.6$_{-0.4}^{+0.4}$ & 2.9$_{-0.5}^{+0.5}$ \\
    19 & 2.1$_{-0.3}^{+0.3}$ & 2.4$_{-0.5}^{+0.5}$ \\
    20 & 2.5$_{-1.2}^{+2.0}$ & 3.3$_{-1.1}^{+1.2}$ \\
    22 & 2.4$_{-0.5}^{+0.5}$ & 2.1$_{-0.6}^{+0.8}$ \\ \hline 
    \end{tabular}
    \caption{
    \emph{Columns 2 and 3:} spectral photon index computed before and after the $T_{\rm break}$ for light curves that are well-fit with a broken power-law. Errors are quoted at the 90\% confidence level.} 
    \label{tab:fitting_para_Tbreak}
\end{table}

\subsection{Galactic origin}\label{sec:galactic}

From our sample, FXTs~16 and 19 are clearly aligned with extended objects, proving an extragalactic origin. FXTs~18, 20, 21, and 22, based on their low random match probabilities, could be associated with potential hosts, supporting an extragalactic association (see Sect. \ref{sec:results} for more details).
In the next paragraphs, similar to Paper~I (see its Sect. 3.4 for more details), we explore any of the FXT candidates could still be associated with magnetically active M- or brown-dwarf flares, which are known to produce X-ray flares on timescales of minutes to hours, with flux enhancements up to two orders of magnitude \citep[not only at X-ray wavelengths;][]{Schmitt2004,Mitra-Kraev2005,Berger2006,Welsh2007}. 

Stellar flares typically show soft thermal spectra with temperatures of the order of $kT{\sim}$\hbox{0.5--1}~keV. M-dwarf stars (brown-dwarfs) have optical and NIR absolute magnitudes in the range of $M_z{\sim}$8--13~AB~mag ($M_z{\sim}$\hbox{13--18}~AB~mag) and $M_{K_s}{\sim}$3--10~AB~mag ($M_J{\sim}$15--25~AB~mag), respectively \citep{Hawley2002, Avenhaus2012, Tinney2014}.

The enhanced X-ray emission of M dwarfs shows flares on the order of $L_X^{\rm M-dwarf}{\sim}$10$^{28}$--10$^{32}$~erg~s$^{-1}$ \citep{Pallavicini1990,Pandey2008,Pye2015}, while brown dwarf flares cover a luminosity range of $L_X^{\rm B-dwarf}{\sim}$10$^{27}$--10$^{30}$~erg~s$^{-1}$ \citep{Berger2006,Robrade2010}. Empirically, the ratio between the X-ray luminosity ($L_X$) and bolometric luminosity ($L_{\rm bol}$) of cool M dwarfs and L dwarfs typically exhibits values no larger than $\log(L_X/L_{\rm bol}){\lesssim}$0.0 and ${\lesssim}-3.0$, respectively \citep[e.g.,][]{Garcia2008,DeLuca2020}. Adopting this limiting ratio, we rule out a stellar flare scenario for FXT candidates.  As in Paper~I, we compute the ratio $\log(L_X/L_{\rm bol})$ considering stellar synthetic models of dwarf stars \citep[taken from][$1000{\lesssim}T_{\rm eff}{\lesssim}3000$~K and $2.5{\lesssim}\log{g}{\lesssim}5.5$]{Phillips2020}, normalised to the deepest photometric upper limits and/or detections (as listed in Table~\ref{tab:photometric_data}), and compute bolometric fluxes by integrating the normalized models. We describe the constraints for each FXT below:

For FXT~15, the $m_g{>}$24.0 and $m_r{>}$23.0~AB~mag limits imply distances to any putative M- and brown dwarfs responsible for the X-ray flares of ${\gtrsim}$0.2--1.7~kpc and ${\gtrsim}$0.02--0.2~kpc, respectively. The corresponding X-ray flare luminosities would be $L_X^{\rm M-dwarf}{\gtrsim}$(7.0--690)$\times$10$^{29}$ and $L_X^{\rm B-dwarf}{\gtrsim}$(7.0--700)$\times$10$^{27}$~erg~s$^{-1}$, respectively. These are not enough to discard a Galactic stellar flare nature. Furthermore, the ratio $\log(F_X/F_{\rm bol}){\gtrsim}-0.9$ to $-1.4$ remains consistent with the extreme spectral type L1 stars \citep[e.g., J0331-27 with $\log(F_X/F_{\rm bol}){\sim}0.0$;][]{DeLuca2020}. Thus, we cannot completely rule out an extreme stellar flare origin for FXT~15.

In the case of FXT~17, the $m_g{>}24.1$ and $m_r{>}22.4$~AB~mag limits imply distances of ${>}$0.2--1.8~kpc and ${>}$0.02--0.2~kpc for M- and brown-dwarfs, respectively, and corresponding X-ray flare luminosities are $L_X^{\rm M-dwarf}{\gtrsim}$(1.5--149)$\times$10$^{31}$ and $L_X^{\rm B-dwarf}{\gtrsim}$(1.5--150)$\times$10$^{29}$~erg~s$^{-1}$, respectively. The X-ray-to-total flux ratio is $\log(F_X/F_{\rm bol}){\gtrsim}-0.1$ to $+0.4$. Based on these, we cannot discard a Galactic stellar flare association for FXT~17.

For FXT~18, due to the large X-ray positional uncertainty, there are several possible optical counterparts. We consider only \emph{source \#1} here, as it is closest, lying inside the 2$\sigma$ \hbox{X-ray} uncertainty position (see Fig.~\ref{fig:image_cutouts}). The $m_g{=}$24.6 and $m_i{=}$24.9~AB~mag DECam detections (see Table~\ref{tab:photometric_data}) implies distances of ${\approx}$0.3--3.1~kpc and ${\approx}$0.03--0.3~kpc for M- and brown-dwarfs, respectively, and corresponding X-ray flare luminosities of $L_X^{\rm M-dwarf}{\approx}$(\hbox{5.9--600})$\times$10$^{33}$ and $L_X^{\rm B-dwarf}{\approx}$(5.9--590)$\times$10$^{31}$~erg~s$^{-1}$, respectively. The ratio $\log(F_X/F_{\rm bol})$ is ${\approx}$2.5 to 3.0. These allow us to rule out robustly any Galactic stellar flare origin.

In the case of FXT~20, we consider \emph{source \#1} for the stellar flaring analysis (see Fig.~\ref{fig:image_cutouts}). The detections $m_g{=}24.3$ and $m_r{=}23.5$~AB~mag equate to distance ranges of ${\approx}$0.2--2.5~kpc and ${\approx}$0.02--0.3~kpc for M- and brown-dwarfs, respectively, and corresponding X-ray flare luminosities of $L_X^{\rm M-dwarf}{\approx}$(1.5--147)$\times$10$^{32}$ and $L_X^{\rm B-dwarf}{\approx}$(1.5--150)$\times$10$^{30}$~erg~s$^{-1}$, respectively. The ratio $\log(F_X/F_{\rm bol})$ is $ {\approx}$0.9 to 1.4.  These allow us to discard a Galactic stellar flare origin for FXT~20.

Finally, for FXT~21, the $m_i{=}$22.7~AB~mag PanSTARRS detection yields distance ranges of ${\approx}$0.4--4.1~kpc and ${\approx}$\hbox{0.04--0.4}~kpc for M- and brown-dwarfs, respectively, and corresponding X-ray flare luminosities of $L_X^{\rm M-dwarf}{\approx}$(3.0--300)$\times$10$^{31}$ and $L_X^{\rm B-dwarf}{\approx}$(3.1--300)$\times$10$^{29}$~erg~s$^{-1}$, respectively. The ratio $\log(F_X/F_{\rm bol})$ is ${\approx}$0.2 to 0.7. Thus, we can rule out FXT~21 as a stellar flare.

In summary, the multi-wavelength photometry indicate that three FXTs (FXTs~18, 20, and 21) appear inconsistent with stellar flaring episodes from Galactic M dwarfs and brown dwarfs, while deeper observations are required to completely rule out this option for FXTs~15 and 17.

\begin{table*}
    \centering
    \scalebox{1.0}{
    \begin{tabular}{lllllllllll}
    \hline\hline
    FXT & RA (deg) & DEC (deg) & Offset & $z$ &  Log(M$_*/M_\odot$) & SFR/($M_\odot$/yr) & $A_V$ (mag) & Ref. \\ \hline
    (1) & (2) & (3) & (4) & (5) & (6) & (7) & (8) & (9) \\ \hline
    \multicolumn{8}{c}{Parameters obtained from the literature} \\ \hline
    16 & 53.07658 & -27.87332 & 0\farcs44 & 0.738 & 9.07 & 0.81 & 0.02 & 1,2 \\ \hline
    \multicolumn{8}{c}{Parameters derived from photometric data using \texttt{BAGPIPES} \citep{Carnall2018}} \\ \hline
    16 & 53.07658 & -27.87332 & 0\farcs44 & 0.738 & 8.91${\pm}$0.04 & 2.98$_{-0.57}^{+0.78}$ & 1.12${\pm}0.12$ & --\\
    18(S1) & 36.7144 & -1.0826 & 2\farcs63 & 0.35$_{-0.15}^{+0.05}$ & 7.87$_{-0.33}^{+0.27}$ & 0.17$_{-0.09}^{+0.38}$ & 0.36$_{-0.21}^{+0.76}$ & --\\ 
    19(S1) & 356.26812 & -42.64566 & 0\farcs45 & 1.44${\pm}0.08$ & 8.67${\pm}0.11$ & 2.59$_{-0.48}^{+0.66}$ & 0.16${\pm}0.10$ & -- \\
    21(S1) & 50.47531 & 41.24695 & 0\farcs46 & 0.85${\pm}0.14$ & 11.20$_{-0.27}^{+0.24}$ & 1.66$_{-1.64}^{+26.53}$ & 1.48$_{-0.77}^{+0.37}$ & -- \\
    22(S1)$^a$ & 207.23646 & 26.66300 & 4\farcs6 & 1.5105 & 10.73${\pm}0.62$ & 35.23$_{-19.65}^{+50.66}$ & 0.63${\pm}0.45$ & 3,4 \\ \hline
    \end{tabular}
    }
    \caption{Parameters obtained from the literature and by our SED fitting to archival photometric data using the \texttt{BAGPIPES} package \citep{Carnall2018}.
    \emph{Column 2 and 3:} Right ascension and declination of the host galaxies. \emph{Column 5:} Angular offset between the transient and the host galaxy. 
    \emph{Column 5:} Host galaxy redshift or distance. 
    \emph{Columns 6 and 7:} Logarithmic values of the stellar mass, and the  SFR from the host galaxies. 
    \emph{Column 8:} Dust attenuation.
    \emph{Column 9:} Literature references.\\
    References: (1) \citet{Xue2019}, (2) \citet{Schlafly2011}, (3) \citet{Jonker2021}, (4) \citet{Andreoni2021} \\
    $^a$ Assuming an association with \emph{source \#1} at $z=1.5105$.
    }
    \label{tab:SED_para}
\end{table*}

\begin{figure}
    \centering
    \hspace{-0.3cm}
    \includegraphics[scale=0.85]{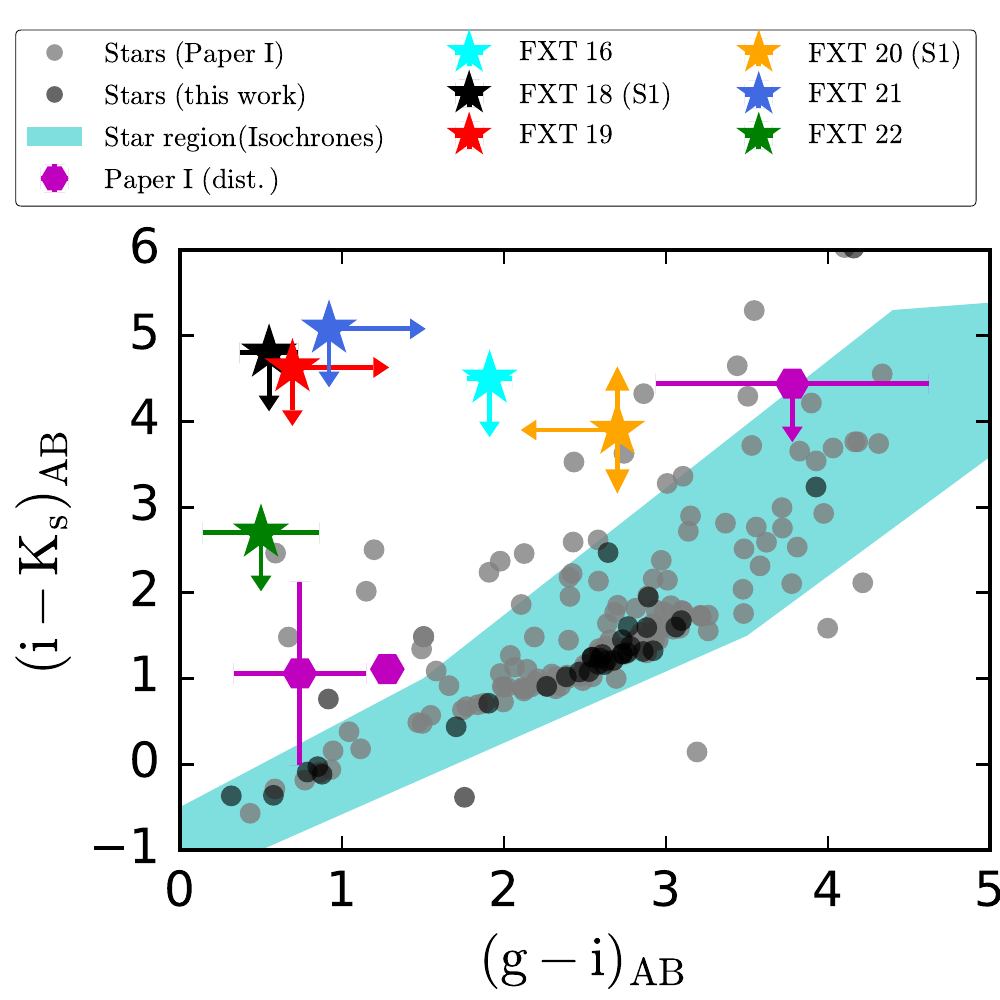}\\
    \hspace{-0.3cm}
    \includegraphics[scale=0.85]{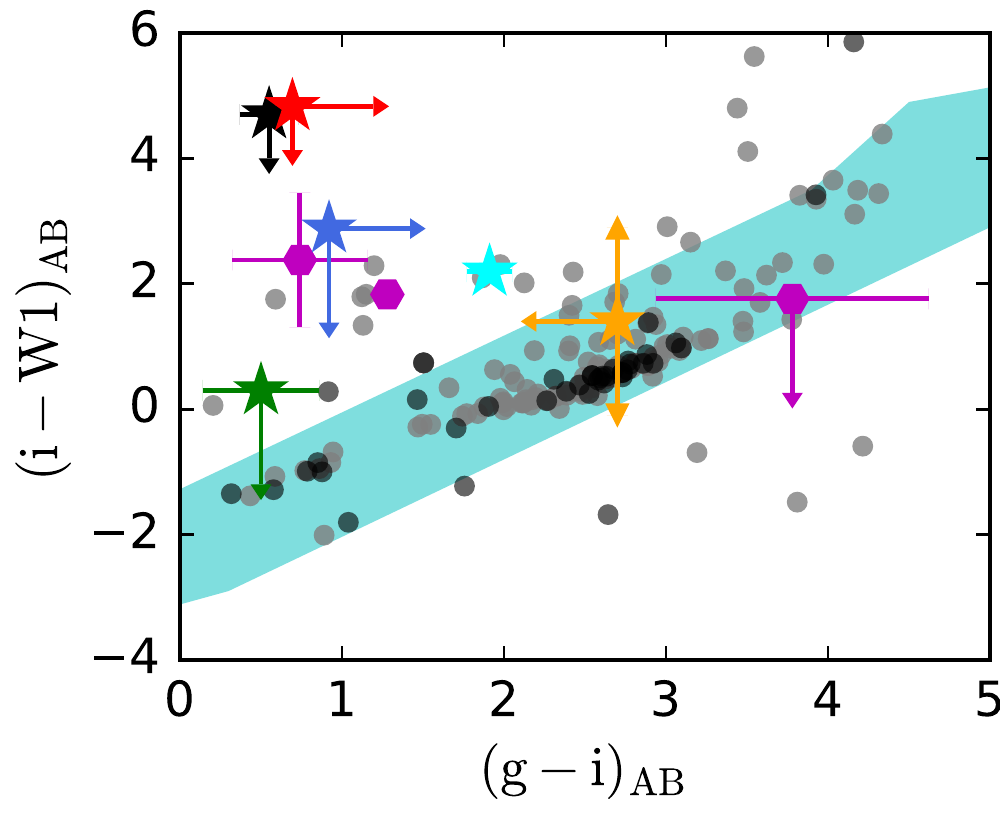}
    \vspace{-0.3cm}
    \caption{Colour-colour diagrams of the potential host galaxies associated with FXTs~16, 18, 19, 20, 21 and 22 (\emph{coloured stars}), ``distant'' FXTs from Paper~I (\emph{magenta hexagons}), and X-ray sources classified as stars in Paper~I (\emph{gray filled circles}) and here according to {Criterion 2} in Sect. \ref{sec:gaia} (\emph{black filled circles}). The expected parameter space of stars with different ages [$\log(\rm{Age}){=}$7.0-10.3], metallicities ($\rm{[Fe/H]}{=}-$3.0--0.5), and attenuations ($A_V{=}$0.0--5.0) taken from the MIST package \citep{Dotter2016, Choi2016} are overplotted as \emph{cyan} regions.}
    \label{fig:colour_colour_2}
\end{figure}

\subsection{One or two populations of FXTs?}\label{sec:population}

In Paper~I, we found that the FXT candidates could be robustly classified into ``\emph{local}'' and ``\emph{distant}'' populations (see their Sect.~3.5), based on the proximity of some sources to local galaxies (distances${\lesssim}$100~Mpc). In contrast, we find no plausible and robust association between our final sample and local galaxies in the current work. Two FXT candidates, XRT~191127 and XRT~210423, are detected at projected distances of ${\approx}$500 and 300~kpc, respectively, from the center of the galaxy cluster Abell 1795 (${\approx}$285.7~Mpc). However, neither is obviously associated with cluster members and physical offsets in this range are not easily explained by any possible physical scenario of FXTs (see Sec.~\ref{sec:flux}).

The lack of local FXTs could be explained by the \emph{Chandra} exposure time spent to observe local galaxies in recent years. Around 26.5\% of the \emph{Chandra} ObsIds (amounting to ${\approx}$0.8~years of exposure and a sky area of ${\approx}$66.7~deg$^2$) analyzed in this work covers local galaxies \citep[based on a match with the GLADE catalog;][]{Dalya2018}. Adopting the \emph{local} FXT rate from Paper~I, $\mathcal{R}_{\rm Local}{=}$53.7$_{-15.1}^{+22.6}$~deg$^{-2}$~yr$^{-1}$, we thus expect ${\approx}$2.8$_{-2.7}^{+5.6}$ \emph{local} FXTs in this work, which remains consistent with our non-detection of \emph{local} FXTs at 3$\sigma$ confidence for Poisson statistics. On the other hand, the \emph{distant} FXTs rate from Paper~I is $\mathcal{R}_{\rm Distant}{=}$28.2$_{-6.9}^{+9.8}$~deg$^{-2}$~yr$^{-1}$, implying ${\approx}$4.5$_{-3.7}^{+7.3}$~sources, which is consistent with our new eight FXT candidates.

\section{Host-galaxy features}\label{sec:counterpart_SED}

The host galaxy or host environment can provide additional information about the nature and origin of this FXT sample. Five FXTs (16, 19, 20, 21, and 22) lie close to extended optical/NIR sources, which are plausible host galaxies (see Fig.~\ref{fig:image_cutouts}). The host galaxies of FXTs~19 and 22 were previously identified, but their properties were not reported so far. For FXT~18, just one faint source (\emph{source \#1}) falls inside the X-ray uncertainty position, but it is not clear whether it is extended.

As a first step, in Fig.~\ref{fig:colour_colour_2}, we explore the nature of the hosts using $i-K_s$ vs. $g-i$ (\emph{top panel}) and $i-W1$ vs. $g-i$ (\emph{bottom panel}) colours, compared to the colours of X-ray sources previously classified as stars both in Paper~I (\emph{gray points}) and this work (\emph{black points}; see Sect. \ref{sec:gaia}), and the expected parameter space for stars (\emph{cyan regions}) with different ages [$\log(\rm{Age/yr}){=}$7.0--10.3], metallicities (from $\rm{[Fe/H]}{=}-3.0$--0.5), and attenuations ($A_V{=}$0.0--5.0~mag) from theoretical stellar isochrones \citep[MIST;][]{Dotter2016,Choi2016}. The vast majority of the X-ray flares with stellar counterparts form a tight sequence (see Fig.~\ref{fig:colour_colour_2}, \emph{cyan region}), with the outliers identified as PNe, YSOs, eruptive variable stars, T Tauri stars, or emission-line stars. Overall, the potential hosts of the FXT candidates appear to reside outside or at the edge (e.g., FXT~16, 18, 19, 21 and 22) of the stellar region, although the limits or large uncertainties (e.g., FXT~20) indicate that the current colour estimates are not the best discriminators by themselves. Thus the spatially resolved nature of the counterparts remains vital to their confirmation as a candidate host galaxy.

\begin{figure*}
    \centering
    \includegraphics[scale=0.58]{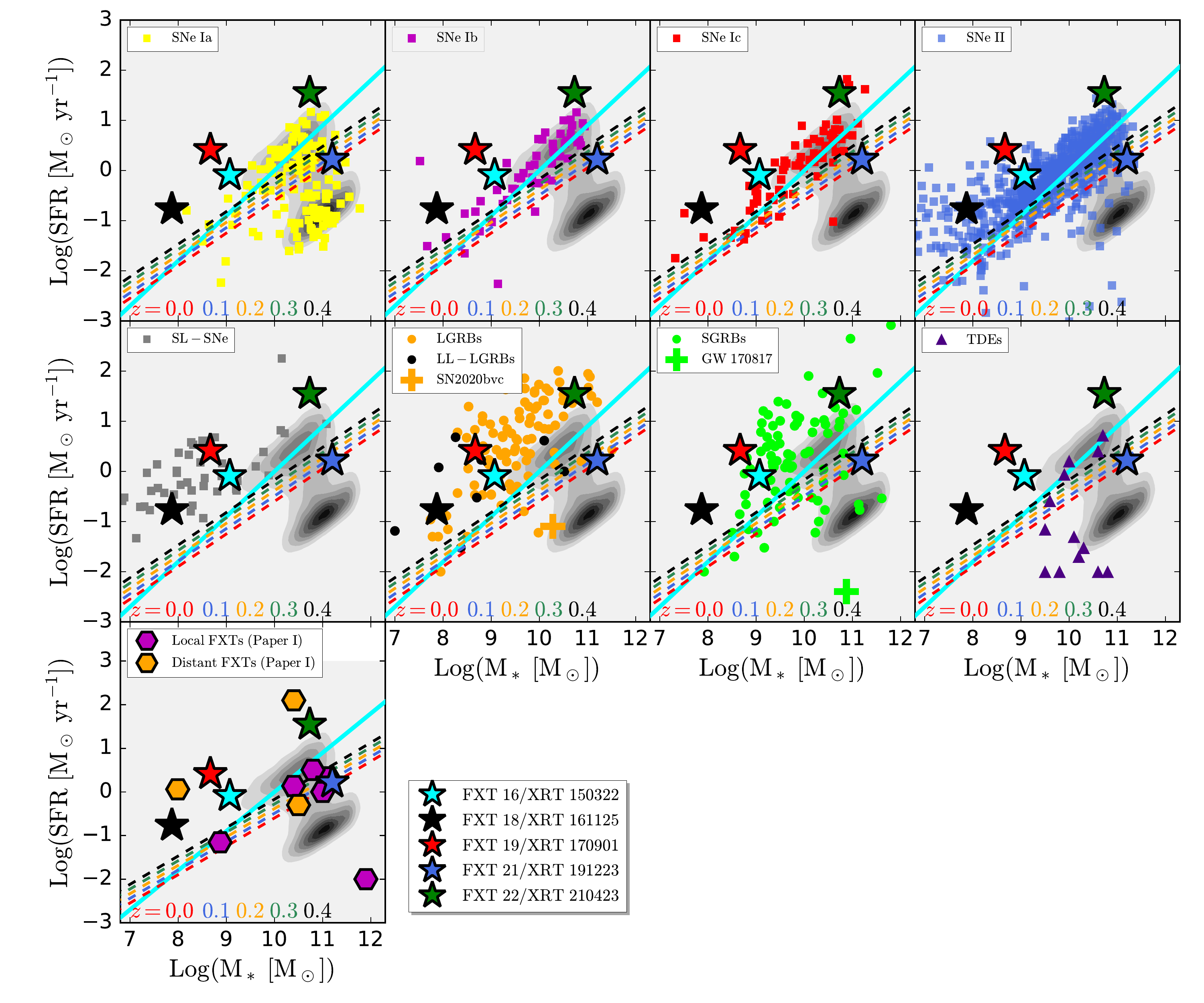}
    \vspace{-0.2cm}
    \caption{Star-forming galaxy main sequence diagram, stellar mass vs. SFR, comparing hosts of FXTs and various other transient classes (one per panel) such as SNe type Ia, Ib, Ic, II \citep{Tsvetkov1993,Galbany2014,Schulze2021}, super-luminous SNe \citep[SL-SNe;][]{Schulze2021}, LGRBs \citep[including SN~2020bvc;][]{Chang2015,Li2016,Izzo2020,Ho2020}, low-luminosity LGRBs \citep[LL-LGRBs; GRB~980425, GRB~020903, GRB~030329, GRB~031203, GRB~050826, GRB~060218, and GRB~171205A;][]{Christensen2008,Michalowski2014,Levesque2014,Kruhler2017,Wiersema2007,Wang2018,Arabsalmani2019}, SGRBs \citep[including GW~170817/GRB~170817A;][]{Li2016,Im2017,Nugent2022}, TDEs \citep{French2020}, and Paper~I FXT candidates (nearby and distant FXTs). \emph{Grayscale contours} denote the SDSS galaxy distribution from \citet{Brinchmann2004}. The \emph{solid cyan lines} show the best-fit local galaxy main sequence relation from \citet{Peng2010}, while the \emph{dashed coloured lines} denote the upward evolution of the boundary separating star-forming and quiescent galaxies as a function of redshift \citep[at $z{=}$0.0, 0.1, 0.2, 0.3, and 0.4;][]{Moustakas2013}.}
    \label{fig:Mass_SFR_plot}
\end{figure*}

\begin{figure*}
    \centering
    \hspace*{-0.5cm}
    \includegraphics[scale=0.6]{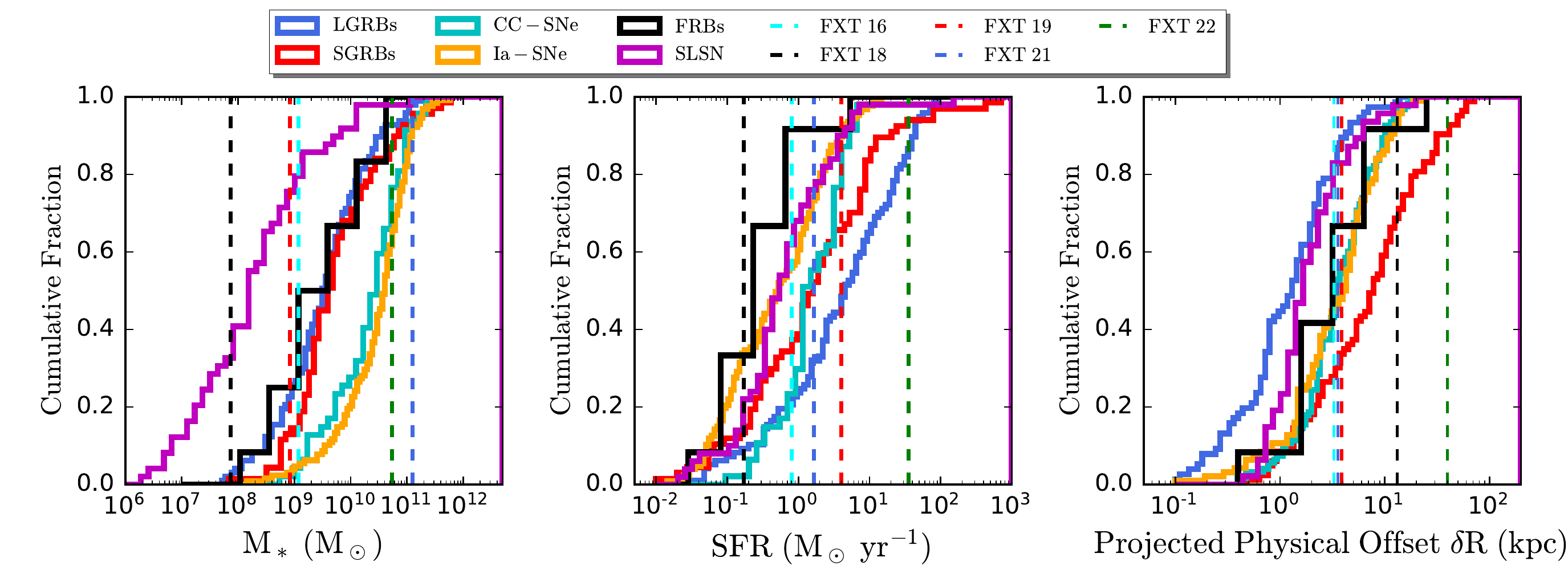}
    \caption{Comparison of the host-galaxy properties of FXTs~16, 19, 21 and 22 (\emph{colour vertical dashed lines}) from Table~\ref{tab:SED_para} with the cumulative distributions of galaxy
    stellar mass (\emph{left panel}),
    star-formation rate (\emph{center panel}),
    and projected physical offset (\emph{right panel})
    for LGRBs \citep[\emph{blue lines};][]{Li2016, Blanchard2016}, 
    SGRBs \citep[\emph{red lines};][]{Fong2010, Fong2012, Fong2013, Margutti2012b, Sakamoto2013, Berger2013b, Fong2022, Nugent2022}, 
    FRBs \citep[\emph{black lines};][]{Heintz2020}, 
    CC-SNe and Ia-SNe \citep[\emph{cyan and orange lines};][]{Tsvetkov1993,Prieto2008,Galbany2014}, and 
    SLSNe \citep[\emph{magenta lines};][]{Schulze2021}.}
    \vspace{-0.1cm}
    \label{fig:host_parameters_1}
\end{figure*}

We further constrain the host properties through spectral energy distribution (SED) model fitting of their existing photometry using \texttt{BAGPIPES} \citep[Bayesian Analysis of Galaxies for Physical Inference and Parameter EStimation;][]{Carnall2018}, which fits broadband photometry and spectra with stellar-population models taking star-formation history and the transmission function of neutral/ionized ISM into account via a \texttt{MultiNest} sampling algorithm \citep{Feroz2008,Feroz2009}. 
In Appendix \ref{app:host_galaxy}, we list the different conditions considered for the SED fitting. Table~\ref{tab:SED_para} provides the best-fit parameters obtained with \texttt{BAGPIPES} for the hosts of FXTs~16, 18, 19, 21 and 22\footnote{FXT~20 only has faint $g$, $r$ and $z$-band DECam detections, which are too few and too loosely constrained to compute a SED photometric redshift.}, while Fig.~\ref{fig:SED_models} shows the 16th to 84th percentile range for the posterior spectrum, photometry, and the posterior distributions for five fitted host-galaxy parameters. Also, Figs.~\ref{fig:Mass_SFR_plot} and ~\ref{fig:host_parameters_1} compare the SFR and stellar masses of the FXT hosts to those of several well-known transient classes (such as CC- and Ia SNe, LGRBs, SGRBs and FRBs) and cumulative functions, respectively.

Overall, in terms of stellar mass and star-formation rate, FXTs~ 16, 18, 19, and 22 are located above the galaxy main sequence, while FXT~21 lies slightly below it (see Fig.~\ref{fig:Mass_SFR_plot}, \emph{solid cyan line}).
The hosts of FXTs~16, 19, and 22 have SFRs,
stellar masses, 
and young stellar populations indicative of star-forming galaxies \citep{Moustakas2013}. 
In terms of SFRs, the FXT hosts broadly lie in the same region populated by SNe type Ib, Ic, II, SL-SNe, and GRBs.
The SFR of the host of FXT~22 
compares more favorably with LGRB (30\%) and SGRB (10\%) hosts over SNe (${\sim}$0\%); meanwhile, its large host stellar mass 
shares little overlap with the LGRB/SGRB (${\approx}$10/15\%) and partially with type-Ia/CC-SN (${\approx}$40/30\%) host populations. 
For FXTs~16 and 19, the overlapping fractions of LGRB, SGRB, and SLSNe host galaxies with galaxy stellar mass ${\lesssim}$10$^{9}$~$M_\odot$ are ${\approx}$20, 15, and 80\%, respectively.
In the particular case of FXT~18, it has a moderate SFR 
and low stellar mass. 
This low stellar mass matches with a very small fraction of LGRB (${\lesssim}$5\%) and SGRB (${\lesssim}$2\%) hosts and some SL-SNe (${\approx}$30\%) hosts.

The host of FXT~21 
has a moderate SFR 
and high stellar mass, 
implying a classification as a quiescent galaxy \citep{Moustakas2013}. Its SFR falls in a region populated by ${\approx}$70 and 50\% of LGRBs and SGRBs, respectively, and ${\approx}$40 and 25\% of CC-SNe and Ia-SNe hosts, respectively, with SFR${\gtrsim}$2.0~$M_\odot$~yr$^{-1}$. Meanwhile, only ${\lesssim}$10\% of SNe and GRBs have similar host galaxy stellar masses ${\gtrsim}$10$^{11}$~$M_\odot$.

Moreover, these sources fall in the same parameter space occupied by the distant FXTs reported in Paper~I (see Fig.~\ref{fig:Mass_SFR_plot}, \emph{lower panel}). Thus, we conclude that a majority of distant FXTs appear to be associated with actively star-forming galaxies (${\gtrsim}$10$^{8}$~M$_\odot$ and ${\gtrsim}$0.5~M$_\odot$~yr$^{-1}$), while a subset is associated with post-starburst ("green valley") galaxies.

Another crucial parameter is the projected physical offset ($\delta R$) between the transient's position and the host galaxy center. Figure~\ref{fig:host_parameters_1}, \emph{right panel}, compares the projected physical offset distribution of FXTs~16, 18, 19, 21 and 22 to several transient classes such as CC-SNe (\emph{cyan}), Ia-SNe (\emph{orange}), SL-SNe (\emph{magenta}), FRBs (\emph{black}), LGRBs (\emph{blue}) and SGRBs (\emph{red}). SGRBs have a physical offset, which is about ${\sim}$4--5 times greater than the median offset for LGRBs \citep{Bloom2002} and SL-SNe \citep{Schulze2021}, and about ${\sim}$1.5 times larger than the median offsets of CC- and Type Ia SNe \citep{Prieto2008} and FRBs \citep{Heintz2020}. In addition, practically no LGRBs and SL-SNe, and only ${\approx}$10\% of CC- and type Ia SNe have offsets ${\gtrsim}$10~kpc, while $\approx$40\% of SGRBs have such offsets. Moreover, ${\approx}$15\% of SGRBs have offsets ${\gtrsim}$20~kpc, while essentially no SL-SNe, CC- and type Ia SNe, or LGRBs exhibit such large offsets.

The physical offsets of FXTs~16 (${\approx}$3.3~kpc), 19 (${\approx}$3.9~kpc), and 21 (${\approx}$3.6~kpc) overlap with the cumulative distributions of CC- and type Ia SNe, and SGRBs at 1$\sigma$ confidence level, although only ${\approx}$10--15\% of SL-SNe and LGRBs have equal or higher offset values. 
In the case of FXT~18, its offset (${\approx}$13.2~kpc) resides well inside the offset distribution of SGRBs (${\approx}$70\% with $\delta R{<}$13~kpc), while just ${\lesssim}$5\% of LGRBs, and CC- and Ia-SNe have equal or higher offset values. Nevertheless, it has a large X-ray positional uncertainty.
In contrast, FXT~22 has a physical offset of ${\approx}$40~kpc, which is just compatible with ${\approx}$10\% of SGRB hosts with equal or higher offsets.

\begin{table}
    \centering
    \scalebox{1.0}{
    \begin{tabular}{lllll}
    \hline\hline
    FXT & $z$ & $L_{\rm X,peak}$ & $z_{\rm max}$ & $V_{\rm max}$  \\ 
     &  & (erg~s$^{-1}$) &  & (Gpc$^3$)  \\\hline
    (1) & (2) & (3) & (4) & (5) \\ \hline
    \multicolumn{5}{c}{Paper~I \citep{Quirola2022}} \\ \hline
    1  &  0.1866$^{\dagger,a}$  &  1.9$\times$10$^{46}$  &  4.30 & 1673 \\
    7  &  1.0$^\dagger$  &  1.3$\times$10$^{46}$  &  3.59 & 1354 \\
    8  &  0.61  &  1.1$\times$10$^{44}$  &  0.52  &  32 \\
    9  &  0.7  &  1.5$\times$10$^{45}$  &  1.49  &  343 \\
    10  &  1.0$^\dagger$  &  4.8$\times$10$^{45}$  &  2.42 & 789 \\
    11  &  0.0216$^{\dagger,b}$  &  2.4$\times$10$^{44}$  &  0.72 & 70 \\
    12  &  1.0$^\dagger$  &  3.0$\times$10$^{45}$  &  2.00 & 584 \\
    13  &  1.0$^\dagger$  &  6.5$\times$10$^{44}$  &  1.08 & 177 \\ 
    14  &  2.23$^c$  &  1.7$\times$10$^{47}$  & 11.01  &  3751 \\ \hline
    \multicolumn{5}{c}{This work} \\ \hline
    15  &  1.0$^\dagger$  &  1.0$\times$10$^{45}$  &  1.30 & 261 \\
    16  &  0.738  &  2.8$\times$10$^{45}$  &  1.94  &  554 \\
    17  &  1.0$^\dagger$  &  6.3$\times$10$^{45}$  &  2.73 & 946 \\
    18  &  0.35  &  1.9$\times$10$^{47}$  &  11.55 & 3867 \\
    19  &  1.44  &  3.7$\times$10$^{46}$  &  5.72  &  2245 \\
    20  &  1.0$^\dagger$  &  8.1$\times$10$^{46}$  &  7.98 & 2987 \\
    21  &  0.85  &  6.9$\times$10$^{45}$  &  2.80  &  978 \\
    22  &  1.51  &  1.3$\times$10$^{46}$  &  3.61 & 1367 \\
    \hline
    \end{tabular}
    }
    \caption{FXT properties from Paper~I and this work used to compute the volumetric density rates (Sect.~\ref{sec:vol_rate}) and X-ray luminosity functions (Sect.~\ref{sec:LF}). 
    \emph{Column 2:} adopted redshift for each FXT. 
    \emph{Column 3:} Peak isotropic X-ray luminosity in cgs units (corrected for Galactic and intrinsic absorption). 
    \emph{Columns 4:} maximum observable redshift. 
    \emph{Columns 5:} maximum comoving volume in Gpc$^{3}$ units.\\
    $^\dagger$ fiducial redshifts.\\
    $^a$ redshift taken from \citet[][$z{=}$0.1866]{Eappachen2022}.\\
    $^b$ redshift taken from \citet[][94.9~Mpc]{Glennie2015}.\\ 
    $^c$ redshift taken from \citet[][$z{=}$2.23]{Bauer2017}.}
    \label{tab:LF_parameters}
\end{table}

\section{Rates}\label{sec:rates}

We update the FXT event rates determined in Paper~I and revisit comparisons with other transients to explore possible interpretations. Specifically, we derive the observed event rates (deg$^{-2}$~yr$^{-1}$; Sect. \ref{sec:event_rate}),  FXT X-ray luminosity function (Sect.~\ref{sec:LF}), and volumetric rates (yr$^{-1}$ Gpc$^{-3}$; Sect. \ref{sec:vol_rate}).

\subsection{Event-rate estimation}\label{sec:event_rate}

We compute FXT event rates following the procedure and assumptions outlined in Paper~I (their sect~6.1 and eqs.~5--7). 
We first estimate the rate independently of Paper~I to confirm consistency. We found eight FXTs
inside ${\approx}$89~Ms of \emph{Chandra} data from 2014 to 2022, yielding $\mathcal{R}_{\rm This\ work}{=}$45.6$_{-14.3}^{+18.2}$~deg$^{-2}$~yr$^{-1}$ (for sources with $F_{\rm X,peak}{\gtrsim}$1$\times$10$^{-13}$~erg~cm$^{-2}$~s$^{-1}$). This rate is consistent with the rates derived by \citet[][$\mathcal{R}_{\rm Yang+19}{\approx}$59$_{-38}^{+77}$]{Yang2019}  and Paper~I ($\mathcal{R}_{\rm Paper~I, distant}{=}$28.2$_{-6.9}^{+9.8}$~deg$^{-2}$~yr$^{-1}$) at the Poisson 1$\sigma$ confidence level and higher than the rate derived by \citet[][${\approx}$3.4~deg$^{-2}$~yr$^{-1}$]{Glennie2015}. As already mentioned in Paper~I, this is not surprising since \citet{Glennie2015} computed the rate for a higher peak flux of $F_{\rm X,peak}{\gtrsim}$1${\times}$10$^{-10}$~erg~cm$^{-2}$~s$^{-1}$. 
Then, considering all 17 distant FXTs (i.e., the nine distant FXTs from Paper~I, also including the ambiguous FXTs~1 and 11 which might be extragalactic sources according to \citealt{Eappachen2022}, and the eight FXTs from this work) detected by \emph{Chandra} (ACIS-I/S) instruments between 2000 and 2022, we estimate a total distant FXT event-rate  of $\mathcal{R}_{\rm Total, distant}{=}$36.9$_{-8.3}^{+9.7}$~deg$^{-2}$~yr$^{-1}$. 
Since the number of FXTs removed erroneously by our selection criteria is ${\ll}$1 (see Sect. \ref{sec:completness}), the estimated event rates are robust results for FXT candidates brighter than $F_X{\gtrsim}$1$\times$10$^{-13}$~erg~cm$^{-2}$~s$^{-1}$ (from the search algorithm developed in Paper~I). 
Finally, we found no new local FXTs among the \emph{Chandra} observations of nearby galaxies included in this work (i.e., 26.5\% of the ObsIds), allowing a revised estimate (considering ObsIds from Paper I and this work) of the nearby FXT event rate of $\mathcal{R}{=}$34.3$_{-10.8}^{+13.7}$~deg$^{-2}$~yr$^{-1}$, consistent with that derived in Paper~I.

\begin{figure*}
    \centering
    \includegraphics[scale=0.65]{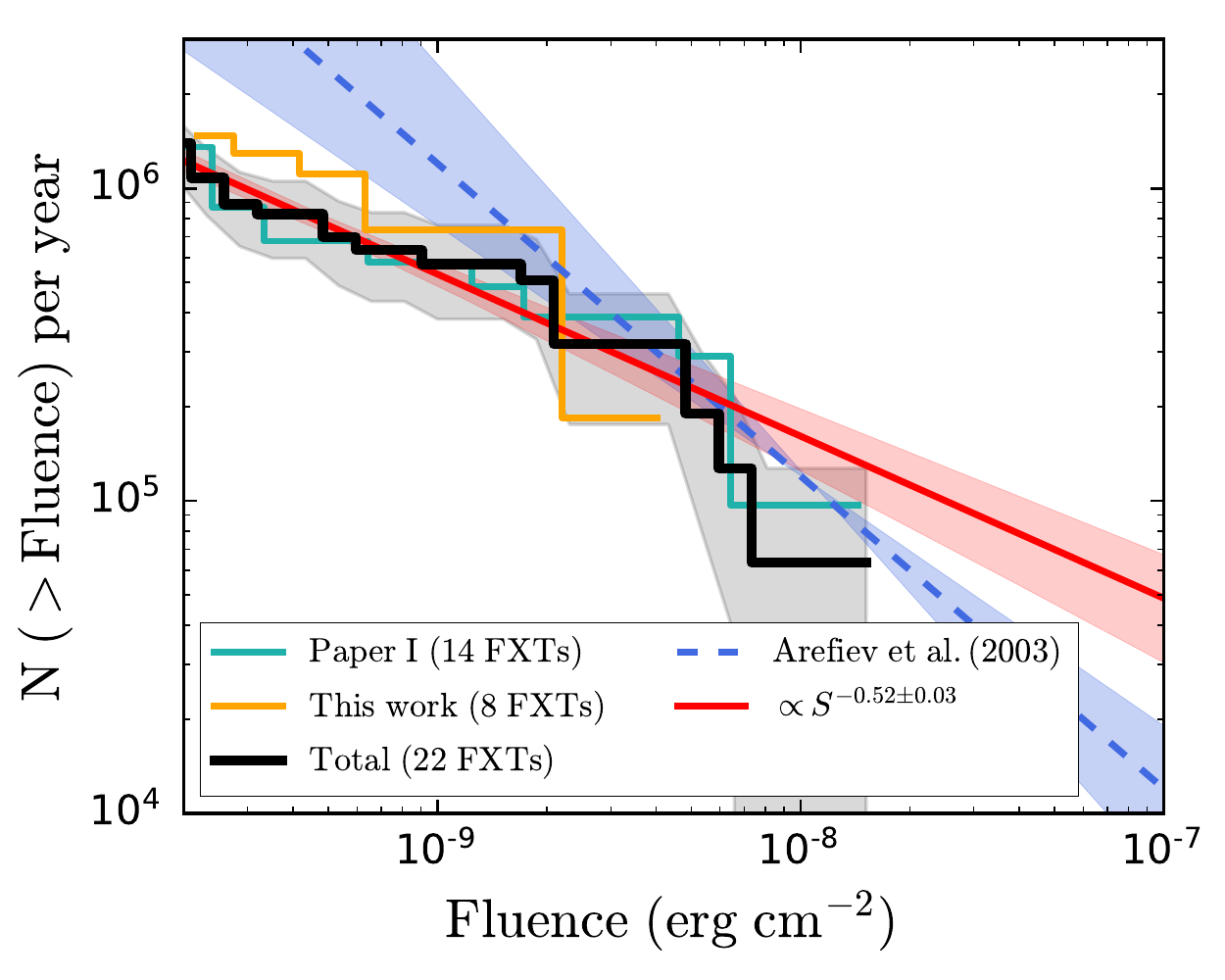}
    \includegraphics[scale=0.65]{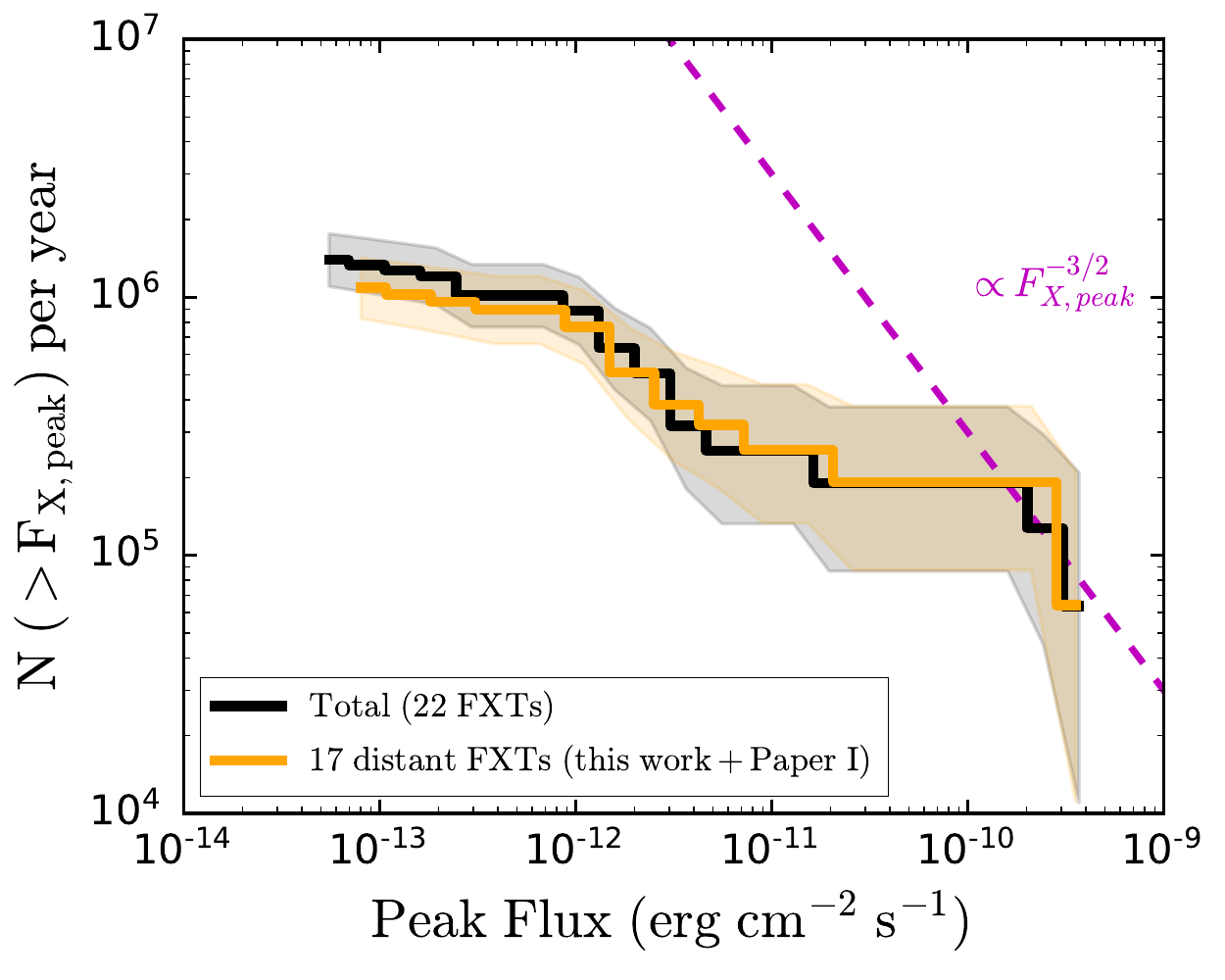}
    \caption{Observed cumulative log$\mathcal{N}$--log$S$ and log$\mathcal{N}$--log$F_{\rm peak}$ distributions. 
    \emph{Left panel:} log$\mathcal{N}$--log$S$ distribution of the sample of extragalactic FXTs analyzed in this work (\emph{orange line}), from Paper~I (\emph{cyan line}), and combined (\emph{black line}), as a function of fluence (in cgs units). Also shown are two PL models, $N({>}Fluence){\propto}S^{-\gamma}$, with slopes $\gamma{=}$0.52 (\emph{red line}) and 1.0 (\emph{blue dashed line}). The $\gamma{=}1$ line represents the best fit and 1$\sigma$ error of \citet{Arefiev2003} based on bright FXTs (including Galactic flares). The brightest sources in our sample appear to be consistent with this bright-end extrapolation, although our fainter sources fall up to ${\sim}$1~dex below, implying a break. For comparison with \citet{Arefiev2003}, we convert all FXT fluences to the 2--10~keV band from their best fits.
    \emph{Right panel:} log$\mathcal{N}$--log$F_{\rm peak}$ distribution of the total sample of extragalactic FXTs analyzed in Paper~I and this work (\emph{black line}) and only distant FXTs (\emph{orange line}). The \emph{magenta dashed line} shows objects uniformly distributed in Euclidean space (${\propto}F_{\rm peak}^{-3/2}$) for comparison. The \emph{color regions} represent the 1-$\sigma$ confidence interval.
    }
    \label{fig:fluence_plot}
\end{figure*}

The event rate as a function of fluence (or peak flux\footnote{Similar to Paper~I, due to the lack of a standardized method to estimate the $F_{\rm X,peak}$, first we find the shortest time interval during which 25\% of the counts are detected, and we compute the count rate during this shortest interval. Next, to convert the peak-count rates to fluxes, we multiply the flux from the time-averaged spectral fits by the ratio between the peak and the time-averaged count rates (i.e., we assume no spectral evolution).}) behaves as a power-law function as $\mathcal{R}{\propto}F_{\rm peak}^{-\gamma}$, where $\gamma$ is a positive value. 

Figure~\ref{fig:fluence_plot}, \emph{left panel}, shows the cumulative log$\mathcal{N}$--log$S$ distribution of the entire sample analyzed in this work (i.e., 8~FXTs; \emph{orange line}), FXTs identified in Paper~I (i.e., 14~FXTs; \emph{cyan line}), and finally combining both samples (i.e., 22~FXTs; \emph{black line}) which appears to follow $\gamma{\approx}0.5$ (\emph{red region/line}). We also plot the extrapolation of the best-fit slope, $\gamma{=}1.0$, based on the estimates of FXTs at bright fluxes (${\gtrsim}10^{-10}$~erg~cm$^{-2}$~s$^{-1}$) from \citet{Arefiev2003}.\footnote{We caution that \citet{Arefiev2003} does not specify an exact energy band and makes no distinction between various potential Galactic and extragalactic classes. It is noteworthy that the sky distribution at these bright fluxes is also isotropic.} The brightest sources in the total sample of 22~FXTs appear to be consistent with the \citet{Arefiev2003} extrapolation at 1-$\sigma$ confidence. In contrast, the fainter sources fall well below it by ${\sim}$1~dex (following a power-law with an exponential index of $\gamma{=}$0.52), indicating a break around a fluence of ${\sim}10^{-8}$~erg~cm$^{-2}$ to our best-fit slope. A similar result was found in Paper~I. Furthermore, we define a BPL model, which fits the results obtained (see Fig.~\ref{fig:fluence_plot}), for the cumulative log$\mathcal{N}$--log$S$ distribution with a break fluence around ${\sim}1{\times}10^{-8}$~erg~cm$^{-2}$ and two power-law indexes of $\gamma_1{=}-0.52$ and $\gamma_2{=}-1.0$ for the faint and bright ends, respectively.

Finally, Fig.~\ref{fig:fluence_plot}, \emph{right panel}, represents the cumulative log$\mathcal{N}$--log$F_{\rm peak}$ curves considering the whole sample of 22 FXTs (\emph{black line}) and just the 17 distant FXTs for Paper~I and this work (\emph{orange line}). The local FXTs identified in Paper~I contribute mostly to low fluxes (compare the \emph{orange} and \emph{black} lines). The log$\mathcal{N}$--log$F_{\rm peak}$ slope appears to be significantly shallower at low $F_{\rm peak}$ than the Euclidean prediction (i.e., ${\propto}S^{-3/2}$, which is expected for astrophysical objects uniformly distributed in a Euclidean space; \emph{dashed magenta line}). A combination of four effects could explain this deviation: $i)$ near the sensitivity threshold of the detector, the number of FXTs depends on the detection efficiency, which affects the log$\mathcal{N}$--log$F_{\rm peak}$ plot; $ii)$ due to the flux being inversely proportional to the square of the luminosity distance, which will differ from the Euclidean distance as $z$ approaches unity (this implies that the FXTs should be cosmological); $iii)$ the sample of FXTs likely has a mix of origins, such that the cosmic event rate density is not constant with redshift; and $iv)$ the sample is dominated by low-number statistical fluctuations, particularly at the bright end, due to the \emph{grasp} (area $\times$ sensitivity) of \emph{Chandra}. New X-ray missions which are focusing on scanning the sky, such as \emph{eROSITA} and \emph{Einstein Probe}, will increase the number of FXTs and improve our statistics.

\begin{figure*}
    \centering
    \includegraphics[scale=0.68]{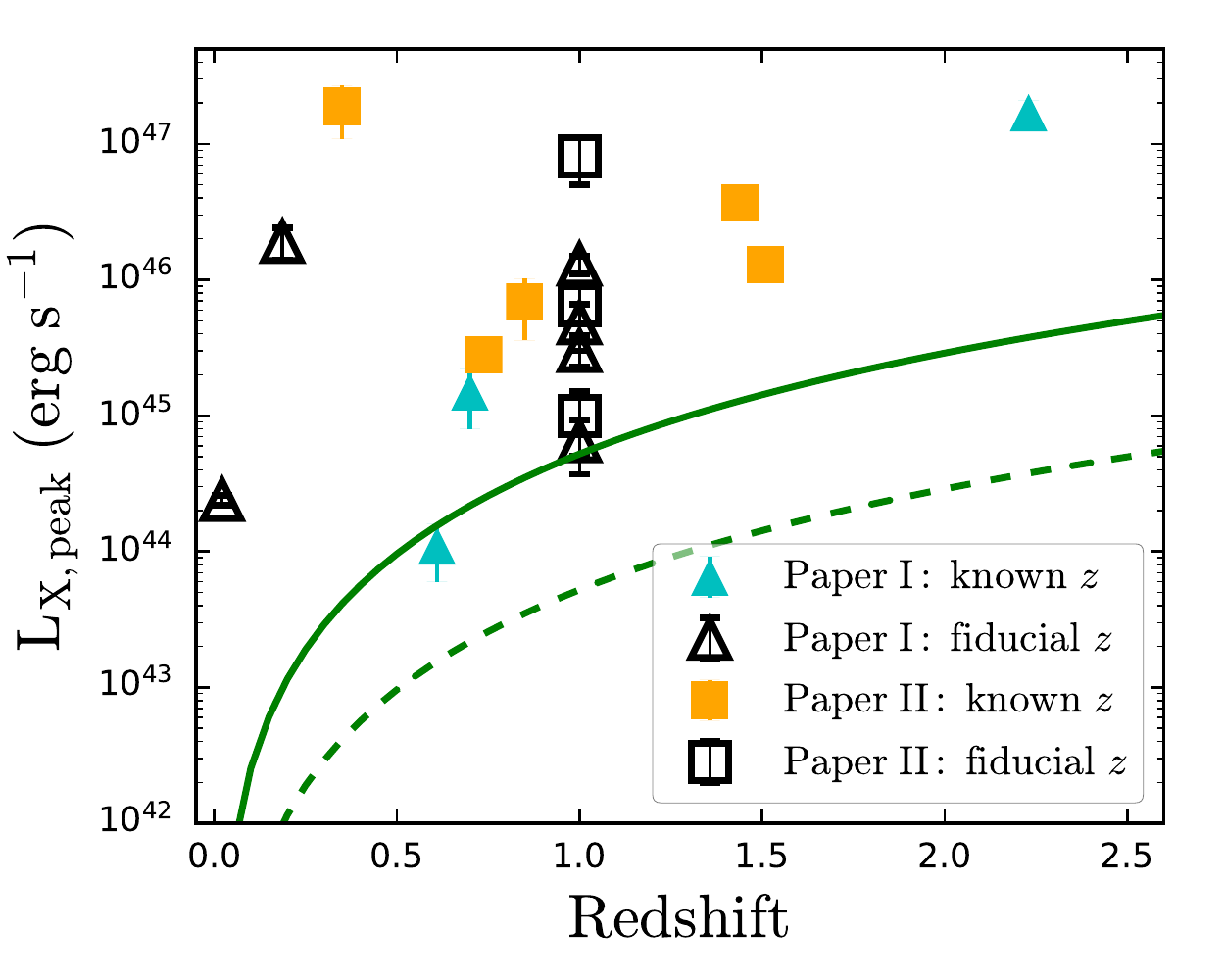}
    \includegraphics[scale=0.68]{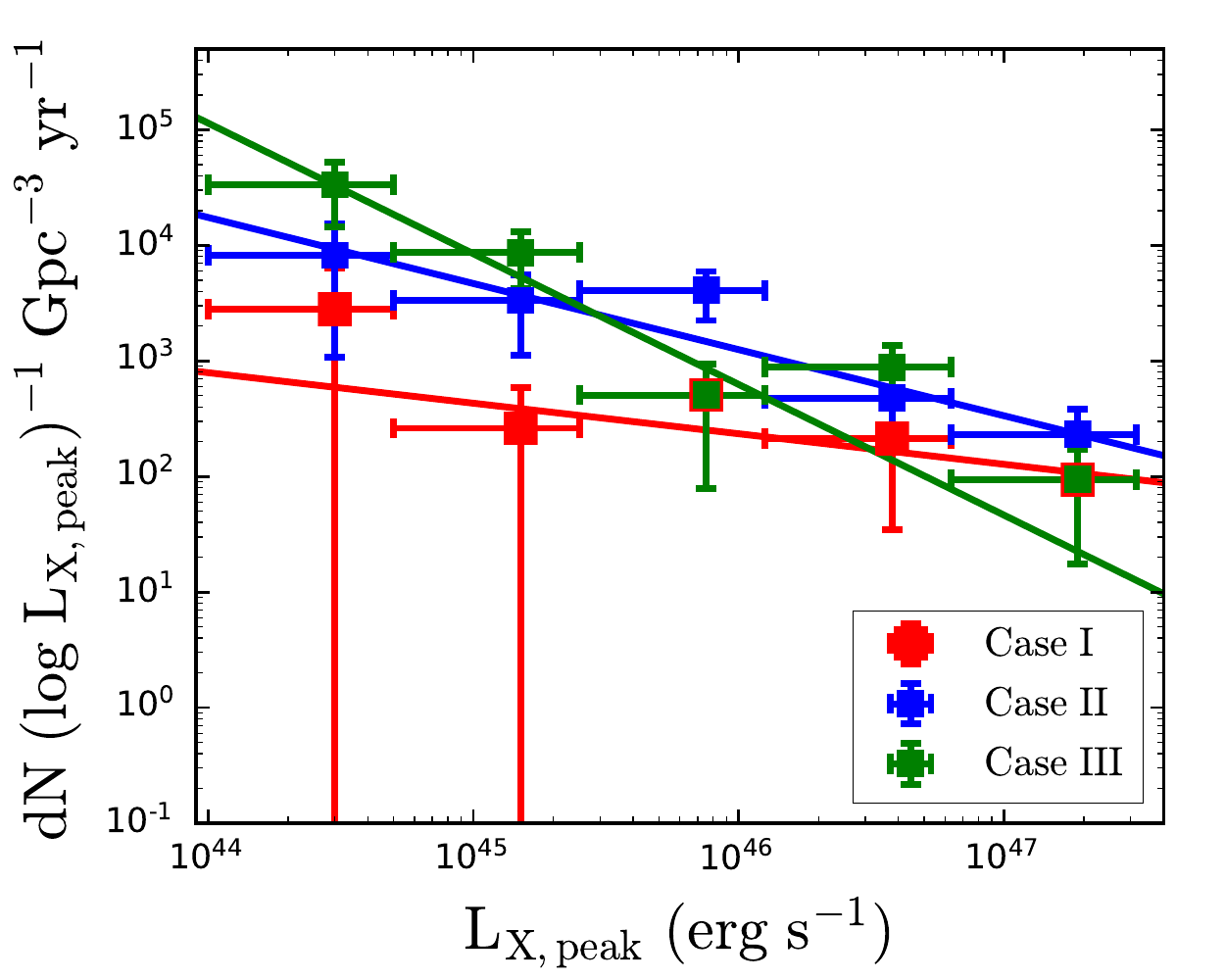}
    \vspace{-0.1cm}
    \caption{\emph{Left panel:} The peak X-ray luminosity of FXT candidates from Paper~I and this work (with known and fiducial redshifts; see Table~\ref{tab:LF_parameters}). The green \emph{solid and dashed} lines indicate peak flux limits of 10${^{-13}}$ (set by our algorithm) and 10$^{-14}$~erg~cm$^{-2}$~s$^{-1}$ (approximate \emph{Chandra} on-axis detector limit), respectively. \emph{Right panel:} FXT X-ray (0.3--10~keV) luminosity function (XLF) of the total sample from Paper~I and this work. The \emph{red squares} represent the XLF using the seven FXTs with known (photometric or spectroscopic) redshifts, while the \emph{blue} and \emph{green squares} show XLFs using all 17 FXTs with known+fiducial redshifts, adopting $z=1.0$ and 0.5 for unknown objects, respectively. The \emph{solid lines} show best-fit power-law models (see Table~\ref{tab:LC_fitting} for values).}
    \label{fig:LF}
\end{figure*}

\begin{table*}
    \centering
    \scalebox{1.0}{
    \begin{tabular}{llll}
    \hline\hline
    Case \# & $\alpha$ & $\beta$ & $\rho_0(L_{\rm X,peak}{>}10^{44}{\rm erg~s^{-1}})$  \\ 
     & & (Gpc$^{-3}$~yr$^{-1}$~dex$^{-1}$) & (Gpc$^{-3}$~yr$^{-1}$)  \\ \hline
     (1) & (2) & (3) & (4) \\\hline
    \emph{Case I} (known redshifts) & $-0.26{\pm}0.13$ & $(7.90{\pm}1.49){\times}10^2$ & $(1.24{\pm}0.35){\times}10^3$ \\
    \emph{Case II} ($z_{\rm fiducial}{=}1$) & $-0.57{\pm}0.11$ & $(1.74{\pm}0.29){\times}10^4$ & $(1.32{\pm}0.11){\times}10^4$ \\
    \emph{Case III} ($z_{\rm fiducial}{=}0.5$) & $-1.13{\pm}0.27$ & $(1.14{\pm}0.45){\times}10^5$ & $(4.38{\pm}0.21){\times}10^4$ \\
    \hline
    \end{tabular}
    }
    \caption{Results of XLF fitting. \emph{Column 1:} \emph{Case}\# considered. \emph{Columns 2 and 3:} best-fitting parameters of the XLF from Eq.~\ref{eq:009}. \emph{Column 4:} FXT volumetric rate from integrating the XLF.}
    \label{tab:LC_fitting}
\end{table*}

\subsection{Luminosity function}\label{sec:LF}

Past works have constructed X-ray luminosity functions (XLFs) for GRBs \citep{Sun2015}, SBOs \citep{Sun2022}, and TDEs \citep{Sazonov2021}. We construct here the XLF of FXTs considering distant sources from Paper~I and this work, using the classical $V_{\rm max}$ method from \citet[][]{Schmidt1968}. We adopt the redshifts and peak X-ray luminosities shown in Table~\ref{tab:LF_parameters} (\emph{columns 2 and 3}, respectively) and Fig.~\ref{fig:LF} (\emph{left panel}), which also plots the limiting luminosity corresponding to a  \emph{Chandra} detection threshold of $F_{\rm X-peak}^{\rm lim}{\sim}$1$\times$10$^{-13}$~erg~cm$^{-2}$~s$^{-1}$ (\emph{green solid line}; representing the conservative detection limit of our search algorithm) and 1$\times$10$^{-14}$~erg~cm$^{-2}$~s$^{-1}$ (\emph{green dashed line}; representing an approximate instrument threshold). All the FXTs except FXT~8 lie above $F_{\rm X,peak}^{\rm lim}{\approx}$1$\times$10$^{-13}$~erg~cm$^{-2}$~s$^{-1}$. 

The XLF, $\Phi(L)$, is the sum of the individual contributions by each source ($j$) in the luminosity range from $\log(L)$ to $\log(L)+d\log(L)$ (with total $\Delta N_L$ sources), i.e.,
\begin{equation}
    \Phi(L)\ d\log(L)=\sum_{j=1}^{\Delta N_L}\frac{4\pi\ dN(L)_j}{[\sum_{i=1}^{N_{\rm obs}}T_i\Omega_i V_{\rm max,i}]_j},
\end{equation}
where $i$ runs over the $i$th observation, $N_{\rm obs}$ is the total number of observations, $T_i$ and $\Omega_i$ are the exposure time and FoV per observation, respectively, $V_{\max,i}$ is the maximum observable volume, and $\Delta N_L$ is the total detectable number of sources.

$V_{\rm max}$ for a given FXT in the sample depends on its intrinsic peak luminosity in the 0.3–10.0~keV energy band, which are both given in Table~\ref{tab:LF_parameters}, assuming a flux limit of $F_{\rm X,peak}^{\rm lim}{\approx}$10$^{-13}$~erg~cm$^{-2}$~s$^{-1}$ from our detection algorithm. To calculate the FXT XLF, we sum the derived $V_{\rm max}^{-1}$ values of individual FXTs in five equal 0.5~dex interval luminosity bins from $\log(L_{\rm X,peak})$ between 44.0 and 47.5. The uncertainty within a given bin depends on the Poisson uncertainty of the number of FXTs per bin and $V_{\rm max}$ (computed as $\sqrt{\sum(V_{\rm max})^{-2}}$, where the summation is done over the objects within that bin).

We estimate three different cases: $I)$ considering just eight FXTs with {\it known} redshifts, $II)$ considering 17 distant FXTs with {\it known} and {\it fiducial ($z{=}1.0$)} redshifts, and $III)$ considering 17 distant FXTs with {\it known} and {\it fiducial ($z{=}0.5$)} redshifts. The computed FXT XLFs are shown in Fig.~\ref{fig:LF} (\emph{right panel}). In \emph{Case~I} (\emph{red squares}), the largest uncertainties are associated with the lowest luminosity bins where there is just one FXT per luminosity bin.
For \emph{Cases~II} and \emph{III} (\emph{blue} and \emph{green squares}, respectively), the uncertainties are somewhat smaller because some luminosity bins have more than one source.

The FXT XLFs all appear to decline with increasing X-ray luminosity and are fit with power-law models as:

\begin{equation}
    \frac{dN}{d\log(L_{\rm X,peak})dVdt} = \beta \times \left(\frac{L_{\rm X,peak}}{10^{44}~{\rm erg}~{\rm s}^{-1}}\right)^\alpha.
    \label{eq:009}
\end{equation}
The best-fit models are plotted in Fig.~\ref{fig:LF} as \emph{red, blue, and green lines} for \emph{Cases~I, II} and \emph{III}, respectively, with the results summarized in Table~\ref{tab:LC_fitting}. Assuming fiducial redshifts of $z{=}1.0$ ($z{=}0.5$) naturally leads to shallower (steeper) XLF slopes. 

Under the implicit assumption of no evolution, we can estimate the average FXT volumetric rate in the $z{\approx}$0--2.2 (see Table~\ref{tab:LF_parameters}) Universe by integrating the XLF over the entire luminosity range used here ($L_{\rm X,peak}{>}10^{44}$~erg~s$^{-1}$). The results are tabulated in Table~\ref{tab:LC_fitting}. For \emph{Case~I}, the average event rate density is ${\approx}1.2{\times}10^3$~Gpc$^{-3}$~yr$^{-1}$. Due to the total galaxy volume density of ${\sim}$2$\times$10$^{7}$~Gpc$^{-3}$ \citep{Bell2003}, this equates to a rate of $R{\sim}6{\times}10^{-5}$~FXTs~yr$^{-1}$ per galaxy. On the other hand, considering the FXTs with known+fiducial redshifts of $z{=}1.0$ (\emph{Case~II}) and $z{=}0.5$ (\emph{Case~III}), the average FXT volumetric rates would be ${\approx}1.3{\times}10^4$ and ${\approx}4.4{\times}10^4$~Gpc$^{-3}$~yr$^{-1}$, respectively, implying rates of $R{\sim}6.5{\times}10^{-4}$ and $2.2{\times}10^{-3}$~FXTs~yr$^{-1}$ per galaxy, respectively. 
The three derived FXT volumetric rates (which could be interpreted as a mean value for \emph{Cases~I} or \emph{II/III}, respectively) remain consistent with previous results computed in Paper~I (${\approx}5{\times}10^3$~Gpc$^{-3}$~yr$^{-1}$), \citealp{Xue2019} (${\approx}1.3{\times}10^4$~Gpc$^{-3}$~yr$^{-1}$), and \citealp{Bauer2017} (${\gtrsim}10^{2}$~Gpc$^{-3}$~yr$^{-1}$ at $z{\lesssim}1$). Notably, these values differ by orders of magnitude from the rates computed previously for SMBH TDEs \citep[${\approx}$210~Gpc$^{-3}$~yr$^{-1}$;][]{Donley2002,Sazonov2021} and marginally with SBOs \citep[${\approx}4.6{\times}$10$^4$~Gpc$^{-3}$~yr$^{-1}$;][]{Sun2022} identified in eROSITA and \emph{XMM-Newton} archival data, respectively. This further helps to exclude the SBOs (considering \emph{Case~I}) and SMBH TDE scenarios, although other transients may remain viable due to the large beaming correction uncertainties.

\begin{figure*}
    \centering
    \hspace{-0.3cm}
    \includegraphics[scale=0.7]{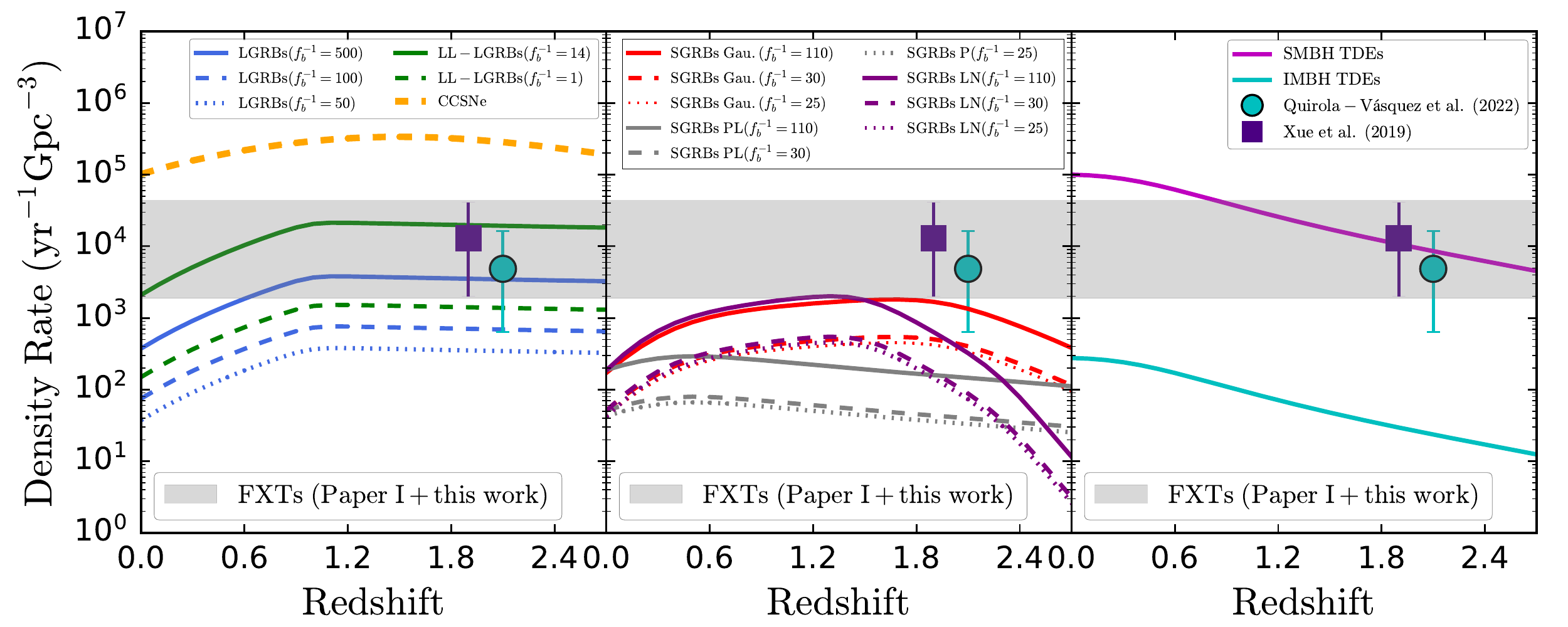}
    \vspace{-0.1cm}
    \caption{Volumetric density rate as a function of redshift comparing FXTs (\emph{gray filled region}, assuming no evolution with redshift and 1-$\sigma$ confidence; see text for details) and other transients. 
    \emph{Left panel:} comparison to massive-star related sources such as CC-SNe \citep[\emph{dashed orange line}; $k_{\rm CC-SNe}{=}0.0068$~$M_\odot^{-1}$ times the cosmic SFR density from][]{Madau2014}, LGRBs (\emph{blue-solid, dashed and dotted lines} show evolution normalized at $z{=}0$ to $\rho_{\rm 0,LGRBs}{=}$0.75~yr$^{-1}$~Gpc$^{-3}$ for $f_b^{-1}{=}$500, 100, and 50, respectively; \citealp{Sun2015,Wanderman2010}), LL-LGRBs (\emph{green-solid and dashed lines} denote evolution normalized at $z{=}0$ to $\rho_{\rm 0,LL-LGRBs}{=}$150~yr$^{-1}$~Gpc$^{-3}$ for $f_b^{-1}{=}$14 and 1, respectively, where $f_b^{-1}$ is the jet beaming correction factor; \citealp{Liang2007,Zhang_book_2018}).
    \emph{Middle panel:} comparison to compact-object binary systems such as SGRBs considering Gaussian (Gau.; \emph{red lines}), Power-law (PL; \emph{gray lines}), and Log-Normal (LN; \emph{purple lines}) merger delay models (\emph{solid, dashed and dotted lines} denote evolution normalized at $z{=}0$ to $\rho_{\rm 0,SGRBs}{=}$1.75~yr$^{-1}$~Gpc$^{-3}$ for $f_b^{-1}{=}$110, 30 and 25, respectively; \citealp{Sun2015,Wanderman2015}).
    \emph{Right panel:} comparison to SMBH-TDEs (\emph{magenta line} indicates evolution normalized at $z{=}0$ to $\rho_{\rm 0,SMBH-TDEs}{=}$1.1$\times$10$^5$~yr$^{-1}$~Gpc$^{-3}$ for luminosities ${\gtrsim}$10$^{44}$~erg~s$^{-1}$, assumed to be emitted isotropically;  \citealp{Sun2015}) and IMBH-TDEs (\emph{cyanline} shows evolution normalized at $z{=}0$ to $\rho_{\rm 0,IMBH-TDEs}{=}$290~yr$^{-1}$~Gpc$^{-3}$ emitted isotropically; \citealp{Bloom2011,Lu2018,Malyali2019,Tanikawa2021}). We also show the estimated rates from Paper~I (\emph{cyan circle}) and \citet{Xue2019} for CDF-XT2-like objects (\emph{purple square}).}
    \label{fig:rate_plot}
\end{figure*}

\subsection{Volumetric Density rate estimation}\label{sec:vol_rate}

In Section~\ref{sec:LF}, we estimated the average FXT volumetric rate in the redshift range $z{\approx}$0--2.2 from the XLF (see Table~\ref{tab:LC_fitting}), assuming zero evolution. Now, we compare the volumetric density rate, in units of yr$^{-1}$~Gpc$^{-3}$, with other known transient classes such as GRBs, SNe/SBOs, or TDEs. First, because only eight of 17 distant FXTs have redshift estimates, we correct the FXT volumetric rate of \emph{Case I} by the inverse of this fraction (i.e., multiply by $17/8$) to account for the fact that we do not include all the sources. We implicitly assume here that the underlying redshift distribution of the sources without redshifts is the same as those with. Without this correction, the luminosity functions are lower limits rather than best estimates. Meanwhile, for \emph{Cases~II and III}, a correction factor is not necessary because both cases adopted fiducial redshifts ($z_{\rm fiducial}{=}1$ and $0.5$, respectively) for all FXTs that lacked estimates. Considering this correction, the volumetric density rate for \emph{Cases I, II and III} ranges between ${\sim}1.9{\times}10^3{-}4.6{\times}10^4$~Gpc$^{-3}$~yr$^{-1}$ at 1$\sigma$ confidence.

The derived density rate as a function of redshift is shown in Fig.~\ref{fig:rate_plot}  (\emph{gray filled region}). Our result is consistent with the rates estimated in Paper~I (${\approx}$4.8$\times$10$^3$~Gpc$^{-3}$~yr$^{-1}$ at $z_{\rm max}{=}$2.1; \emph{cyan circle}) or CDF-S-XT2-like sources \citep[\emph{purple square};][]{Xue2019}.
Each panel in Fig.~\ref{fig:rate_plot} represents a comparison with transients related to massive stars (LGRBs, LL-LGRBs, and CC-SNe; \emph{left panel}), compact binary mergers (SGRBs; \emph{middle panel}), and tidal disruption events (SMBH and IMBH TDEs; \emph{right panel}).

As CC-SNe progenitors are massive, short-lived stars, the event rates should reflect ongoing star formation at different cosmological epochs \citep[][and references therein]{Madau2014}. Thus to build the cosmic density rate shown in Fig.~\ref{fig:rate_plot} (\emph{left panel}), we use the star-formation history derived by \citet{Madau2014}, weighted by the number of stars that explode as SNe per unit mass $k_{\rm CC-SNe}{=}0.0068$~$M_\odot^{-1}$ \citep{Madau2014}, adopting a Salpeter initial mass function (\emph{orange-dashed lines}).

One caveat for CC-SNe, however, is that we do not expect strong X-ray emission from all types of SBO CC-SNe. Thus, we analyze the expected rates for different sub-samples of CC-SNe. The local event rate density of all CC-SNe types is ${\sim}10^5$~Gpc$^{-3}$~yr$^{-1}$ \citep[see Fig.~\ref{fig:rate_plot}, \emph{left panel};][]{Smartt2009,Li2011,Madau2014}. Around ${\sim}$59\% of CC-SNe are Type II-P SNe from red supergiant star (RSG) progenitors. This means that the local rate of SBO from RSGs \citep[which peak in the UV at a few eV;][]{Alp2020,Sun2022} should be ${\sim}6{\times}10^{4}$~Gpc$^{-3}$~yr$^{-1}$, which is slightly higher than our result for FXTs (${\sim}2{\times}10^3{-}4.5{\times}10^4$~Gpc$^{-3}$~yr$^{-1}$). Meanwhile, around ${\approx}$1--3\% of CC-SNe are Type II SNe from blue supergiant star progenitors \citep[BSGs;][]{Arnett1989,Pastorello2005}, and ${\sim}$30\% are Type Ib/c SNe from Wolf-Rayet star (WR) progenitors. SBOs from BSGs and WRs are expected to peak in the soft X-rays, 0.3 and 3~keV, respectively \citep{Matzner1999,Nakar2010,Sapir2013}. Thus, the local rates of SBOs related to BSGs and WRs are ${\sim}2{\times}10^{3}$ and ${\sim}6{\times}10^{2}$~Gpc$^{-3}$~yr$^{-1}$, respectively. The derived event rate density of FXTs falls especially close to the expected rate of BSGs.

For LGRBs, we adopt a cosmic evolution rate following \citet{Sun2015}, which we normalize to the local universe value to characterize the cosmic density rate, as shown in Fig.~\ref{fig:rate_plot} (\emph{left panel}). LGRBs have an isotropic luminosity of ${\sim}10^{49}{-}10^{54}$~erg~s$^{-1}$, and an observed local density rate above $10^{50}$~erg~s$^{-1}$ of $\rho_{0,\rm LGRBs}{\sim}0.5-1.0$~Gpc$^{-3}$~yr$^{-1}$ \citep{Zhang_book_2018}. We additionally consider a jet beaming correction factor of $f_b^{-1}{\sim}$500 (\emph{blue-solid line}), which corresponds to a mean jet opening angle $\theta_j^{\rm LGRBs}{\sim}3.6^\circ$ \citep{Frail2001}. However, the beaming factor for LGRBs carries some uncertainties and various authors claim lower correction factors of ${\approx}$50--100 \citep[\emph{blue-dotted and dashed lines}, respectively;][]{Piran2004,Guetta2005}. At $z{\lesssim}0.6$, the FXT volumetric rate exceeds the nominal LGRB rate by up to a factor of ${\sim}$7 (for the most favorable $f_b^{-1}{\sim}$500), while they appear consistent beyond $z{\gtrsim}0.6$. The FXT rate does not appear consistent with LGRB rates that adopt lower jet beaming correction factors (e.g., $f_b^{-1}{\sim}$50 or 100).

LL-LGRBs have relatively low isotropic luminosities of ${\sim}5{\times}10^{46}$--$10^{49}$~erg~s$^{-1}$, limiting our ability to see them out to large distances, and hence comprise only a small fraction of observed LGRBs. As such, they have a much higher local density rate of $\rho_{0,\rm LL-LGRBs}{\sim}$100--200~Gpc$^{-3}$~yr$^{-1}$ \citep{Zhang_book_2018}, and generally do not show strong evidence of collimation, implying a much wider jet opening angle, or even that the emission is essentially isotropic, i.e., $f_b^{-1}{\sim}$1  \citep{Virgili2009,Pescalli2015}. Normalizing the adopted cosmic evolution rate from \citet{Sun2015} to this value, we show the cosmic LL-GRB density rate as the \emph{green lines} in Fig.~\ref{fig:rate_plot} (\emph{left panel}). Following \citet{Liang2007}, we also consider a LL-LGRB jet beaming correction factor of $f_b^{-1}{\sim}$14 denoted by the \emph{green-solid line} and the isotropic case (i.e., $f_b^{-1}{\sim}$1) denoted by the \emph{green-dashed line}. The derived FXT volumetric rate is consistent with the more strongly beamed LL-LGRB rate, while it is slightly higher than lower beamed LL-LGRBs, especially at $z{\gtrsim}1$.

For SGRBs, the cosmic density rate is then shown in Fig.~\ref{fig:rate_plot} (\emph{middle panel}), considering Gaussian (\emph{red lines}), power-law (\emph{gray lines}) and log-normal (\emph{purple lines}) merger delay evolution models.\footnote{The merger delay is defined as the time elapsed between the formation of the binary star system and the merger, which is dominated by the timescale for gravitational wave losses during the compact binary phase \citep{Anand2018}.}, adopting an observed local density rate above $10^{50}$~erg~s$^{-1}$ of $\rho_{0,\rm SGRBs}{\sim}$\hbox{0.5--3.0}~Gpc$^{-3}$~yr$^{-1}$ \citep{Wanderman2015,Sun2015}. Additionally, it is known that at least some short GRBs are collimated \citep{Burrows2006,Soderberg2006,DePasquale2010}, with a mean jet opening angle $\theta_j^{\rm SGRBs}{\gtrsim}10^\circ$ \citep[e.g.,][]{Berger2014,Fong2022,Escorial2022}, translating to a mean value of $f_b^{-1}{\sim}$25 \citep[\emph{dotted lines};][]{Fong2015}. Nevertheless, the opening angle is not well-constrained, and other authors have suggested a wider range of $f_b^{-1}{\approx}$110--30 \citep[\emph{solid and dashed lines}, respectively;][]{Berger2014,Fong2022,Escorial2022}. From different delay models, we have distinct outcomes. For instance, from the delay merger Gaussian and log-normal models, the FXT volumetric rates between cosmic epochs $z{\approx}0.8$--2.0 appear slightly higher than the most extreme beaming correction case ($f_b^{-1}{\sim}$110). In contrast, under the power-law model, the FXT volumetric rates are higher than even the most extreme beaming correction case.

Finally, in Fig.~\ref{fig:rate_plot}, (\emph{right panel}), we consider both SMBH and IMBH TDEs, adopting the analytical cosmic density rate evolution of \citet{Sun2015}. For SMBH TDEs, the model is normalized to the local value of $\rho_{0,\rm SMBH-TDE}{\sim}$\hbox{(0.7--1.4)}${\times}10^5$~Gpc$^{-3}$~yr$^{-1}$ \citep[\emph{magenta line};][]{Sun2015}. Moreover, we assume that IMBHs have grown in a similar way to SMBHs, and adopt same cosmic evolution,  with a local density normalization of $\rho_{0,\rm IMBH-TDE}{\sim}$\hbox{75--500}~Gpc$^{-3}$~yr$^{-1}$ \citep[\emph{cyan-dotted line};][]{Malyali2019,Tanikawa2021}. Like SMBH TDEs \citep[e.g., \emph{Swift} 1644$+$57;][]{Bloom2011,Levan2011} and IMBH-WD TDEs could be capable of launching luminous jets which can be detected by current satellites, until reaching X-ray luminosities as large as ${\sim}10^{48}$~erg~s$^{-1}$ \citep{MacLeod2014,MacLeod2016}.
The FXT volumetric rate is generally lower than the rate of SMBH TDEs at $z{\lesssim}0.8$, while it matches with them at $z{\gtrsim}0.8$.
On the other hand, the FXT rate is much higher than our estimate of IMBHs, albeit with many untested assumptions \citep{Malyali2019,Tanikawa2021}. Of course, inconsistencies in several other parameters rule out an SMBH-TDE channel for several FXTs.

In Sect.~\ref{sec:flux}, we use the volumetric rate of FXTs to understand the most likely progenitors of the different sources.

\begin{figure*}
    \centering
    \includegraphics[scale=0.8]{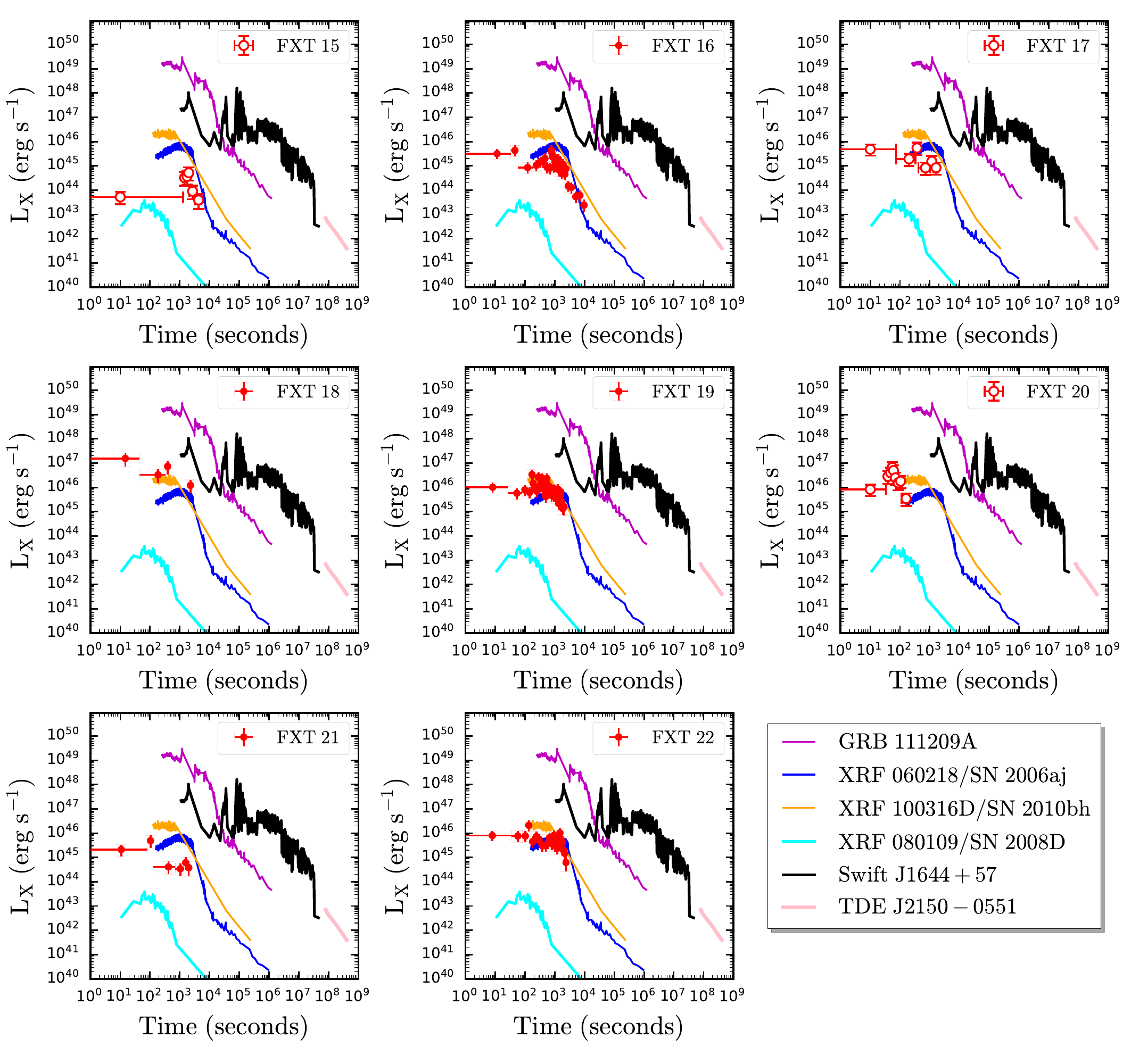}
    \vspace{-0.3 cm}
    \caption{Light curves of the eight FXTs in 0.3--10 keV luminosity units (converted from 0.5--7 keV light curves assuming best-fit spectral models in Sect.~\ref{sec:X-ray_fitting}). 
    Several individual transients are overplotted: 
    XRF~080109/SN~2008D \citep[the low-luminosity supernova SBO; \emph{solid cyan lines}, 27~Mpc;); XRF~060218/SN~2006aj (\emph{solid blue lines}, 145~Mpc;); XRF~100316D/SN~2010bh (\emph{solid orange lines}, 263~Mpc;][]{Barniol2015,Starling2011,Modjaz2009,Evans2009,Soderberg2008,Evans2007,Campana2006}; 
    GRB 110328A/\emph{Swift}~J1644$+$57 \citep[relativistically beamed TDE; \emph{solid black lines}, $z{=}0.3543$;][]{Bloom2011,Levan2011}; J2150$-$0551 \citep[unbeamed TDE; \emph{solid pink line}, $z{=}0.055$;][]{Lin2018}. For FXTs~15, 17, 18 and 20 (\emph{open symbols}), we assume $z{=}$1.0, and for FXTs~16, 19, 21, and 22 we take the redshift values from Table~\ref{tab:SED_para}.}
    \label{fig:flux_comparison}
\end{figure*}

\begin{figure*}
    \centering
    \includegraphics[scale=0.8]{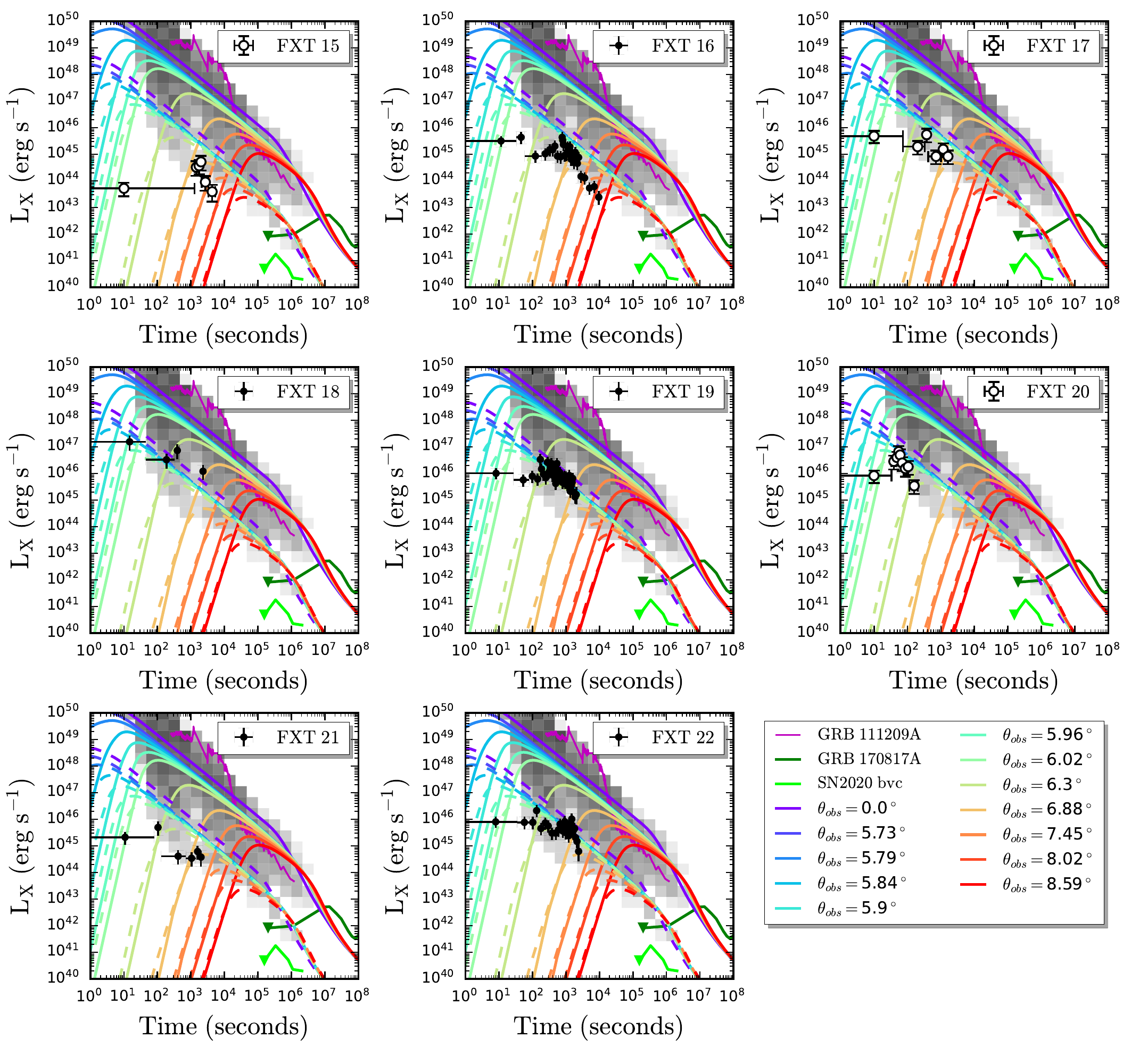}
    \vspace{-0.3 cm}
    \caption{Similar to Fig.~\ref{fig:flux_comparison}.
    The X-ray afterglow light curves of 64 LGRBs plus 32 SGRBs \citep[taken from][]{Bernardini2012,Lu2015} are shown as a 2D histogram, while several individual transients are overplotted: 
    GRB~111209A \citep[ultra-long duration LGRB; \emph{solid magenta line}, $z{=}0.677$;][]{Levan2014}; 
    GRB~170817A \citep[off-axis SGRB, multiplied by $\times$1000; \emph{solid dark green line};][]{Nynka2018,DAvanzo2018,Troja2020,Troja2022}; 
    SN~2020bvc \citep[the first off-axis LGRB candidate; \emph{solid light green line};][]{Izzo2020}, 
    and theoretical off-axis GRB afterglows at different viewing angles $\theta_{\rm obs}$ \citep[\emph{solid and dashed colour lines} represents afterglows with isotropic-equivalent energy and circumburst density of $10^{53}$~erg, 1~cm$^{-3}$ and $10^{51}$~erg, $0.15$~cm$^{-3}$, respectively;][]{Berger2014,Ryan2019,Chrimes2022}. For FXTs~15, 17, 18, and 20 (\emph{open symbols}), we assume $z{=}$1.0, and for FXTs~16, 19, 21, and 22, we take the redshift values from Table~\ref{tab:SED_para}.}
    \label{fig:off-axis_GRBs}
\end{figure*}

\section{Possible interpretations}\label{sec:flux}

To assess the nature of our final sample of FXTs, we compare them with other well-known transients. For FXTs~16, 18, 19, 21, and 22, we adopt their best-fit photometric or spectroscopic redshifts in Table~\ref{tab:SED_para}. For FXTs~15,  and 17, which lack clear host associations in optical and NIR images, and 20, which only has three detections in DECam images and poor photometric redshift constraints, we assume a fiducial distance of $z{=}1$, consistent with the average known redshift distribution. From the best-fit PL spectral model (see Table~\ref{tab:spectral_para}), we compute the peak X-ray flux (corrected for Galactic and intrinsic absorption; $F_{\rm X,peak}$), the associated intrinsic peak X-ray luminosity ($L_{\rm X,peak}$), and the Eddington mass (defined as $M_{\rm Edd}{=}7.7{\times}10^{-39}L_{\rm X,peak}$ in solar mass units); values are reported in Table~\ref{tab:F_L_para} in the 0.3--10.0~keV band.

FXTs~16, 18, 19, 21 and 22 reach peak X-ray luminosities of $L_{\rm X,peak}{\approx}$2.8$\times$10$^{45}$, 1.9$\times$10$^{47}$, 3.7$\times$10$^{46}$, 6.9$\times$10$^{45}$  and 1.3$\times$10$^{46}$~erg~s$^{-1}$, and 
have isotropic energies of $E_{\rm X}^{\rm iso}{\approx}$3.6${\times}$10$^{48}$, 1.7${\times}$10$^{50}$, 1.3${\times}$10$^{49}$, 1.8${\times}$10$^{48}$, and 1.0${\times}$10$^{49}$~erg, respectively (see Table~\ref{tab:F_L_para}).
Likewise, adopting $z{=}1$, FXTs~15, 17, and 20 would have peak \hbox{X-ray} luminosities of $L_{\rm X,peak}{\approx}$1.0$\times$10$^{45}$, 6.3$\times$10$^{45}$  and 8.1$\times$10$^{46}$~erg~s$^{-1}$ and isotropic energies of $E_{\rm X}^{\rm iso}{\approx}$7.5${\times}$10$^{47}$, 3.4${\times}$10$^{48}$, and 3.2${\times}$10$^{48}$, respectively (see Table~\ref{tab:F_L_para}).
This luminosity range automatically excludes lower luminosity ($L_{\rm X,peak}{\lesssim}$10$^{42}$~erg~s$^{-1}$) X-ray flaring transients such as X-ray binaries (including ultra-luminous X-ray sources), soft gamma repeaters, quasi-periodic eruptions, and anomalous X-ray pulsars \citep[e.g.,][]{Colbert1999,Kaaret2006,Woods2006,Miniutti2019}. Below, in Sects.~\ref{sec:SBOs}--\ref{sec:TDEs} we investigate the SBO, GRB, and TDE scenarios as origins of this FXT sample. In Sect.~\ref{sec:FXTs_P1} we compare these FXTs with those identified in Paper~I.

\subsection{Supernova Shock Breakouts (SBOs)}\label{sec:SBOs}

One intriguing explanation for FXTs is related to the SBO from a CC-SNe. An initial flash of thermal UV or soft X-ray radiation is expected when a CC-SNe shock wave emerges from the stellar surface of the progenitor \citep{Falk1977,Klein1978,Matzner1999,Schawinski2008,Ganot2016,Waxman2017}. The physical features of an SBO depend mainly on the density structure of the progenitor star and the explosion energy driving the shock wave \citep{Chevalier2011,Gezari2015}, which means that the temperature and duration of SBOs might cover a range of \hbox{${\sim}$10$^5$--5$\times$10$^6$}~K and ${\approx}$100--5,000~seconds, respectively \citep{Ensman1992,Tominaga2011}, leading to a bolometric peak luminosity of order ${\sim}$10$^{44}$--10$^{45}$~erg~s$^{-1}$. In the \hbox{0.5--7}~keV band, we would expect to observe a soft thermal spectrum, potential spectral softening with time, and peak luminosities at least 1 dex lower than the bolometric values.

Until now, just one SBO has been detected conclusively in multi-wavelength observations, XRT~080109/SN~2008D, serendipitously discovered during \emph{Swift}-XRT observations of SN~2007uy in NGC 2770 \citep[][]{Soderberg2008,Modjaz2009,Waxman2017}. Recently, a dozen further SBO candidates were reported among \emph{XMM-Newton} archival data by \citet{Alp2020} and \citet{Novara2020}. This subsample of FXTs has luminosities that fall within the ranges predicted by models and observations of SBOs \citep[\hbox{$L_{\rm X,peak}^{\rm SBOs}{\sim}\times$10$^{42}$--10$^{44}$~erg~s$^{-1}$};][]{Soderberg2008,Modjaz2009,Waxman2017,Alp2020,Novara2020}; however, at least four of these FXTs are associated with energy releases that are two orders of magnitude higher than SBO model predictions \citep[e.g.,][]{Waxman2017} or observations \citep[e.g., XRT~080109/SN~2008D had $E_{\rm X}{\sim}2\times$10$^{46}$~erg;][]{Soderberg2008}. 

Based on the energetics (the luminosity peaks are higher than those expected for SBOs, $L_{\rm X,peak}^{\rm SBOs}{\sim}10^{42}-10^{44}$~erg~s$^{-1}$) and light curves (which are much brighter than the SBO XRT~080109/SN~2008D, see Fig~\ref{fig:flux_comparison}), we rule out an SBO origin for FXTs~16, 18, 19, 21, and 22. Due to the natural relation between SBOs and both CC-SNe and super-luminous SNe (SL-SNe), we expect them to share similar host-galaxy properties. The host properties of FXTs~16, 18, and 19 fall in regions populated by SNe type II and SL-SNe hosts, but lie at the edges of SNe type Ib and Ic host distributions (see Fig.~\ref{fig:Mass_SFR_plot}). FXTs~21 and 22 reside at the edges of the SNe type Ib, Ic, and II, and outside the SL-SNe host distributions (see Fig.~\ref{fig:Mass_SFR_plot}). Thus, the FXT hosts do not show a robust link with SNe host galaxies, reinforcing the previous results from the energetics.

Similarly, for the FXTs which lack hosts (15,17) or redshift (20), the SBO scenario is ruled out at the fiducial $z{=}1$ values. The sources would need to lie at redshifts of $z{\lesssim}$0.37, 0.10, and 0.05, 
respectively, to comply with the expected energetic limits. At these redshifts, the apparent $r$-band magnitudes or limits would imply hosts with $M_{r}{\gtrsim}-18.5$, $-15.9$, and $-13.0$, respectively; only the host of FXT 15 lies at the faint end of regular galaxies, while the rest fall in the broad range of dwarf galaxies. FXTs~15 and 20 have BPL light curves with break times at ${\approx}$3.7 and 0.1~ks, respectively, followed by PL decays, $F_X{\propto}t^{-2.7}$, that are accompanied by possible softening (see Table~\ref{tab:fitting_para_Tbreak}) and photon indices ($\Gamma{=}$2.1 and 3.0, respectively; see Table~\ref{tab:spectral_para}), similar to the SBO XRF~080109/SN~2008D \citep[$\Gamma{\approx}2.3$;][]{Soderberg2008}. Finally, we note that contemporary optical time domain surveys would have detected an observable SNe associated with FXTs~17 and 20, if they were at $z{\lesssim}0.1$. In summary, we do not find evidence in support of an SBO origin for FXTs 17 or 20, but cannot discard it completely for FXT~15.

Comparing the FXT rate with the much larger total CC-SNe rate, it is clear that only a small fraction of SBOs can lead to FXTs (see Fig.~\ref{fig:rate_plot}, \emph{left panel}). After analyzing the volumetric rate of different massive star progenitors, we conclude that just some stars are more consistent with FXTs (see Sect.~\ref{sec:vol_rate}). Although the derived event rate density of FXTs falls especially close to the expected rate of BSGs (${\sim}2{\times}10^{3}$~Gpc$^{-3}$~yr$^{-1}$), such an association is largely ruled out by other characteristics such as energetics and host-galaxy properties. Thus, we conclude that this sample of FXTs is unlikely to be associated with SBOs from normal CC-SNe.

\subsection{Gamma-ray bursts (GRBs)}\label{sec:GRBs}

GRBs are characterized by an average emission timescale of ${\approx}$20~seconds for LGRBs and ${\approx}$0.2~seconds for SGRBs \citep{Meegan1996,Meszaros2006}. Currently, the accepted model of GRBs consists of a relativistically expanding fireball with associated internal and external shocks \citep{Meszaros1997}. Once the gamma-ray emission is generated, the expanding jetted fireball interacts with and shocks the surrounding interstellar medium, producing a broadband X-ray-to-radio afterglow. 
When the Doppler boosting angle of the decelerating fireball exceeds the jet aperture angle, it produces a steepening in the light curve known as the ``jet break'' \citep{Sari1999,Rhoads1999,Zhang2004}. The majority of LGRBs arise from the core-collapse of massive stars associated with hydrogen-poor, high-velocity type Ic supernovae \citep{Hjorth2003,Cano2013,Levan2016}. On the other hand, the current model of SGRBs is linked to the merger of a compact NS--NS or NS--BH binary \citep[e.g.,][]{Eichler1989,Narayan1992}, induced by angular momentum and energy losses due to gravitational wave (GW) emission and leading to a GW burst \citep{Abbott2016b}. The NS--NS channel could produce as a remnant either a millisecond magnetar \citep[e.g.,][]{Zhang2013,Sun2017} or a BH surrounded by a hyper-accreting debris disk. The NS--BH channel may also generate a debris disk, if the NS is disrupted outside the BH event horizon by tidal forces \citep{Rosswog2007,Metzger2019}. Once it happens, both the high accretion rate and rapid rotation yield energy extraction, thus allowing the launching of a relativistic jet, via either neutrino-antineutrino annihilation or magneto-hydrodynamic processes \citep[e.g.,][]{Blandford1977,Rosswog2002,Lee2007}.
The accretion event could produce an isotropic thermal supernova-like emission on timescales of ${\approx}$10$^4$--10$^6$~seconds called ``kilonova'' \citep[e.g.,][]{Berger2013,Tanvir2013,Gao2015,Sun2017,Pian2017,Arcavi2017,Metzger2019}.

No contemporaneous gamma-ray counterparts are detected near the X-ray trigger times for any FXTs in our sample, ruling out an on-axis GRB scenario. 
The intrinsic light curves of all FXTs, except 18, are flatter and fainter than the vast majority of on-axis X-ray afterglows over the same timescales (2D shaded histogram in Fig.~\ref{fig:off-axis_GRBs}), with initial luminosities ${\approx}$1--2~dex below the luminosity range $L_{\rm X,peak}^{\rm GRBs}{\gtrsim}$10$^{47}$~erg~s$^{-1}$ observed for GRBs. Beyond ${\sim}$10$^{2}$--10$^{3}$~seconds, however, most FXTs do begin to overlap energetically with the low-luminosity on-axis X-ray GRB afterglows. 

Overall, GRBs have \emph{canonical} light curves which can be split into up to five different components \citep{Zhang2006}, although not all X-ray afterglows necessarily exhibit all of them \citep[e.g.,][]{Nousek2006,Willingale2007,Zhang2007,Evans2007,Liang2007b,Liang2009,Evans2009}. The light curves components are (from the earliest to the latest): $i)$ steep decay phase (it is the tail of the prompt emission, from $F_X{\propto}t^{{\sim}-3}$ to ${\propto}t^{{\sim}-10}$); $ii)$ shallow decay or plateau phase (it could be interpreted invoking a continuous energy injection by a central engine, from $F_X{\propto}t^{{\sim}-0.7}$ to ${\propto}t^{{\sim}0.0}$); $iii)$ normal decay phase (it is the typical value predicted in the standard external forward shock model, $F_X{\propto}t^{{\sim}-1}$); $iv)$ jet break phase (it is a geometrical effect, $F_X{\propto}t^{{\sim}-2}$); and $v)$ X-ray flares (the GRB central engine directly powers them).

We note that FXTs~17, 18, and 21 exhibit PL decay light curves, similar to the normal decay phase of GRBs, while the other FXTs follow BPLs. However, FXTs~17, 18, and 21 as $F_X\,{\propto}\,t^{-0.3}$ and ${\propto}\,t^{-0.5}$, which is much shallower than the characteristic normal and jet-break phases \citep{Evans2009,Evans2007,Racusin2009}, but could be consistent with a shallow decay or plateau phase \citep{Troja2007,Rowlinson2010}.
Notably, FXT~21 exhibits temporal flaring behavior, which is potentially comparable to the strong X-ray flaring episodes seen in the tails of the X-ray afterglow in some GRBs \citep{Barthelmy2005,Campana2006,Chincarini2010,Margutti2011}, while its best-fit X-ray spectral slope of $\Gamma_{\rm FXT~21}{=}$3.1$\pm$0.6 is consistent with that of the standard afterglow distribution \citep[$\Gamma_{\rm GRBs}{=}$1.5--3.0;][]{Berger2014,Wang2015,Bauer2017} at the 1$\sigma$ confidence level.

On the other hand, FXTs~16, 19, and 22 show light curves consistent with a ${\approx}$2.1--3.7~ks plateau phase followed by a power-law decay ($F_X\ {\propto}\,t^{-1.9/-3.0}$; see Fig~\ref{fig:models_BPL} and Table~\ref{tab:fitting_para}) accompanied by likely spectral softening, especially for FXT~16 (see Fig~\ref{fig:models_BPL} and Table~\ref{fig:hardness}). Spectral softening has been seen previously in some GRBs afterglows \citep[e.g., GRB~130925A;][]{Zhao2014}. FXTs~16, 19 and 22 have photon indices ($\Gamma{\approx}$2.1--2.3; see Table~\ref{tab:spectral_para}) similar to GRB afterglows \citep[$\Gamma_{\rm GRBs}{\approx}$1.5--3.0;][]{Berger2014,Wang2015}.
Notably, some subsets of LGRBs \citep[e.g.,][]{Lyons2010} and SGRBs \citep[e.g.,][]{Rowlinson2010,Rowlinson2013,Gompertz2014} exhibit plateau phases, although only $<$10\% have plateau luminosities ${\lesssim}$10$^{47}$~erg~s$^{-1}$ which would be consistent with FXTs~16, 19 and 22 at their redshifts.

Finally, FXTs~15 and 20 have BPL light curves with break times at ${\approx}$3.7 and 0.1~ks, respectively, followed by PL decays, $F_X{\propto}t^{-2.7}$, accompanied by possible softening (see Table~\ref{tab:fitting_para_Tbreak}) and photon indices ($\Gamma{=}$2.1 and 3.0, respectively; see Table~\ref{tab:spectral_para}) similar to GRB afterglows \citep[$\Gamma{\approx}$1.5--3.0;][]{Berger2014,Wang2015} at a 1$\sigma$ confidence level.  However, their early rise phases ($F_{\rm X}\,{\propto}\,t^{0.4}$ and ${\propto}\,t^{1.0}$ for FXTs~15 and 20, respectively, see Table~\ref{tab:fitting_para}) are incongruent with the typical decays of on-axis GRBs X-ray afterglows (from $F_{\rm X}^{\rm GRBs}\,{\propto}\,t^{-1.5}$ to ${\propto}\,t^{-2.0}$). FXT~20's light curve shows many similarities to FXT~14/CDF-XT1 (see Fig.~\ref{fig:flux_comparison_2}), where its X-ray luminosity reaches a value of $L^{\rm XT1}_{\rm X,peak}{\approx}$10$^{47}$~erg~s$^{-1}$ without a clear softening in the spectra. The nature of FXT~14/CDF-XT1 is  still unknown, although several scenarios have been proposed recently by \citet{Sun2019}, \citet{Peng2019} and \citet{Sarin2021}.

We introduce the option of FXTs being associated with X-ray flashes (XRF), which may be related to shock breakout from choked GRB jets \citep[see Fig.~\ref{fig:flux_comparison};][]{Campana2006,Bromberg2012,Nakar2012}. We compare the light curves of the XRF~060218/SN~2006aj \citep{Pian2006,Campana2006} and XRF~100316D/SN~2010bh \citep{Starling2011} related with LL-LGRBs. We note that the plateau phases of FXTs~16, 19, and 22 have similar luminosities to those of XRF~060218 and XRF~100316D ($L_{\rm X,peak}{\sim}$10$^{45}$--10$^{46}$~erg~s$^{-1}$), and the break and late-time light curves also appear to match reasonably well. FXT~18 has higher luminosities (${\approx}10^{47}$~erg~s$^{-1}$) at early times than XRF~060218 and XRF~100316D, but it matches with them at later times. On the other hand, FXT~21, with its known redshift, looks inconsistent by a factor of ${\gtrsim}$5 compared to the XRFs khown. Finally,  the light curves of FXTs~15, 17, and 20 could be consistent with  those of XRFs, but the lack of constrained redshifts does not permit a proper intrinsic comparison of energetics.
Importantly, XRF~060218 and XRF~100316D show significant soft thermal components ($kT{\sim}$0.1--0.2~keV) which become dominant beyond ${\sim}$1000~s \citep{Campana2006,Starling2011,Barniol2015}. We find that only FXT~16 shows comparable spectral behavior; FXTs~15, 19, 20, and 22 do not exhibit any similar robust trend while FXTs~17, 18, and 21 actually appear to marginally harden at late times.

We also consider the option of FXTs being off-axis components of GRB afterglows (see Fig.~\ref{fig:off-axis_GRBs}). To explore this scenario, we use a numerical model, called \texttt{afterglowpy} \citep[developed by][]{Ryan2019}, to calculate synthetic light curves in X-rays. We generated synthetic X-ray afterglow for a range of viewing angles (from 0 to 8.6~deg) and assuming an isotropic-equivalent energy of $10^{53}$ ($10^{51}$)~erg and a circumburst density of $1.0$ ($0.15$)~cm$^{-3}$, respectively (Fig.~\ref{fig:off-axis_GRBs}, \emph{solid} [\emph{dashed}] lines), which represent the typical parameters for LGRBs (SGRBs) \citep[e.g.,][and references therein]{Berger2014,Chrimes2022}. For instance, LGRBs and SGRBs occur in high- and low-density environments, respectively, while SGRBs are less energetic than LGRBs.\footnote{In both cases, we consider a half-opening angle of $5.7$~deg, an electron energy distribution index of 2.2, a thermal energy fraction in electrons of $\epsilon_e{=}$0.1, and a thermal energy fraction in the magnetic field of $\epsilon_B{=}$0.01.} The light curves have a rise (from ${\sim}$1 to $10^4$~s, see Fig.~\ref{fig:off-axis_GRBs}) before reaching a peak luminosity, followed by an afterglow consistent with the on-axis GRB trend. Figure~\ref{fig:off-axis_GRBs} shows that an off-axis afterglow under small viewing angles could match the light curves of FXTs~15 and 20. In contrast, sources such as FXTs~16, 19, and 22, which have a plateau phase, cannot match the expected fast rise and curvature at early times of the slightly off-axis GRB cases. On the other hand, the light curves of some FXTs do appear to crudely match certain off-axis angle cases of SGRBs (\emph{dashed lines}), because of their lower luminosity. Finally, we compare FXTs with the potential high inclination off-axis LGRB SN~2020bvc \citep[with viewing angle $\theta_{\rm obs}{\approx}23$~deg;][]{Izzo2020}, and SGRB GRB~170817A \citep[with viewing angle $\theta_{\rm obs}{\approx}23$~deg;][]{Nynka2018, DAvanzo2018, Troja2020, Troja2022} in Fig.~\ref{fig:off-axis_GRBs}. Notably, SN~2020bvc and GRB~170817A are much less luminous ($L_{\rm X,peak}{\lesssim}$3$\times$10$^{41}$~erg~s$^{-1}$) than the sample of FXTs, by at least ${\sim}$5 orders of magnitude. 
In general, at high off-axis angles, we can expect later onsets, fainter light curves, lack of decay phases at early times, and peak luminosities at later times \citep[e.g.,][]{Granot2002,Ryan2019,Oganesyan2020,Ascenzi2020}. Overall, this comparison suggests that an association of FXTs with high inclination angle afterglows of GRBs is unlikely, although a mildly off-axis SGRB scenario remains plausible.

In terms of host galaxies (see Sect.~\ref{sec:counterpart_SED} for more details), based on the host stellar mass and SFR of FXTs~16, 18, 19, and 22, the galaxies lie above the galaxy-main sequence, in a parameter-space region populated mainly by GRBs (see Figs.~\ref{fig:Mass_SFR_plot}). Nevertheless, it remains difficult to disentangle an association with LGRBs or SGRBs from the current data. In contrast, FXT~21's host is below the galaxy-main sequence and shares properties more similar to SGRB hosts (especially the stellar mass).

Due to the physical offsets of FXTs~16, 19, and 21 (Fig.~\ref{fig:host_parameters_1}, \emph{right panel}) overlapping with the cumulative distributions of CC- and type Ia SNe, and SGRBs at 1$\sigma$ confidence level (see Fig.~\ref{fig:host_parameters_1}), the projected physical offsets are not enough to confirm or rule out the different scenarios.
Although the offset distance of FXT~18 suggests a unique and apparent association with SGRBs, the considerable associated X-ray positional uncertainty does not permit us to consider its offset as a robust discriminator.
Finally, FXT~22 has a sizeable physical offset which strongly disfavors a robust association with LGRBs, CC-, type Ia SNe, leaving only a SGRB association as a possible scenario. For instance, the dynamical evolution of the BNS due to a kick velocity \citep[the formation of each compact object is associated with one supernova explosion;][and references inside]{Fong2013c,Berger2014} could explain the significant offset of FXT~22 (${\approx}40$~kpc).

In the case of FXT~16, its light curve (the plateau and power-law decay as $F_X{\propto}t^{-2}$), spectral softening trend, \hbox{host-galaxy} offset distance, and host-galaxy properties are consistent with a compact star merger origin, following \citet{Xue2019}. \citet{Sun2019} explain the X-ray emission assuming a magnetar remnant after a BNS merger observed at a slightly off-axis viewing angle. Although FXTs~19 and 22 do not follow the same spectral trend, they share similar timing properties (a plateau phase in their light curves) and belong to star-forming host galaxies, as does FXT~16. However, FXT~22's host is one of the most massive galaxies of the sample.

The volumetric rates reinforce some previous conclusions from the timing, spectra, and host properties (for more details, see Sect.~\ref{sec:vol_rate}).
In the case of LGRBs (see in Fig.~\ref{fig:rate_plot}, \emph{left panel}), the FXT volumetric rate is higher than the LGRB rate by up to a factor of ${\sim}$7 at $z{\lesssim}0.6$ (even for $f_b^{-1}{\sim}$500), but appears consistent beyond $z{\gtrsim}0.6$ just for the case $f_b^{-1}{\sim}$500. In this sense, LGRBs with higher beaming corrections remain a potential progenitor for FXTs, while an association with LGRBs with lower jet beaming factors (e.g., $f_b^{-1}{\lesssim}$200) seems unlikely. However, the lower luminosity of FXTs becomes challenging to explain under this context. Moreover, we identified that the FXT volumetric rate is well-matched to the LL-LGRB rate considering a moderate beaming correction ($f_b^{-1}{\sim}14$), while it is slightly higher than lower beamed LL-LGRBs ($f_b^{-1}{\sim}1$) beyond $z{\gtrsim}1$ (see Fig.~\ref{fig:rate_plot}, \emph{left panel}). Thus, based on volumetric rates and luminosities, we conclude that LL-LGRBs remain a viable channel to explain FXTs. However, host properties do not align completely with this statement.

Finally, in the case of SGRBs, the volumetric rates give possible clues about an association between FXTs and SGRBs. From the delayed merger Gaussian and log-normal models, the FXT volumetric rates at $z{\gtrsim}0.8$ and $z{\lesssim}2$ appear slightly higher than even the case of $f_b^{-1}{\sim}$110 (see Fig.~\ref{fig:rate_plot}, \emph{middle panel}). Meanwhile, FXT remains a factor of ${\sim}4$ higher than SGRB rates assuming lower beaming correction values (i.e., $f_b^{-1}{\sim}$30-25). Thus, a link with SGRBs remains plausible, although it requires relatively strong beaming corrections, which unfortunately remain poorly constrained. This result agrees with the low luminosity of FXTs and the host galaxy properties.

\subsection{Tidal disruption events}\label{sec:TDEs}

Another potential FXT progenitor scenario is related to TDEs \citep{Rees1988,Phinney1989,Burrows2011,Saxton2021}. TDEs occur when a red giant (RG), main-sequence (MS) star or WD (${\approx}$0.008--0.02~$R_\odot$; ${\approx}$1~$M_\odot$) passes so close to a SMBH or IMBH that it undergoes tidal forces which exceed its self-gravity, causing it to be disrupted. A substantial fraction of the tidal debris will fallback onto the BH, leading to luminous thermal emission at soft X-ray through optical wavelengths, either by the accretion of this gas onto the BH and/or the initial shocks due to colliding stellar debris streams \citep{Guillochon2015}. A delay between the disruption and the accretion of gas onto the black hole may cause a delay between the optical and X-ray emission \citep[e.g.,][]{Hayasaki2021}. The debris fallback rates can range from strongly (${\sim}10^{4}$) super-Eddington to strongly (${\sim}10^{-3}$) sub-Eddington, with respective peak timescales from $<$1 day to more than 100 years \citep[e.g.][]{LawSmith2017}. This is confirmed by an observed empirical correlation between the peak light curve emission time and IMBH/SMBH mass \citep{vanVelzen2020}, such that IMBH-WD TDEs are expected to rise to peak within minutes/hours, while SMBH/IMBH-MS TDEs (depending on the mass and spin) take  roughly months to years \citep{Krolik2011,Haas2012,Kawana2018}.

For MS stars disrupted by a SMBH (10$^6$--10$^8$~$M_\odot$) or IMBH (10$^3$--10$^5$~$M_\odot$), the radiation should peak around $T_{\rm eff}{\sim}10^{4}$--$10^{6}$~K, i.e., at UV to soft X-ray wavelengths. Mild cooling is predicted, although substantial variations are seen empirically \citep{Gezari2021}.
For BHs exceeding ${\sim}$10$^5$~$M_\odot$, a  WD would be swallowed whole, leaving no expected emission signature \citep{Clausen2012,Kawana2018}. However, a high spin rate that increases the Hills mass for BHs with masses ${\lesssim}10^6$~$M_\odot$ could enable IMBHs to potentially disrupt more white dwarfs, but, overall, it is difficult to explain all the X-ray flares as solely due to WD TDEs \citep{Maguire2020}.
In addition to the possibility of exceeding the Eddington rate by large factors, emission from relativistic jets is also possible, particularly if the disruption involves a strongly magnetic WD \citep[e.g.,][]{Cenko2012, Brown2015,Sadowski2016}. Relativistic beaming from jetted TDEs such as \emph{Swift}~J1644+57 \citep[\emph{black solid lines} in Fig.~\ref{fig:flux_comparison};][]{Bloom2011,Levan2011,Saxton2021} can generate much higher luminosities ($L_{\rm X,peak}{\sim}$10$^{48}$~erg~s$^{-1}$), rapid and strong variability, and harder X-ray spectra  \citep[$\Gamma{=}$1.6--1.8;][]{Levan2011}, although the photon index softens with decreasing flux \citep{Bloom2011}. However, some sources, such as the TDE AT2021ehb, show a hardening spectral trend with time, which is interpreted as the gradual formation of a magnetically dominated corona \citep{Yao2022}.

TDEs involving SMBHs should occur in the centers of more massive galaxies. Thus,  we can automatically discard an association of FXTs~16, 19, 20, and 22 with SMBH TDEs because the sources are offset from the centers of their host galaxies. In the case of FXT~18, although it is offset from its host galaxy candidate, the significant X-ray positional uncertainty does not allow us to rule out the association with SMBH TDEs. We arrive at a similar conclusion for FXT~21 and the hostless events FXTs~15 and 17. In contrast, TDEs involving IMBHs may occupy a larger range of possibilities, e.g., occurring near the centers of dwarf galaxies or in crowded stellar systems such as globular clusters \citep[e.g.,][]{Jonker2012,Reines2013}. Thus, the offset of FXTs~16, 19, 20, and 22 remain consistent with a possible IMBH-WD TDE association. Moreover, given the short durations of the FXTs and exclusive detection in the X-ray band to date, the IMBH-WD TDE scenario seems to be most applicable. Below, we will explore the association with IMBH TDEs.

The BPL light curves of FXTs~15 and 20 most closely follow the expected light curve shape for IMBH--WD TDE candidates \citep[e.g.,][]{MacLeod2014,Malyali2019,Peng2019}, with a fast rise and exponential decline (see Fig~\ref{fig:models_BPL} and Table~\ref{tab:fitting_para}). Although their late-time spectral slopes are relatively soft, their initial slopes are much harder than expected for TDEs \citep[$T_{\rm bbody}{\approx}$0.02--0.13 keV;][]{Gezari2021}. The nominal peak luminosities of $L_{\rm X,peak}{\sim}$\hbox{10$^{45}$--10$^{47}$~erg~s$^{-1}$}, respectively, at fiducial redshifts of $z=1$, are several orders of magnitude larger than the expected Eddington limits for IMBH-WD TDEs or what is observed from local candidates \citep[e.g., IMBH TDE candidate TDE~J2150-05 has a $L_{\rm X,peak}{\sim}$10$^{43}$~erg~s$^{-1}$;][]{Lin2018}, requiring invocation of extreme super-Eddington accretion or relativistic beaming to explain them under a TDE scenario. The peak luminosities are more in line with beamed TDEs \emph{Swift}~J1644+57 \citep[$L_{\rm X,peak}{\sim}$10$^{46}$--10$^{48}$~erg~s$^{-1}$; see Fig.~\ref{fig:flux_comparison};][]{Bloom2011,Levan2011}, but it is related with an SMBH TDE emission \citep[although some authors claim by an association with IMBH TDEs, e.g.,][]{Krolik2011}. We cannot exclude an IMBH TDE explanation for FXTs~15 and 20, although they would clearly require special conditions.

FXTs 17, 18, and 21 show PL declines from the very start with relatively soft spectral slopes. The lack of any detectable rise appears inconsistent with expected TDE light-curve shapes. However, the soft X-ray spectral shapes, particularly in the case of FXT~18, are potentially consistent with the properties of some IMBH TDEs \citep[e.g., ][]{MacLeod2014,Malyali2019}. Moreover, we do not have sufficient counts to resolve the fast rise times which it may expect for some IMBH-WD TDEs \citep[e.g.,][]{MacLeod2016}. Again, the peak luminosities (adopting a fiducial redshift of $z{=}1$ and photometric redshift of $0.35$ for FXTs~17 and 18, respectively) are a few orders of magnitude larger than the expected Eddington limits for IMBH-WD TDEs or what is observed from local candidates, requiring super-Eddington accretion or relativistic beaming to explain them under an IMBH TDE scenario. Subsequent observations for FXT~17 can rule out any extended bright, long-term variability, 
however, it is not the case for FXT~18, which has not been revisited by X-ray observatories. For these reasons, we disfavor a TDE explanation for FXTs~17 and 21.

The light curves of FXTs~16, 19, and 22 show ${\approx}$2.1--3.7~ks plateaus with subsequent power-law decay (from $F_X{\propto}t^{-1.9}$ to ${\propto}t^{-3.0}$), accompanied by robust spectral softening in the case of FXT~16 (Fig.~\ref{fig:models_BPL}, and Tables~\ref{tab:fitting_para} and \ref{fig:hardness}). Although not commonly observed in X-ray emission from TDEs as yet \citep{Gezari2021}, some eccentric fallback or reprocessing scenarios could potentially explain this behavior. On the other hand, the overall spectra of FXTs~16, 19 and 22 are best-fit with photon indices of $\Gamma{\approx}$2.1--2.3 (see Table~\ref{tab:spectral_para}), which are much harder than expected for TDEs, while their peak luminosities ($L_{\rm X,peak}{\approx}$\hbox{3$\times$10$^{45}$--7$\times$10$^{46}$}~erg~s$^{-1}$ are also generally much higher than candidate IMBH-TDEs identified to date.
A relativistic-beamed IMBH-TDE scenario could better explain some of the X-ray properties (luminosities, spectral slopes) of FXTs~16, 19, and 22 (e.g., \citealp{Peng2019} argue for an IMBH--WD TDE scenario for FXT~16), although subsequent observations indicate that none shows extended durations or variability evolution such as seen in \emph{Swift}~J1644+57. 
Finally, FXTs~16, 19, and 22 are all significantly offset from the nuclei of their associated hosts by ${\approx}$\hbox{0\farcs4--4\farcs6} (or physical distances of ${\approx}$3.3--40~kpc; see Fig.~\ref{fig:image_cutouts}), requiring an ejected IMBH scenario, or in the stripped nucleus of an infalling galaxy \citep[such as TDE~J2150-05;][]{Lin2018}, in order for the TDE scenario to remain viable. For these reasons, we disfavor a TDE explanation for FXTs~16, 19, and 22, although relativistically beamed emission from an IMBH--WD TDE scenario cannot be ruled out.

Regarding the host galaxy properties, we might expect to find IMBHs near the centers of dwarf galaxies, in globular clusters at large offsets in massive galaxies, or ejected via 3-body interactions \citep{Komossa2008,Jonker2012,Reines2013}. This, we might naively expect to identify FXTs associated with IMBH-WD TDEs in any type of host galaxy, and with a wide range of projected offsets. FXTs~16, 18 and 19 have hosts with $M_*{\lesssim}10^9$~M$_\odot$, while FXT~21 and 22 hosts have larger stellar masses ($M_*{\sim}10^{11}$~M$_\odot$). 
Thus, we cannot discard an IMBH-WD TDEs scenario for FXTs for any event.

FXTs rate is only lower than the rate of SMBH TDEs for $z{\lesssim}0.8$ (see Fig.~\ref{fig:rate_plot}, \emph{right panel}). In contrast, in the case of IMBH TDEs, the FXT rate is much higher during the cosmic time but potentially consistent with just a fraction of FXTs (because likely we have a mix of FXTs origins). Another possibility could be the different energetics between FXTs and IMBH TDEs (discarding the beaming case, which occurs just in a small fraction of events). Moreover, based on inconsistencies in several other parameters (such as the offset from transient X-ray position and host galaxies) we can rule out an SMBH-TDE channel for several FXTs.

Finally, the partial consistency between volumetric rates of FXTs and different transients classes at different redshifts (see Sect.~\ref{sec:rates} for more details), timing and spectral parameters (see Sect.~\ref{sec:time_spectra_prop}), and host properties (see Sect.~\ref{sec:counterpart_SED}),  may suggest that the overall sample of FXTs arise from a heterogeneous set of progenitors. Detection of contemporaneous EM counterparts from future FXTs remains crucial to disentangle these multiple formation channels. Nonetheless, we strongly caution the reader not to overinterpret the consistency or lack thereof between FXTs and many of the transient classes, as we have implicitly assumed no density evolution in our calculations (which there easily could be) and the density evolution assumed for several of the other transient classes is not well-constrained. Thus, some of the previously mentioned discrepancies at low or high redshift could be no more than artifacts of these assumptions.

\subsection{FXTs discovered in Paper~I}\label{sec:FXTs_P1}

The FXTs discovered here share many similarities with the previous distant FXTs identified in Paper~I, in terms of their timing (Fig.~\ref{fig:flux_comparison_2}), spectral (Figs.~\ref{fig:X-ray-params} and \ref{fig:hardness}), and host-galaxy properties (Fig.~\ref{fig:host_parameters_total_sample}). 
Unfortunately, the lack of host-galaxy detections for many FXTs identified here and in Paper~I does not permit more detailed comparisons of energetics among the two samples.
It is clear that according to the properties of the hosts we do detect, there is no single unifying class of galaxies (in terms of SFR and stellar mass) that could harbor a unique kind of transient. We conclude that the FXTs reported here likely have $z{\gtrsim}0.2$, i.e., they are not related to local galaxies (see Fig.~\ref{fig:host_parameters_total_sample}, \emph{bottom panel}), and presumably span a wide distance range.

\section{Expected sources in current and future missions}\label{sec:event_future}

Based on the event rate computed in Sect. \ref{sec:event_rate}, we explore the expected number of FXTs that should be detectable in other ongoing and future X-ray missions. The expected event rate of another (\emph{New}) mission (called $\mathcal{R}_{\rm New}$) regarding our results ($\mathcal{R}_{\rm Total}$) is
\begin{equation}
    \mathcal{R}_{\rm New}=\left[\frac{\mathcal{N}({>}S_{\rm New,lim})}{\mathcal{N}({>}S_{\rm CXO,lim})}\right]\mathcal{R}_{\rm Total},
    \label{eq:010}
\end{equation}
where $\mathcal{R}_{\rm New}$ and $\mathcal{N}({>}S_{\rm New,lim})$ are the event rate and X-ray fluence limit of the {new mission} (taken from Sect.~\ref{sec:event_rate}), respectively, and $\mathcal{N}({>}S_{\rm CXO,lim})$ represents the fluence limit of \emph{Chandra} (taken from Sect.~\ref{sec:event_rate}). As we explain in Sect.~\ref{sec:event_rate}, the event rate behaves as a BPL function. 
Then, the expected total number of sources must be
\begin{equation}
\begin{split}
    \mathcal{N}_{\rm New}&{=}\Omega_{\rm New} T_{\rm New}\mathcal{R}_{\rm New}\times\frac{EA_{\rm New}}{EA_{\rm CXO}}\\
    &{=}\Omega_{\rm New} T_{\rm New}\left[\frac{\mathcal{N}({>}S_{\rm New,lim})}{\mathcal{N}({>}S_{\rm CXO,lim})}\right]\mathcal{R}_{\rm Total}\times\frac{EA_{\rm New}}{EA_{\rm CXO}},
    \label{eq:011}
\end{split}
\end{equation}
where $\Omega_{\rm New}$ and $T_{\rm New}$ are the FoV and the operational time of a {new mission}, respectively. Also, we consider an {\it ad hoc} term, $EA_{\rm New}/EA_{\rm CXO}$, which is a correction factor defined as the ratio between the integrated effective area of the {new mission} and \emph{Chandra} in the energy range of 0.5--7.0~keV.\footnote{We take into account the effective area per instrument from public data.} It is important to realize that Eq.~\ref{eq:011} considers the ratio between the {new mission} ($F_{\rm New,lim}$) and \emph{Chandra} (the limit imposed by our method $F_{\rm CXO,lim}{=}$1$\times$10$^{-13}$~erg~cm$^{-2}$~s$^{-1}$, which is roughly 10$\times$ the nominal point source detection sensitivity in a 10\,ks window; see sect. 2.1 in Paper~I) X-ray flux limits, respectively. We estimate numbers adopting a limit 10$\times$ above the {new mission's} nominal 10-ks flux limit to avoid the Poisson fluctuations in our calculations. Given the low-count statistics, we quote estimates incorporating the Poisson 1$\sigma$ errors. 
\newcommand{\xmm}{6\ [4--7]}
\newcommand{\xrt}{0.8\ [0.7--1.1]}
\newcommand{\erosita}{3\ [2--4]}
\newcommand{\ep}{13\ [10--16]}
\newcommand{\starx}{180\ [140--235]}
\newcommand{\axis}{50\ [39--63]}
\newcommand{\athena}{460\ [357--581]}

\begin{table}[]
    \centering
    \scalebox{0.8}{
    \begin{tabular}{lrrcc}
    \hline\hline
    Mission & FoV & $T_{\rm avail}$ & $F_{\rm lim}$ & FXTs \\ 
     & (deg$^2$) & (yr) & (erg~cm$^{-2}$~s$^{-1}$) & (\# yr$^{-1}$) \\ \hline
    (1) & (2) & (3) & (4) & (5) \\ \hline
    \multicolumn{5}{c}{Ongoing missions} \\ \hline
    \emph{XMM-Newton}-EPIC & 0.25 & ${\sim}$15.1 & ${\sim}$1$\times$10$^{-13}$ & \xmm \\ 
    \emph{Swift}-XRT       & 0.15 &  ${\sim}$11.3 & ${\sim}$3$\times$10$^{-13}$ & \xrt \\ 
    \emph{SRG}–eROSITA     & 0.80 &  ${\sim}$4.0 & ${\sim}$4$\times$10$^{-13}$ & \erosita \\ 
    \hline 
    \multicolumn{5}{c}{Future missions} \\ 
    \hline
    \emph{Einstein Probe}-WXT & 3600 & ${\gtrsim}$3.0$^\dagger$ & ${\sim}$5$\times$10$^{-10}$ & \ep \\
    \emph{STAR-X}-XRT         & 1.00 & ${\gtrsim}$5.0$^\dagger$ & ${\sim}1{\times}10^{-14}$ & \starx \\
    \emph{AXIS}               & 0.12 & ${\gtrsim}$4.0$^\dagger$ & ${\sim}$3$\times$10$^{-14}$ & \axis \\
    \emph{Athena}-WFI         & 0.40 & 0.6 & ${\sim}$5$\times$10$^{-15}$ & \athena \\
    \hline
    \end{tabular}}
    \caption{Expect number of FXTs for different X-ray missions. \emph{Column 1:} Mission and instrument. \emph{Column 2:} Nominal field-of-view. \emph{Column 3:} Available exposure time considered, actual (launch to date) or nominal (mission lifetime, denoted by $^\dagger$). \emph{Column 4:} Assumed FXT detection X-ray flux limit (to avoid Poisson fluctuations, we adopt 10$\times$ the nominal source detection flux limit). \emph{Column 5:} Predicted FXT number per year.}
    \label{tab:Expected_number}
\end{table}

We begin with estimates for ongoing missions (\emph{XMM-Newton}, \emph{Swift}--XRT, and \emph{SRG}-eROSITA), and then future observatories (\emph{Athena}, \emph{Einstein Probe}, \emph{STAR-X} and \emph{AXIS}) which are expected to have enough flux sensitivity and time in orbit to detect similar FXTs as those found here. Table~\ref{tab:Expected_number} shows a summary of the assumed FoV, $T_{\rm new}$ and $F_{\rm new,lim}$ used to calculate the expected number of FXTs per year for each mission and our results; we give the expected number of FXTs per year to allow the reader to compute more readily the number of FXTs for any exposure time or mission length given. For all missions, we adopt a spectral slope of $\Gamma{=}1.7-2.3$, typical of FXTs. More accurate estimates involving Monte Carlo techniques go beyond the scope of this work.

The European Photon Imaging Camera (EPIC) on board the \emph{XMM-Newton} telescope has a FoV${\approx}$0.25~deg$^2$, 10-ks flux sensitivity of ${\approx}$10$^{-14}$~erg~cm$^{-2}$~s$^{-1}$ in the 0.15--12~keV band, and has accumulated ${\approx}$476~Ms total exposure over ${\sim}$20 years in orbit in full-frame mode \citep[mean value between pn and MOS cameras;][]{Ehle2003,Webb2020}. We adopt a correction factor to account for the contribution of background flares (assuming that 30--40\% of exposure time is affected by them) and a flux cutoff of $F_{\rm XMM,lim}{\approx}$10$^{-13}$~erg~cm$^{-2}$~s$^{-1}$, we predict ${\approx}$\xmm~FXTs~yr$^{-1}$ and means ${\approx}60-106$ FXTs.

\emph{Swift}--XRT has a FoV${\approx}$0.15~deg$^2$, a 10-ks flux sensitivity of ${\approx}$3${\times}$10$^{-14}$~erg~cm$^{-2}$~s$^{-1}$ in the 0.2--10~keV band, with around ${\approx}$~357~Ms of archival data over ${\sim}$14 years in orbit \citep{Hill2000,Burrows2003,Evans2023}. Considering a flux limit of $F_{\rm XRT,lim}{\approx}$8${\times}$10$^{-13}$~erg~cm$^{-2}$~s$^{-1}$, the expected number of FXTs per year is ${\approx}$\xrt~FXTs~yr$^{-1}$, which means ${\approx}7-12$ FXTs.

The Spectrum-Roentgen-Gamma (\emph{SRG})--eROSITA mission, launched in July 2019, is scanning the entire sky in the \hbox{0.2--10}~keV band with a FoV${\approx}$0.833~deg$^2$. \emph{SRG}--eROSITA's all sky survey has an official 4-year survey phase, and is expected to reach 10-ks flux limits of ${\approx}$10$^{-14}$ and ${\approx}$10$^{-13}$~erg~cm$^{-2}$~s$^{-1}$ in the 0.5--2 and 2--10~keV bands, respectively. However, eROSITA scans the entire sky every six months, leading to eight seasons after the nominal 4-year planned lifetime \citep{Predehl2021}. At present, eROSITA has completed four epochs before entering an extended hibernation mode.

One strong limitation for FXTs to be detected by eROSITA comes from the individual 40-second drift-scan exposures during each 4-hour rotation period \citep{Predehl2021}, such that an equatorial field will be visited during only three consecutive passes, or $\approx$12hr, over a 6-month span, while higher declination fields would experience higher numbers of consecutive passes; the net effect is that the light curves of possible FXTs will only be covered sparsely, if at all. Given the typical duration of the extragalactic FXT candidates characterized here and in Paper I, we thus expect an FXT to be detectable only during a single 40-s pass and undetectable in the previous or subsequent pass, i.e., 14400\,s (4\,hr) before or after. 
We must also consider that the 40-s window will almost never catch an FXT at peak, and thus we adopt the average flux (which is typically a factor of 10 lower than the peak). Thus, we consider an FXT 0.5--2~keV flux limit of $F_{\rm eROSITA,lim}{\approx}$3${\times}$10$^{-12}$~erg~cm$^{-2}$~s$^{-1}$ (to avoid Poisson noise), 
which yields an expected number of FXTs detected by \emph{SRG}--eROSITA of ${\approx}$\erosita~FXTs~yr$^{-1}$. However, this estimate should be considered an upper limit due to the short snapshot observations. For example, while FXTs such as CDF-S\,XT2 should be detectable in most situations by comparing excess in one or more snapshots due to their long duration (${\gtrsim}$10~ks), we will miss many sources with shorter burst times (${\lesssim}$1--5~ks) such as FXT~1/XRT~000519, FXT~14/CDF-S\,XT1, and FXT~20/XRT~191127 that occur between passes. Moreover, sources caught in the first or last pass of a 6-month season will remain ambiguous due to the poorly constrained light curves.

The above results suggest that an important number of FXTs await discovery inside the \emph{XMM-Newton}, \emph{Swift}--XRT and eROSITA archives. Until now, a few projects have developed systematic searches to identify FXTs. For instance, the systematic searches made by \citet{Alp2020}, the "Exploring the X-ray transient and variable sky" (EXTraS) project \citep{DeLuca2021} and the EPIC-pn \emph{XMM-Newton} Outburst Detector (EXOD) search project \citep{Pastor2020} have reported 12, 136 and ${\approx}$2500 candidates to date, respectively; the large numbers from the latter two are strongly dominated by Galactic stellar flares, cataclysmic variables, type I X-ray bursts, supergiant flares, as well as extragalactic AGN and SBOs. 

In the case of future missions, {the Advanced Telescope for High ENergy Astrophysics (\emph{Athena})} will characterize the hot and energetic universe from the mid-2030s. It will cover the 0.2--12~keV band with a 1.4\,m$^{2}$ effective area at 1~keV, and have a nominal lifetime of five years \citep[although it could be extended for 10 years depending on consumables;][]{Nandra2013,Barret2013,Barret2023}. The {Wide Field Imager (WFI)} will have a spectral resolution of $\Delta E{<}170$~eV at 7~keV, a spatial resolution of $\leq$10~arcsec PSF on-axis, and FoV of 0.44~deg$^2$ \citep{Rau2016}. To estimate the number of FXTs, we assume a flux limit ${\times}$10 higher than the nominal 10~ks limit of $F_{\rm WFI,lim}{\approx}$5$\times$10$^{-15}$~erg~cm$^{-2}$~s$^{-1}$ for the WFI deep fields. Thus, the expected number of FXTs detected by \emph{Athena} will ${\approx}$\athena~FXTs~yr$^{-1}$. Assuming that the WFI observations will be spread evenly during the mission and that those observations will be performed during the \emph{Athena} ground contact, approximately one-sixth of the sources (${\approx}$60--97~FXTs~yr$^{-1}$) could have \emph{Athena} alerts with latencies ${<}$4~hours. This could permit the investigation of the multi-wavelength properties of FXTs via coordinated campaigns with ground and space telescopes in other energy ranges.

The \emph{Einstein Probe} (EP) mission will explore high-energy transient and variable phenomena in the 0.5--4.0~keV band \citep{Yuan2015,Yuan2017,Yuan2018,Yuan2022}, with a scheduled launch by the end of 2023 and a 3-year operational lifetime \citep[and 5-year goal;][]{Yuan2017}. EP will harbor two scientific instruments, the Wide-field X-ray Telescope (WXT) with a large instantaneous FoV of 3600~deg$^{2}$ and a narrow-field Follow-up X-ray Telescope, and a fast alert downlink system \citep{Yuan2015,Yuan2018}. To compute the expected number of FXTs, we consider only the WXT instrument with a threshold sensitivity of $F_{\rm WXT,lim}{\approx}$5$\times$10$^{-10}$~erg~cm$^{-2}$~s$^{-1}$ at 1~ks, yielding ${\approx}$\ep~FXTs~yr$^{-1}$.

\emph{STAR-X} is a proposed equatorial low-earth orbit NASA mission comprised of an X-ray telescope (XRT) and a UV telescope \citep[UVT;][]{Saha2017,Saha2022}. It aims to conduct a time-domain survey and respond rapidly to transient sources discovered by other observatories such as LIGO, Rubin LSST, Roman, and SKA. XRT will have a ${\approx}$2\farcs5 half-power diameter PSF, an on-axis effective area of ${\gtrsim}$1,800~cm$^2$ at 1~keV, 1~deg$^2$ FOV, low particle background, and an on-board transient event alert capability of $\sim$5 min \citep{Saha2017,Saha2022}. Thus \emph{STAR-X} will be at least 1~dex more capable and more sensitive than \emph{Chandra} and \emph{Swift}-XRT to find and study transient sources in the 0.2--6~keV band.
To compute the potential expected number of FXTs, we again consider a 10~ks threshold sensitivity of $F_{\rm STAR-X,lim}{\approx}1{\times}10^{-15}$~erg~cm$^{-2}$~s$^{-1}$ (at 0.5--2~keV), which to avoid Poisson fluctuations we multiply by 10, yielding ${\approx}$\starx~FXTs~yr$^{-1}$. 
However, during its nominal 2-yr mission, \emph{STAR-X} will observe the extragalactic sky primarily through two time domain surveys, called {\it Deep} and {\it Medium} modes, which invoke different observing strategies. The {\it Deep} ({\it Medium}) mode will have a daily (weekly) cadence, individual exposures of 1.5 (0.5)~ks, a total exposure time of ${\sim}$13.1 (15.6)~Ms over 12 (300)~deg$^2$, and a single-epoch flux limit of $F_{\rm STAR-X,lim}{\approx}1{\times}10^{-14}$ ($3{\times}10^{-14}$)~erg~cm$^{-2}$~s$^{-1}$; which to avoid Poisson fluctuations we again multiply by 10, yielding expected FXT numbers of ${\approx}$18--30 (12--20). As with eROSITA, we should consider these as upper limits, since the relatively short visits will hinder identifying shorter FXTs and lead to poor characterizations  of FXT X-ray properties, especially for the {\it Medium} survey. On the other hand, the simultaneous UVT observations  should strongly constrain possible simultaneous or delayed lower-wavelength emission.

Finally, the \emph{Advanced X-ray Imaging Satellite} (AXIS) is a {NASA Probe Mission Concept} designed to be the premier high angular resolution X-ray mission of the 2020s (${\sim}$1\farcs0 on-axis and ${\sim}$2\farcs0 at 15\farcm0 off-axis). AXIS will cover an energy range of \hbox{0.2--10}~keV, and have an effective area 5600~cm$^2$ at 1~keV, energy resolution $\sim$150~eV at 6~keV, FoV diameter of 24\farcm0, and detector background 4--5 times lower than \emph{Chandra}. To estimate the expected number of FXTs, we consider an FXT threshold sensitivity of $F_{\rm AXIS,lim}{\approx}$3$\times$10$^{-14}$~erg~cm$^{-2}$~s$^{-1}$ (at 1~ks), producing ${\approx}$\axis~FXTs~yr$^{-1}$.

\section{Conclusions}\label{sec:conclusion}

In this work we searched for extragalactic FXTs present in \emph{Chandra} data from 2014 to 2022. We applied an algorithm developed by \citealp{Yang2019} and \citet[][hereafter Paper~I]{Quirola2022} to X-ray sources with $|b|{>}$10~deg (i.e., 3899 \emph{Chandra} observations, totaling $\approx$88.8~Ms and 264.4~deg$^2$). Considering additional criteria (analyzing further X-ray observations taken by \emph{Chandra}, \emph{XMM-Newton}, \emph{Swift}--XRT, \emph{Einstein,} and \emph{ROSAT}) and other astronomical catalogs (e.g., \emph{Gaia}, NED, SIMBAD, VHS, DES, Pan-STARRS), we identify eight FXTs consistent with an extragalactic origin.
We rediscover all (three) previously reported \emph{Chandra} sources: 
XRT~150322 \citep[previously identified by][]{Xue2019},
XRT~170901 \citep[previously identified by][]{Lin2019,Lin2022},
and XRT~210423 \citep[previously identified by][]{Lin2021}.

We analyzed the timing and spectral properties of this new sample of FXTs. Overall, the X-ray spectra are well-fitted by power-law models with a median slope of $\Gamma{=}$2.6 and an overall range $\Gamma{\approx}$2.1--3.4 (excluding the very soft $\Gamma{\gtrsim}$6.5 outlier XRT~161125). We observe significant spectral softening for FXT~16/CDF-XT2 with time, similar to other sources such as FXT~7/XRT~030511 and FXT~12/XRT~110919 (Paper~I), while FXTs~15 and 20 show similar albeit marginal spectral softening trends. Regarding the X-ray timing properties, the light curves of five FXTs (15, 16, 19, 20, and 22) show broken power-law behavior, of which three FXTs (16, 19, and 22) exhibit plateaus with durations of ${\sim}$3--5~ks, followed by PL decays with slopes ranging from ${\sim}$2.0 to 3.8. Only in the case of FXT~16/CDF-XT2 do we simultaneously see spectral softening coincident with the plateau and decay phase (at 90\% confidence), reinforcing the results obtained by \citet{Xue2019}. Finally, three FXTs (FXTs~17, 18, and 21) show simple power-law decays in their light curves.

We compute an event rate for the eight FXTs analyzed in this work of $\mathcal{R}_{\rm This\ work}{=}$45.6$_{-14.3}^{+18.2}$~deg$^{-2}$~yr$^{-1}$. If we also consider the nine FXTs  classified as ``distant" (i.e., ${\gtrsim}$100~Mpc) from Paper~I, the combined event rate is $\mathcal{R}_{\rm Total}{=}$36.9$_{-8.3}^{+9.7}$~deg$^{-2}$~yr$^{-1}$.

Additionally, we constructed the X-ray luminosity function (XLF) in the range from 10$^{44}$ to 10$^{47.5}$~erg~s$^{-1}$, the first of its kind. The XLF clearly shows that the FXT volumetric rate decreases with increasing X-ray luminosity. A power-law model describes this trend with best-fit slopes of $-0.26{\pm}0.13$, considering just eight FXTs with known redshift, or $-0.57{\pm}0.11$ ($-1.13{\pm}0.27$) considering 17 FXTs with known + fiducial redshifts of $z{=}$1.0 (0.5). Finally, we derive the volumetric rate based on the XLF (sources from Paper~I and this work), which covers a range of ${\sim}1.9{\times}10^3-4.6{\times}10^4$~Gpc$^{-3}$~yr$^{-1}$ in the redshift range of $z{\approx}0-2.2$. These values are in good agreement with the values derived by Paper~I and \citet{Xue2019} at similar redshifts ($z_{\rm max}{\approx}$2.1 and 1.9, respectively), and appear broadly consistent with several other transients classes (LL-LGRBs, LGRBs, SGRBs, and IMBH TDEs) across a broad redshift range.

Six FXTs are associated with optical and NIR extended sources; however, only five (FXTs~16, 18, 19, 21, and 22) are sufficiently bright to derive galaxy properties using photometric archival data (at least four photometric points). For FXT~20, its potential host galaxy is detected weakly in just two photometric bands, which does not allow us to derive host properties. The host galaxies appear to cover a wide range in redshift ($z_{\rm phot/spec}{\approx}$0.3--1.5), stellar mass ($M_*{\approx}$10$^{7.9}$--10$^{11}$~$M_\odot$), and SFR (${\approx}$0.2--35~$M_\odot$~yr$^{-1}$). At the assumed distances, the peak luminosities, energetics, and spectro-temporal properties for all five sources robustly rule out an SBO origin, but potentially remain consistent with origins as on-axis LL-LGRBs, off-axis GRBs, or IMBH-WD TDEs.

For the three FXTs (FXTs~15, 17, and 20) without optical and NIR host detections, interpretations are broader and less clear. An SBO scenario remains possible at low redshifts, $z{\lesssim}$0.4, as long as potential hosts are extremely low-mass, low-SFR dwarf galaxies. Nevertheless, at fiducial redshifts of ${\approx}$1.0, an SBO association is ruled out due to their high estimated X-ray luminosities ($L_{\rm X,peak}{\gtrsim}$10$^{44}$~erg~s$^{-1}$). A highly off-axis GRB scenario, similar to SN~2020bvc ($L_{\rm X,peak}{\sim}$10$^{41}$~erg~s$^{-1}$) or GRB~170817A ($L_{\rm X,peak}{\sim}$10$^{39}$~erg~s$^{-1}$), does not appear viable due to the relatively low expected redshifts $z{\lesssim}$0.02. However, off-axis GRBs afterglow (showing a rise of 1--$10^4$~s, before reaching a peak luminosity, followed by an afterglow consistent with the on-axis GRB trend) under a small range of viewing angles (from 0 to 8.6~deg) could match the light curves of some FXTs. An on-axis GRB scenario is possible at high redshifts ($z{\gtrsim}$1.0) and naturally explains the non-detection of faint host galaxies by existing optical and NIR facilities. However, their light curves at early times look inconsistent with on-axis X-ray afterglows, and the lack of gamma-ray detection is a weakness in this interpretation. Just the LL--LGRB scenario at moderate-high redshift could explain the non-detection of faint hosts and the lack of gamma-ray counterparts. Finally, an unbeamed IMBH-WD TDE scenario is possible only up to a redshift of $z{\approx}$0.14 (assuming a luminosity of $L_{\rm X,peak}{\sim}$10$^{43}$~erg~s$^{-1}$ such as TDE~J2150-0551). To reach higher luminosities beyond a fiducial redshift of $z{\approx}$1.0 ($L_{\rm X,peak}{\gtrsim}$10$^{45}$~erg~s$^{-1}$) requires a strongly beamed TDE scenario. Unfortunately, the few counts and the lack of host and EM counterparts do not permit us to analyze this scenario in detail.

All the above, together with the broad range of properties, suggests that this novel population of FXTs has a mix of origins.

The eight FXT candidates discovered or re-discovered in this work and the previous 14 sources from Paper~I establish a novel sample of sources that opens a new window into the poorly explored world of X-ray transients. Unfortunately, the lack of well-determined distances and host properties leaves many questions about their nature unanswered. Given that so few FXTs have firm host detections and distances, concerted resources are needed to identify and follow up their associated host galaxies through photometric and spectroscopic techniques, in order to place extragalactic FXTs in an appropriate cosmic context compared to previous well-studied transients. Moreover, the lack of simultaneous detections across the electromagnetic spectrum has thus far severely limited our understanding of their emission process and progenitor channels. It is not only important to increase the number of detected FXTs, but also to improve efficient strategies for (onboard) detection and alert generation to trigger follow-up campaigns while the FXTs are still active in X-rays and likely other wavelengths. Future narrow and wide-field missions such as \emph{Athena}, \emph{STAR-X}, and \emph{EP} will enhance our detection capabilities and potential for alerts to follow-up in other energy bands. In contrast, missions such as \emph{AXIS} will allow us to accurately catch transient positions to identify host galaxies and offset distances. 

We leave as future work (Quirola-Vasquez et al. in prep.) an account of the ongoing efforts to acquire and analyze imaging and spectroscopy at optical and NIR wavelengths to identify the host galaxies of FXTs and thereby constrain their energetics and host properties.

\begin{acknowledgements}

We acknowledge support from: ANID grants Programa de Capital Humano Avanzado folio \#21180886 (J.Q--V), CATA-Basal AFB-170002 (J.Q--V, F.E.B.), 
FONDECYT Regular 1190818 (F.E.B.), 1200495 (F.E.B.)
and Millennium Science Initiative ICN12\_009 (J.Q--V, F.E.B.); this project was (partially) funded by NWO under grant number 184.034.002 (P.G.J.);
NSF grant AST-2106990 and \emph{Chandra} X-ray Center grant GO0-21080X (W.N.B.); the National Natural Science Foundation of China grant 11991053 (B.L.); support from NSFC grants 12025303 and 11890693 (Y.Q.X.); support from the George P.\ and Cynthia Woods Mitchell Institute for Fundamental Physics and Astronomy at Texas A\&M University, from the National Science Foundation through grants AST-1614668 and AST-2009442, and from the NASA/ESA/CSA James Webb Space Telescope through the Space Telescope Science Institute, which is operated by the Association of Universities for Research in Astronomy, Incorporated, under NASA contract NAS5-03127 (G.Y.). 
The scientific results reported in this article are based on observations made by the \emph{Chandra} \hbox{X-ray} Observatory. This research has made use of software provided by the \emph{Chandra} \hbox{X-ray} Center (CXC).
This research uses services or data provided by the Astro Data Lab at NSF's National Optical-Infrared Astronomy Research Laboratory. NOIRLab is operated by the Association of Universities for Research in Astronomy (AURA), Inc. under a cooperative agreement with the National Science Foundation.

\end{acknowledgements}


\bibliography{mnras_template.bbl}



\begin{appendix} 

\section{Spatial location and duration of X-ray sources}

\begin{figure*}
    \centering
    \includegraphics[scale=0.8]{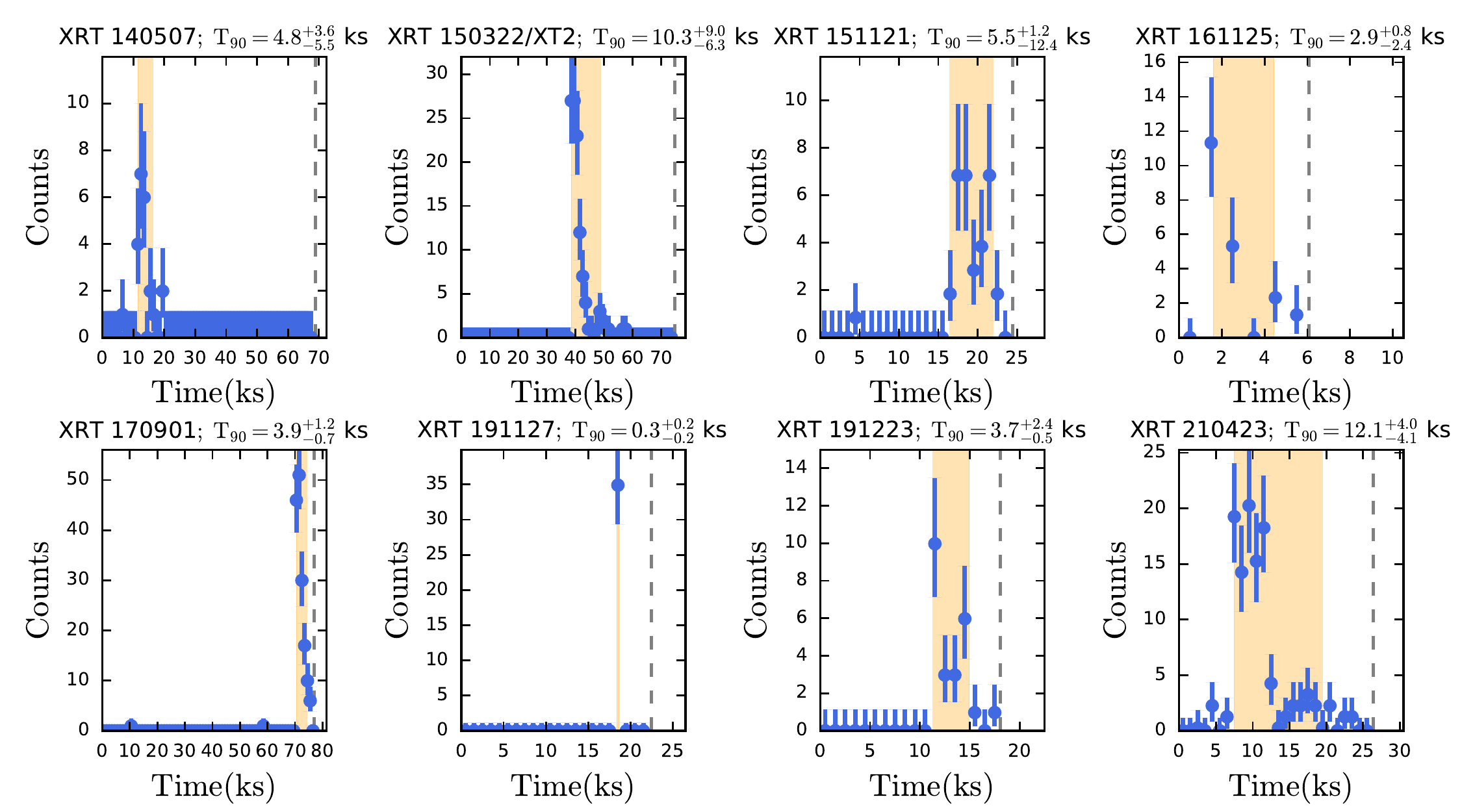}
    \vspace{-0.2cm}
    \caption{
    FXT 0.5--7.0~keV light curves in units of counts in 1~ks bins. The $T_{90}$ duration for each source is denoted by the \emph{orange region}, and listed above. The \emph{gray dashed line} represents the end of the \emph{Chandra} observation.}
    \label{fig:duration_t90}
\end{figure*}

We estimate the duration of the FXTs using the $T_{90}$ parameter, which measures the time over which the source emits from 5\% to 95\% of its total measured counts (in the 0.5--7.0~keV band in our case). Figure~\ref{fig:duration_t90} shows the 0.5--7.0~keV light curves in unit of counts with 1~ks bins. The $T_{90}$ duration for each source is shown as the \emph{orange region}.

\begin{figure*}
    \centering
    \includegraphics[width=17cm,height=22cm]{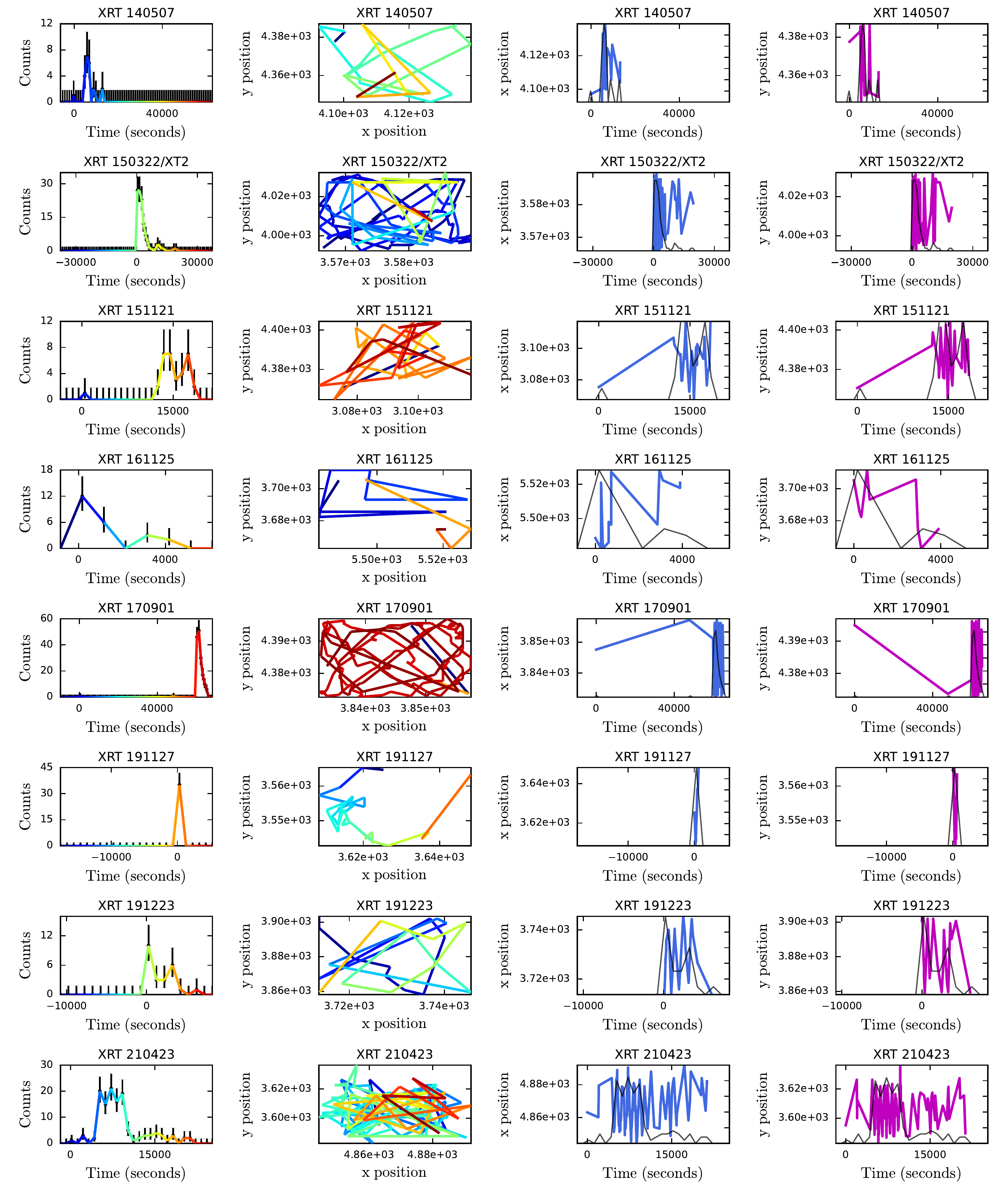}
    \vspace{-0.3cm}
    \caption{Lissajous dither pattern in detector coordinates. \emph{First column:} FXT 0.5--7.0~keV light curves in count units, colour-coded as a function of time. 
    \emph{Second column:} \emph{Chandra} 0.5--7.0~keV images in detector coordinates, with the same colour-coding as a function of time, demonstrating the temporal movement of the source on the detector in response to the Lissajous dither pattern. A flaring pixel would appear as a point on these plots. \emph{Third and fourth columns:} x (\emph{blue}) and y (\emph{purple}) detector coordinates, respectively, of the detected X-ray photons from the FXTs as a function of time, with the candidate light curves superimposed as solid dark gray lines.}
    \label{fig:lissajaus}
\end{figure*}

Furthermore, Fig.~\ref{fig:lissajaus} confirms that the final sample of FXT candidates are real celestial sources in the sky rather than detector artifacts. Due to \emph{Chandra}'s Lissajous dither pattern, executed during observation, the X-ray photons of the FXTs are distributed over dozens to hundreds of individual pixels on the detector. The \emph{first column} of the figure shows the light curves, colour-coded by the phase in the light curve evolution. The \emph{second column} shows the spatial location in $x$ and $y$ chip detector coordinates, also colour-coded by time, tracing out a sinusoidal-like evolution in $x$ and $y$ coordinates over time. The \emph{third and fourth columns} show the $x$ and $y$ position changes (in \emph{blue} and \emph{purple}, respectively, over time, with the light curve superimposed in dark gray.

\begin{table*}
\centering
\begin{tabular}{ccc}
\hline\hline
FXT & Visible & Instruments \\ \hline
 (1) & (2) & (3) \\ \hline
15 & Yes & $n6$, $n7$, $n9$, $b1$, $nb$ \\
16 & Yes & $n4$, $n5$, $b0$ \\
17 & Yes & $n6$, $n7$, $n8$, $n9$, $b1$, $nb$ \\
18 & No & -- \\
19 & No & -- \\
20 & Yes & $n3$, $n4$, $n5$, $b0$ \\
21 & Yes & $n6$, $n7$, $n8$, $n9$, $b1$, $nb$ \\
22 & No & -- \\ 
 \hline
\end{tabular}
\caption{Visibility of the FXTs by the \emph{Fermi}-GBM instruments. \emph{Column 1:} FXT number. \emph{Column 2:} visibility of the sources (if they are behind the Earth) around the X-ray trigger ($T_0{\pm}50$~s). \emph{Column 3:} \emph{Fermi}-GBM instruments covering the field of sources around the X-ray trigger time ($T_0{\pm}50$~s) at a distance of $\lesssim$60~degrees.}
\label{tab:GBM}
\end{table*} 


\section{Forced photometry upper limits}

From the flux density measurements of ZTF and ATLAS, we derive upper limits closer to the X-ray trigger time for FXTs~17, 18, 19, 20, 21, and 22. We compute the upper limits AB magnitude by taking 3 times the uncertainty from the forced photometry flux density. The AB magnitude upper limits inferred from the closest observation in time to the transient are given in Table~\ref{tab:upper_limits} for the available FXTs and filters.

\begin{table*}
\centering
\begin{tabular}{cccccccc}
\hline\hline
FXT & Time & ZTF filters &  ZTF & ${\sim}\Delta T_{\rm ZTF}$ & ATLAS filters & ATLAS & ${\sim}\Delta T_{\rm ATLAS}$ \\
 & (MJD) & & (3$\sigma$) & (days) & & (3$\sigma$) & (days) \\ \hline
 (1) & (2) & (3) & (4) & (5) & (6) & (7) & (8) \\ \hline
17 & 57347 & -- & -- & -- & $o$ & ${>}20.3$ & $15.3$ \\
18 & 57717 & -- & -- & -- & $o$ & ${>}19.3$ & $18.4$ \\
19 & 57997 & -- & -- & -- & $c,o$ & ${>}20.8$,${>}19.1$ & $15.5$,$1.5$ \\
20 & 58813 & $g$,$r$,$i$ & ${>}17.8$,${>}18.3$,${>}19.1$ & $15.5$,$15.5$,$14.5$ & $c$,$o$ & ${>}20.0$,${>}20.5$ & $2.6$,$8.6$ \\
21 & 58840 & $g$,$r$ & ${>}19.4$,${>}19.2$ & $6.1$,$5.3$ & $c$,$o$ & ${>}21.2$,${>}21.1$ & $4.4$,$5.4$ \\
22 & 59327 & $g$,$r$ & ${>}18.1$,${>}18.4$ & $1.0$,$1.0$ & $c$,$o$ & ${>}20.5$,${>}20.0$ & $12.4$,$2.3$ \\
 \hline
\end{tabular}
\caption{Simultaneous counterpart 3$\sigma$ upper limits (AB mag) inferred from forced photometry. \emph{Columns 1 and 2:} FXT number and X-ray trigger time, respectively. \emph{Columns 3, 4 and 5:} ZTF filters used, 3$\sigma$ upper limit per filter (if available), and detection time from the X-ray trigger, respectively. \emph{Columns 6, 7, and 8:} Atlas filters used, 3$\sigma$ upper limit per filter, and detection time from the X-ray trigger, respectively.}
\label{tab:upper_limits}
\end{table*} 


\section{Host-galaxy SED fitting}\label{app:host_galaxy}

\begin{figure*}
    \centering
    \includegraphics[scale=0.4]{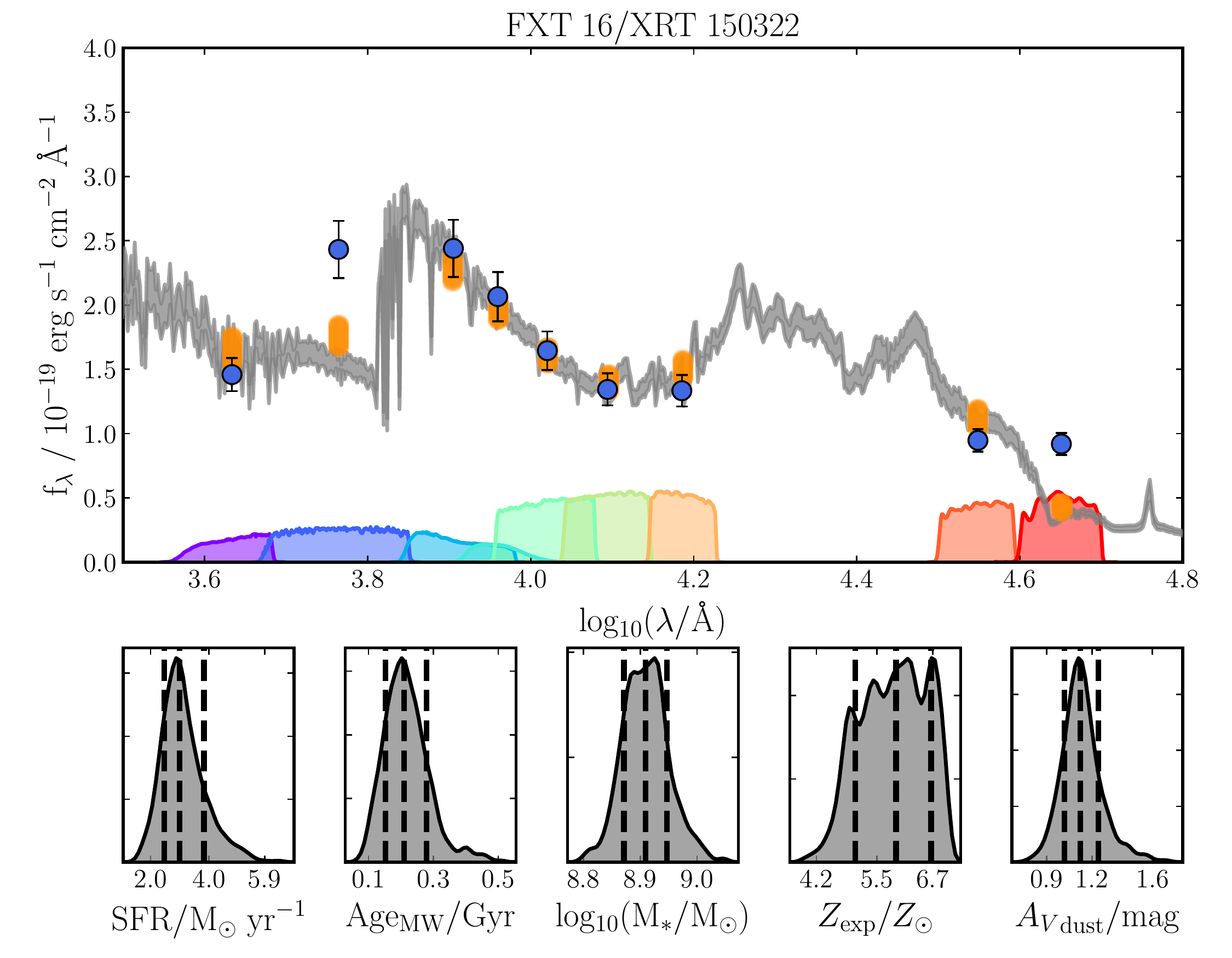}
    \includegraphics[scale=0.4]{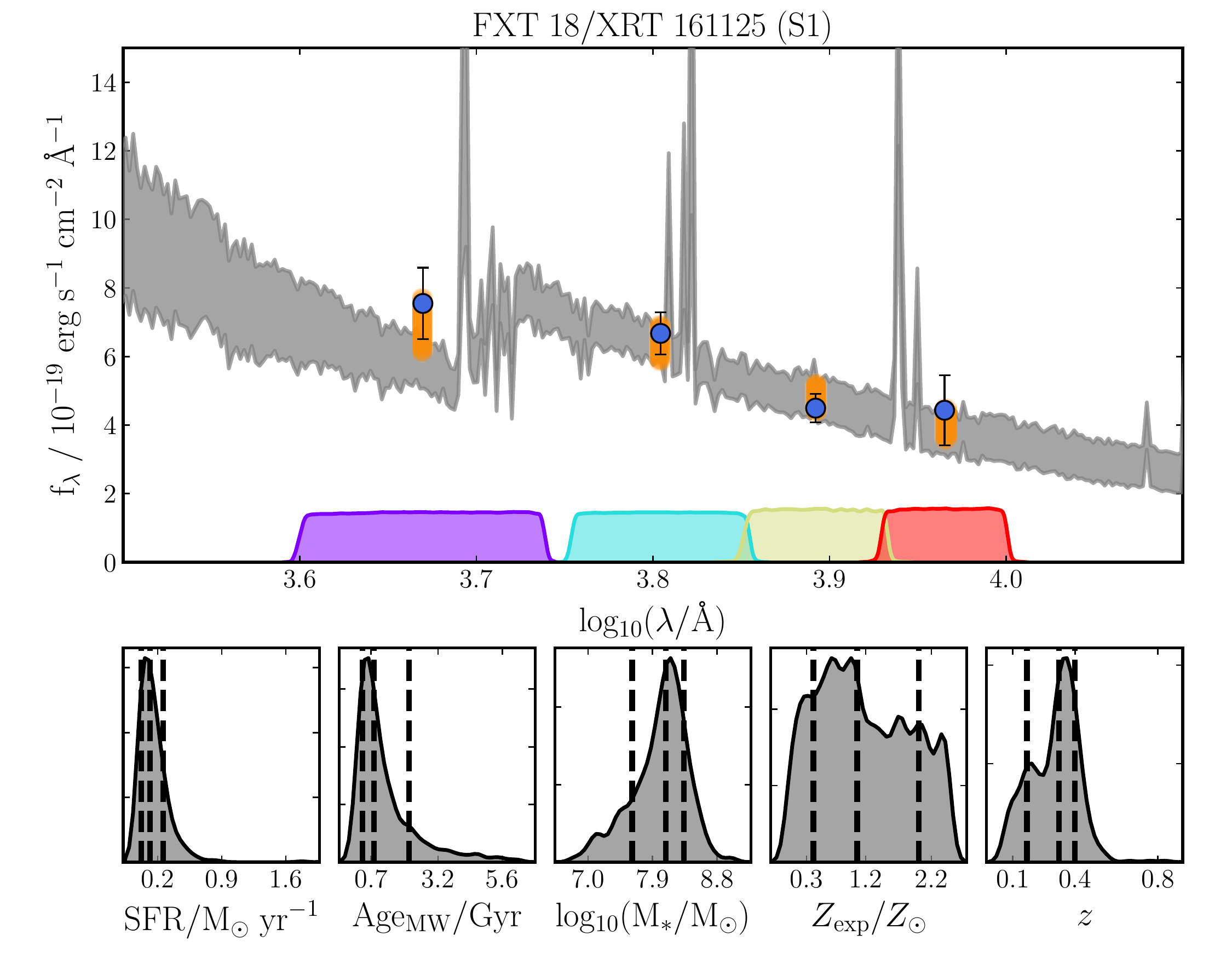}
    \includegraphics[scale=0.4]{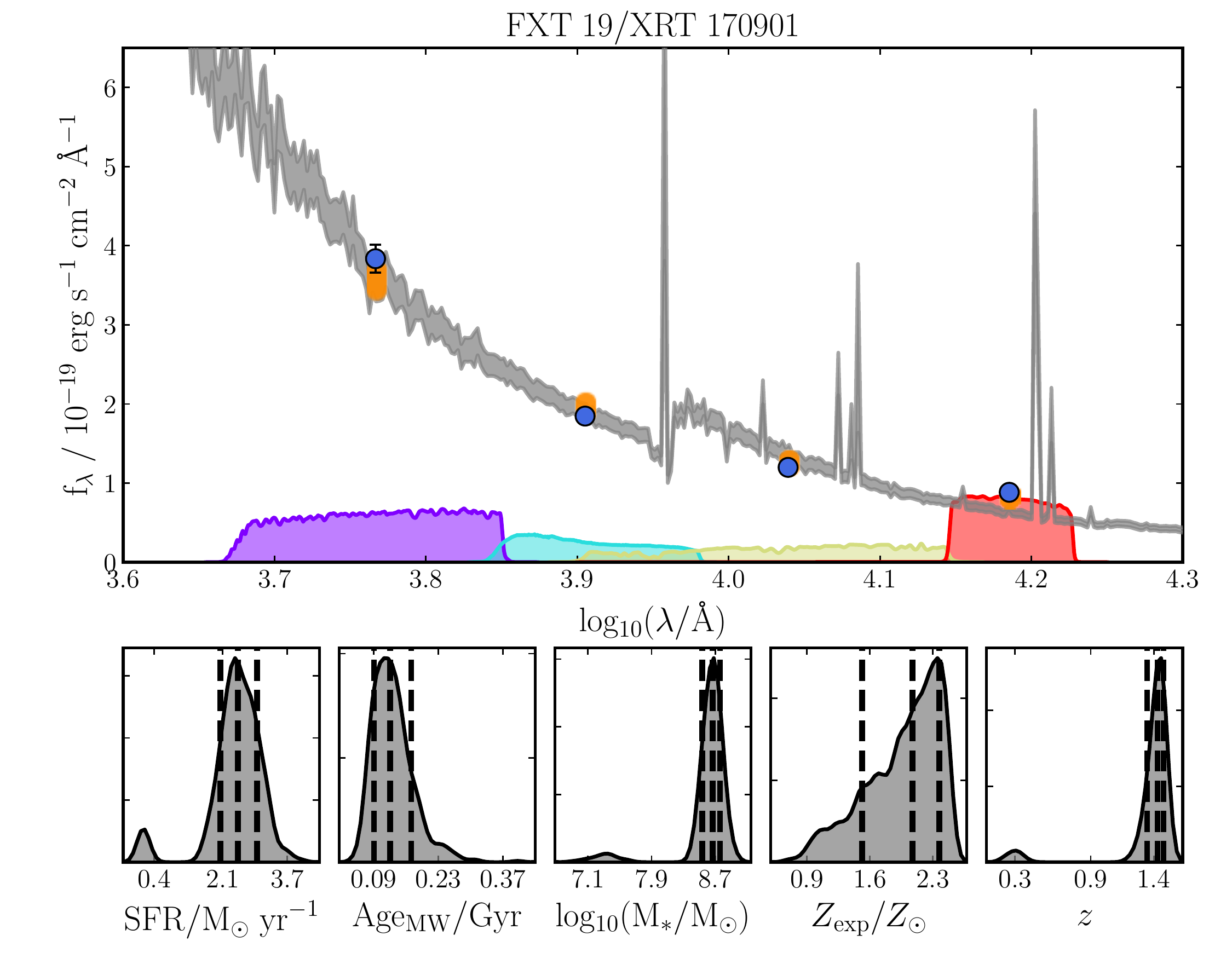}
    \includegraphics[scale=0.4]{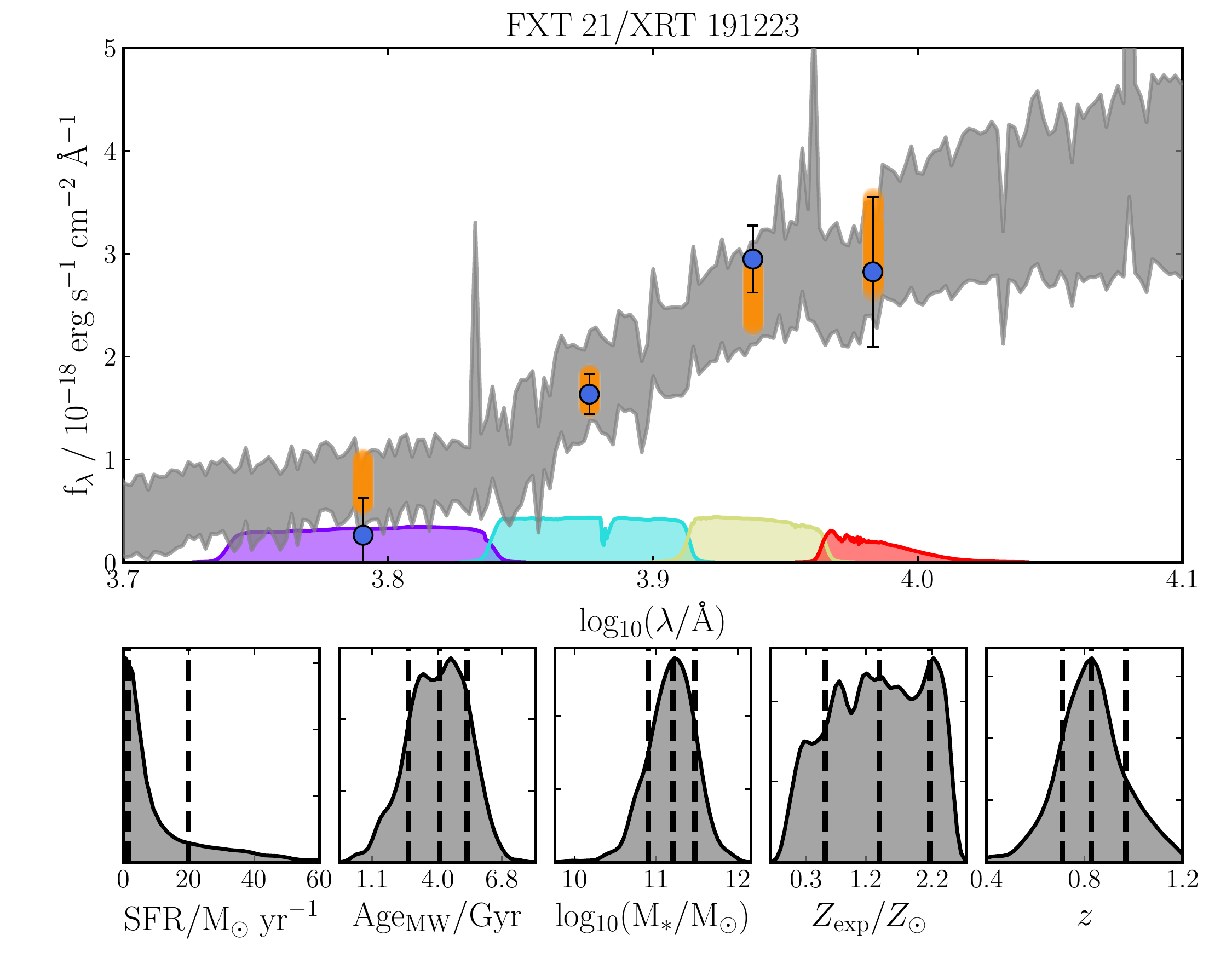}\\
    \includegraphics[scale=0.4]{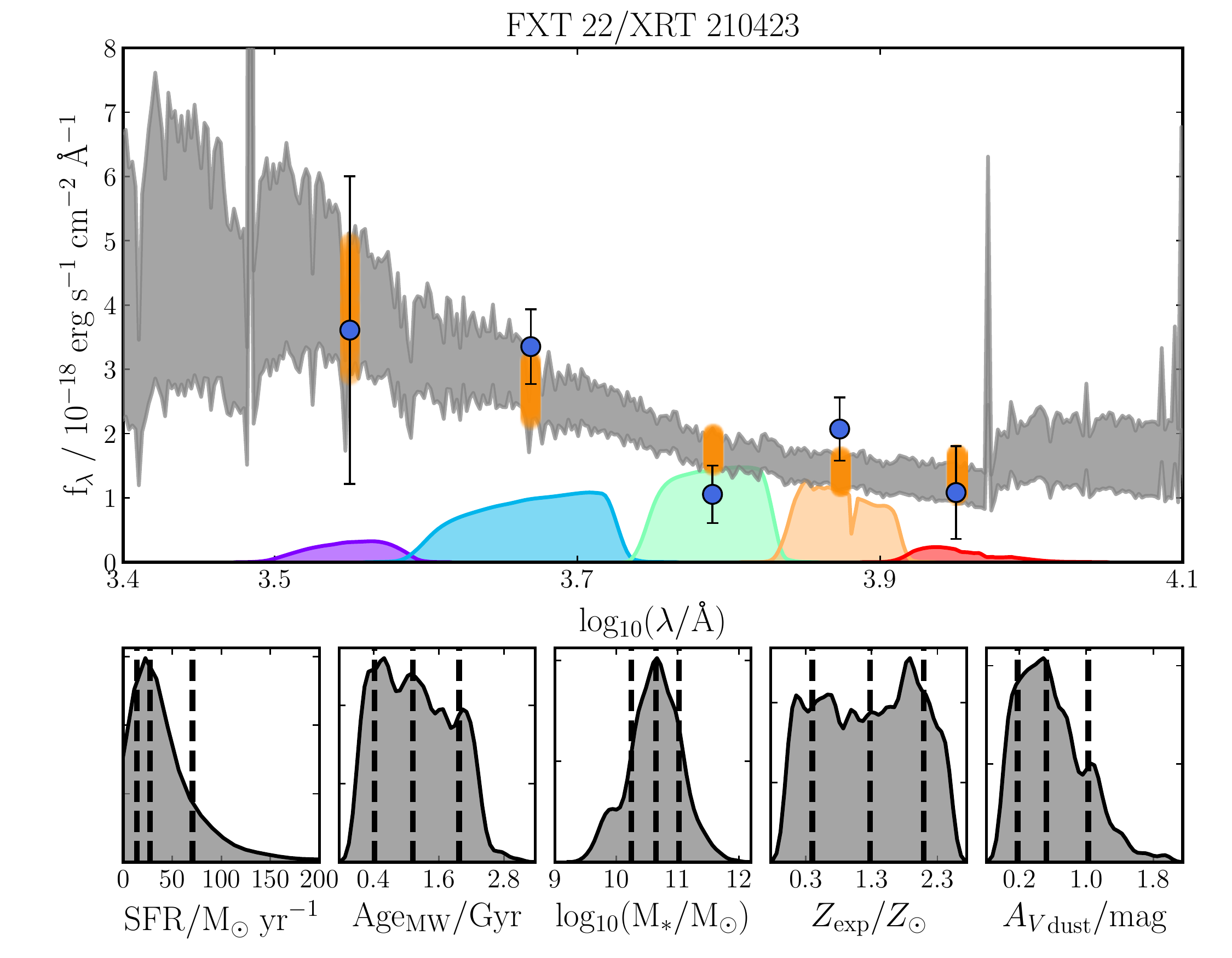}\\
    \vspace{0.1cm}
    \caption{\emph{Top panels:} The best-fitting SED model obtained from \texttt{BAGPIPES} \citep{Carnall2018} for each source \#1 associated with FXTs~16, 18, 19, 21 and 22, and the relative transmission functions of the different filters used in the fitting process (\emph{coloured curves}). The 16th to 84th percentile range for the posterior spectrum (shaded \emph{gray} region) and predicted photometry points (\emph{orange} markers) are shown. The actual photometric data and their uncertainties are given by the \emph{blue markers}.
    \emph{Bottom panels:} the posterior distributions for the five fitted parameters (star-formation rate, age, galaxy stellar mass, metallicity, and redshift) are shown. The 16th, 50th, and 84th percentile posterior values are indicated by the vertical \emph{dashed black lines}.}
    \vspace{-0.2cm}
    \label{fig:SED_models}
\end{figure*}

To derive the host galaxy parameters, we used the existing photometry and the spectral energy distribution (SED) models from the \texttt{BAGPIPES} package \citep[Bayesian Analysis of Galaxies for Physical Inference and Parameter EStimation;][]{Carnall2018}. It fits broadband photometry and spectra with stellar-population models taking star-formation history and the transmission function of neutral/ionized ISM into account via a \texttt{MultiNest} sampling algorithm \citep{Feroz2008,Feroz2009}. \texttt{BAGPIPES} provides posterior distributions for the host-galaxy redshift ($z$), age, extinction by dust ($A_V$), star-formation rate (SFR), metallicity ($Z$), stellar mass ($M_*$), and specific star formation rate. To fit the SEDs, we consider a star-formation history (SFH) described by an exponentially decreasing function with a timescale parameter $\tau$ \citep[which is probably the most commonly applied SFH model;][]{Simha2014,Carnall2019}. The models implemented within \texttt{BAGPIPES} are constructed using an \citet{Kroupa2002} initial mass function (IMF). To model the dust attenuation in the SEDs, we used the theoretical framework developed by \citet{Calzetti2000}, where $A_V$ is a free parameter within the range of 0.0 to 3.0~mag. For the fitting process, we assumed an exponentially declining star formation history function parametrized by the star formation timescale (free parameter). Fig.~\ref{fig:SED_models} shows the 16th to 84th percentile range for the posterior spectrum, photometry, and the posterior distributions for five fitted host-galaxy parameters.

Below, we describe their most important properties and features one by one:

FXT~16 (CDF-S~XT2) is associated with a $z_{\rm spec}{=}0.738$ host galaxy with a relatively flat SED. Fitting its photometry with \texttt{BAGPIPES} at the known redshift, we find that the host galaxy appears to have a low stellar mass and modest star formation rate, consistent with the ones reported in the literature.

For FXT~18, SED fits of the photometric data indicate that the host galaxy has low stellar mass, moderate age, with a low star formation rate and a photometric redshift of ${\approx}0.35$.

The field of FXT~19 was observed by \emph{HST} on 2014-07-10 (three years before the X-ray trigger) in the F606W, F814W, F110W, and F160W filters. The blue host appears to have a low stellar mass, modest star formation rate, and a photometric redshift of $z_{\rm phot}{=}$1.44$\pm$0.08. This host galaxy was observed with the \emph{WFC3} G102 and G141 grisms from the \emph{WFC3 Infrared Spectroscopic Parallel (WISP) Survey}, but unfortunately, no significant features are detected in the 2D spectra, aside from strong contamination due to the zero-order spectra of a neighboring star \citep{Atek2010,Budavari2012,Whitmore2016}.

FXT~20 only has faint $g$, $r$ and $z$-band DECam detections, which are too few and too loosely constrained to compute an SED photometric redshift. 

For FXT~21, SED fits of the photometric data indicate that the host galaxy is likely massive and relatively old, with a highly uncertain star formation rate and a photometric redshift of 0.85$\pm$0.14.

Finally, FXT~22 is located near the extended $z_{\rm spec}{=}1.5105$ galaxy SDSS J134856.75+263946. SED fitting of the photometric data at the known redshift indicates that the host is likely a massive post-starburst galaxy, with a large but highly uncertain star formation rate.

\begin{table*}
    \centering
    \scalebox{0.95}{
    \begin{tabular}{llllll}
    \hline\hline
    FXT & $z$ & $F_{\rm peak}$ & $L_{\rm X,peak}$ & $M_{\rm Edd}$ & $E_X^{\rm iso}$  \\ 
     &  & (erg~cm$^{-2}$~s$^{-1}$) & (erg~s$^{-1}$) & ($M_\odot$) & (erg)  \\\hline
    (1) & (2) & (3) & (4) & (5) & (6) \\ \hline
    15 & 1.0 & (1.9$\pm$0.9)${\times}$10$^{-13}$ & (1.0$\pm$0.5)${\times}$10$^{45}$ & (8.1$\pm$3.8)${\times}$10$^{6}$ & 7.5${\times}$10$^{47}$ \\
    16 & 0.738 & (1.1$\pm$0.2)${\times}$10$^{-12}$ & (2.8$\pm$0.6)${\times}$10$^{45}$ & (2.2$\pm$0.5)${\times}$10$^{7}$ & 3.6${\times}$10$^{48}$ \\
    17 & 1.0 & (1.2$\pm$0.5)${\times}$10$^{-12}$ & (6.3$\pm$2.6)${\times}$10$^{45}$ & (5.0$\pm$2.1)${\times}$10$^{7}$ & 3.4${\times}$10$^{48}$ \\
    18 & 0.35 & (4.5$\pm$1.8)${\times}$10$^{-10}$ & (1.9$\pm$0.8)${\times}$10$^{47}$ & (1.5$\pm$0.6)${\times}$10$^{9}$ & 1.7${\times}$10$^{50}$ \\
    19 & 1.44 & (2.7$\pm$0.5)${\times}$10$^{-12}$ & (3.7$\pm$0.7)${\times}$10$^{46}$ & (2.9$\pm$0.6)${\times}$10$^{8}$ & 1.3${\times}$10$^{49}$ \\
    20 & 1.0 & (1.5$\pm$0.6)${\times}$10$^{-11}$ & (8.1$\pm$3.1)${\times}$10$^{46}$ & (6.4$\pm$2.4)${\times}$10$^{8}$ & 3.2${\times}$10$^{48}$ \\
    21 & 0.85 & (1.9$\pm$0.9)${\times}$10$^{-12}$ & (6.9$\pm$3.3)${\times}$10$^{45}$ & (5.5$\pm$2.6)${\times}$10$^{7}$ & 1.8${\times}$10$^{48}$ \\
    22 & 1.5105& (8.4$\pm$1.8)${\times}$10$^{-13}$ & (1.3$\pm$0.3)${\times}$10$^{46}$ & (1.0$\pm$0.2)${\times}$10$^{8}$ & 1.1${\times}$10$^{49}$ \\
    \hline
    \end{tabular}
    }
    \caption{Energetics of the FXT sample.
    \emph{Column 2:} Redshift taken.
    \emph{Column 3 and 4:} X-ray peak flux and isotropic luminosity in cgs units. Fluxes are corrected for Galactic and intrinsic absorption and calculated over the energy range 0.3--10~keV. Redshifts are taken from Table~\ref{tab:SED_para} or assumed to be $z{=}1$ (denoted by $\dagger$ in Column 2). 
    \emph{Column 5:} Eddington mass (defined as $M_{\rm Edd}{=}7.7{\times}10^{-39} L_{\rm X,peak}$) in solar mass units ($M_\odot$).
    \emph{Column 6:} Isotropic energy (computed from integrating the light curves) in cgs units.}
    \label{tab:F_L_para}
\end{table*}


\section{Comparison with Paper~I FXTs}

\begin{figure*}
    \centering
    \includegraphics[scale=0.85]{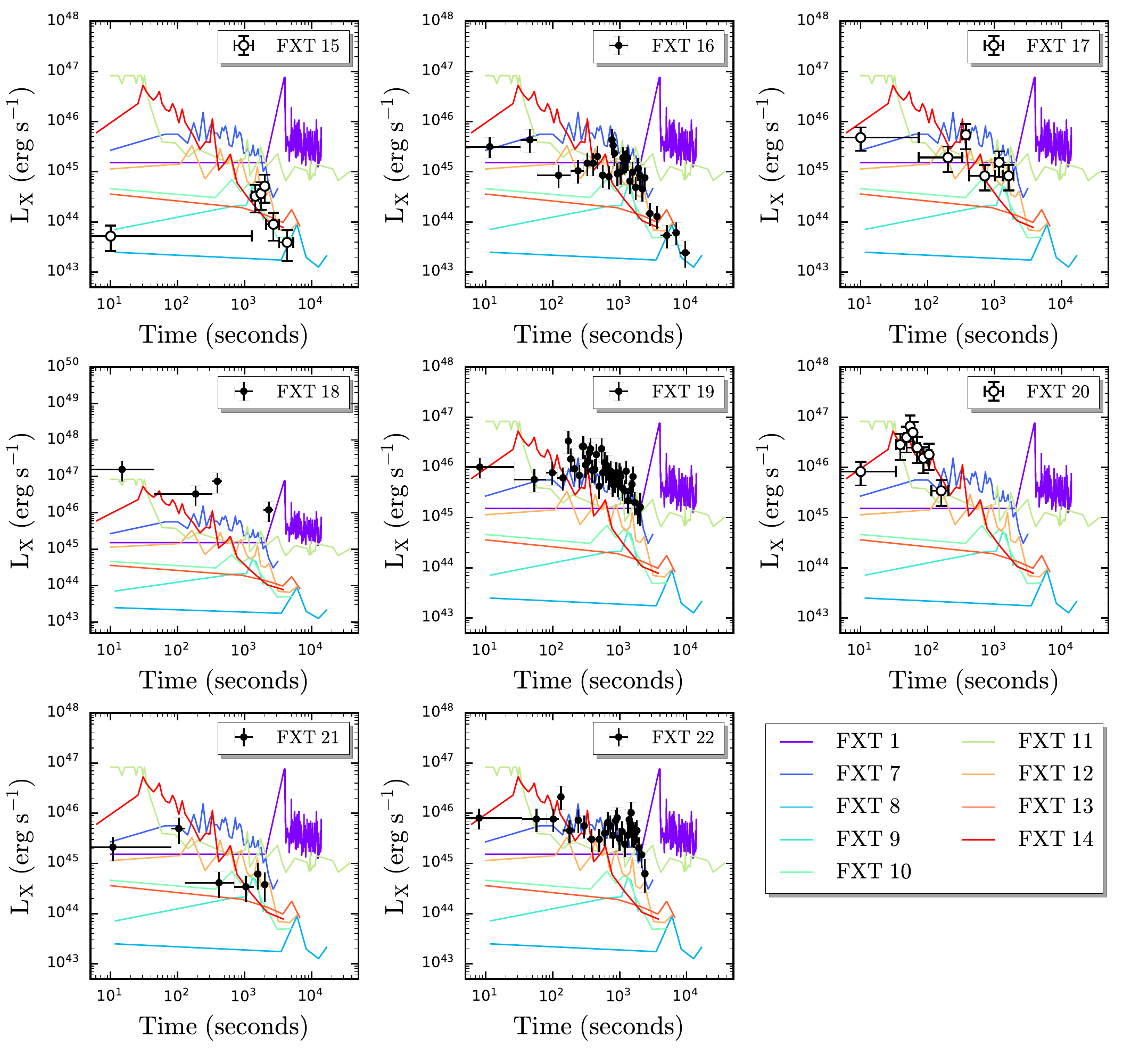}
    \vspace{-0.3 cm}
    \caption{Light curves of the eight FXTs in 0.3--10 keV luminosity units (converted from 0.5--7 keV light curves assuming best-fit spectral models in Sect.~\ref{sec:X-ray_fitting}). Light curves of distant FXTs from Paper~I are shown for comparison. We adopt the redshifts listed in Table~\ref{tab:LF_parameters}.}
    \label{fig:flux_comparison_2}
\end{figure*}

\begin{figure}
    \centering
    \hspace*{-0.5cm}
    \includegraphics[scale=0.8]{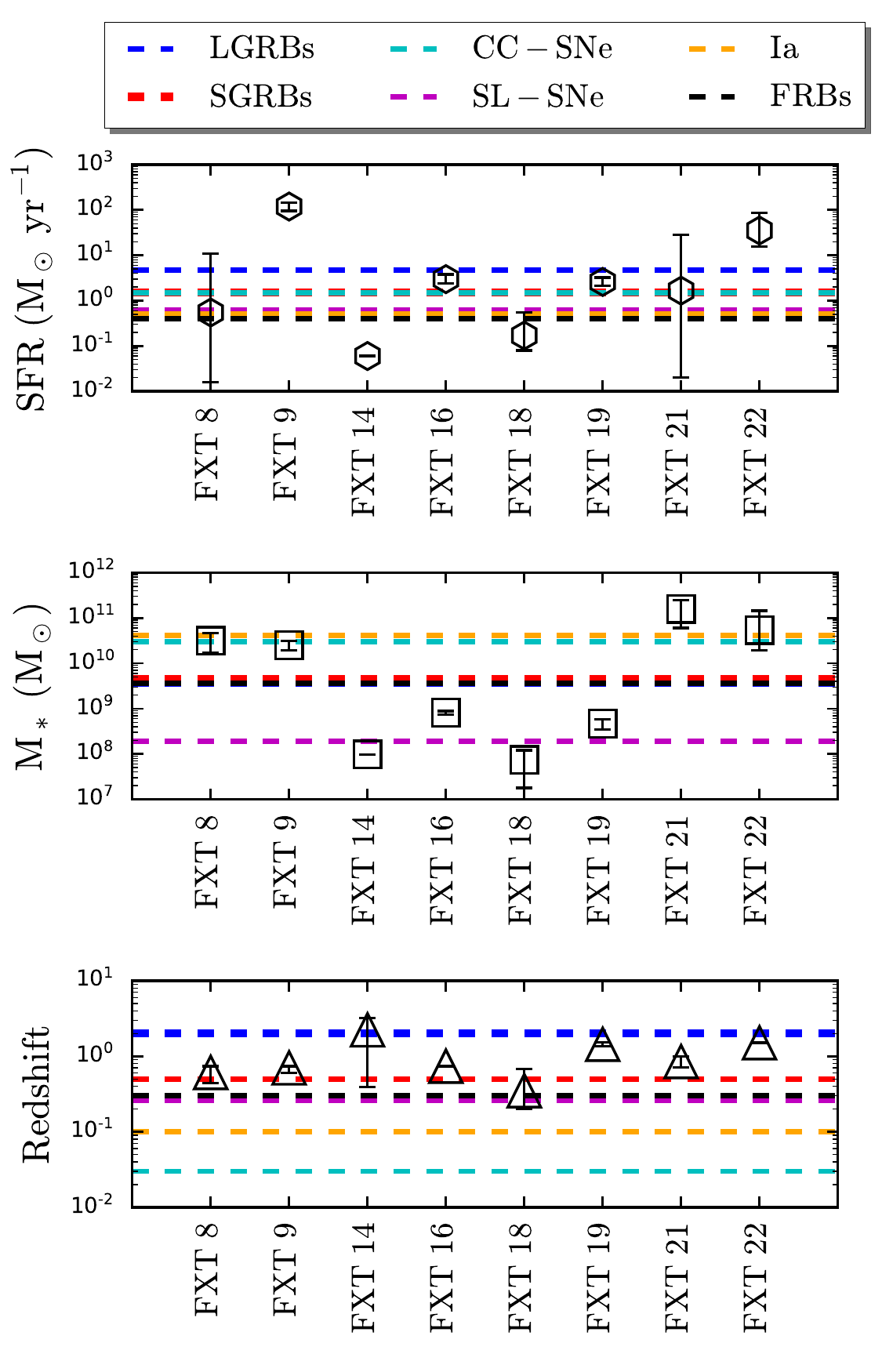}
    \caption{Comparison of the star-formation rates (\emph{top panel}), stellar masses (\emph{middle panel}), and redshifts (\emph{bottom panel}) of FXT hosts identified in Paper~I (FXTs~8, 9 and 14) and this work (FXTs~16, 19, 21 and 22). The mean SFRs, stellar masses, and redshifts from samples of LGRBs (\emph{dashed blue line}), SGRBs (\emph{dashed red line}), CC- (\emph{dashed cyan line}), and Type Ia (\emph{dashed orange line}) SNe, SL-SNe, and FRBs (\emph{dashed black line}) are also plotted \citep{Tsvetkov1993,Prieto2008,Li2016,Galbany2014,Blanchard2016,Heintz2020,Schulze2021,Fong2022,Qin2022}.}
    \vspace{-0.1cm}
    \label{fig:host_parameters_total_sample}
\end{figure}

This new sample of FXTs discovered in this paper shares timing and host-galaxy property similarities with the previous distant FXTs identified in \citet{Quirola2022} (or Paper~I) and other X-ray transients. In that way, Figs~\ref{fig:flux_comparison_2} and \ref{fig:host_parameters_total_sample} compare the light curves of FXTs identified in this work and in Paper~I and the galaxy properties with other transients such as LGRBs, SGRBs, CC-SNe, SNe-Ia, SL-SNe, and FRBs, respectively.

\label{lastpage}
\end{appendix}
\end{document}